%% file: MSPESIII.tex
\documentclass[preprint2]{aastex}
\usepackage{amsmath}
\def\etal{\it et~al.}

\shorttitle{Nulling.}
\shortauthors{Basu \etal}

\begin{document}

\title{Meterwavelength Single-pulse Polarimetric Emission Survey III: The
Phenomenon of Nulling in Pulsars.}

\author{Rahul Basu\altaffilmark{1,*}, Dipanjan Mitra\altaffilmark{2,3,1}, George I. Melikidze\altaffilmark{1,4}}

\altaffiltext{1}{Janusz Gil Institute of Astronomy, University of Zielona G\'ora, ul. Szafrana 2, 65-516 Zielona G\'ora, Poland}
\altaffiltext{2}{National Centre for Radio Astrophysics, Ganeshkhind, Pune 411 007, India}
\altaffiltext{3}{Physics Department, University of Vermont, Burlington VT 05405}
\altaffiltext{4}{Abastumani Astrophysical Observatory, Ilia State University, 3-5 Cholokashvili Ave., Tbilisi, 0160, Georgia}
\altaffiltext{*}{Present address: Inter-University Centre for Astronomy and Astrophysics, Pune 411007, India}
\email{rahulbasu.astro@gmail.com}

\begin{abstract}
\noindent A detailed analysis of nulling was conducted for the pulsars studied in the Meterwavelength Single-pulse Polarimetric
Emission Survey. We characterized nulling in 36 pulsars including 17 pulsars where the phenomena were reported for the first
time. The most dominant nulls lasted for a short duration, less than five periods. The longer duration nulls extending to
hundreds of periods were also seen in some cases. A careful analysis showed the presence of periodicities in the transition from
the null to the burst states in 11 pulsars. In our earlier work fluctuation spectrum analysis showed multiple periodicities in 6
of these 11 pulsars. We demonstrate that the longer periodicity in each case was associated with nulling. The shorter
periodicities usually originate due to subpulse drifting. The nulling periodicities were more aligned with the periodic amplitude
modulation indicating a possible common origin for both. Most prevalent nulling lasts for a single period and can be potentially
explained using random variations affecting the plasma processes in the pulsar magnetosphere. On the other hand, the longer
duration nulls require changes in the pair production processes that need an external triggering mechanism for the change. The
presence of periodic nulling puts an added constrain on the triggering mechanism which also needs to be periodic.
\end{abstract}


\section{\large Introduction} \label{sec:intro}
\noindent The radio emission from pulsars show variability over multiple timescales. These variations can either be due to
intrinsic changes in the emission process or result of external effects like interaction of the emission with the intervening
medium. The presence of nulling was first reported by \citet{bac70} where the emission ceased within one rotation period. Nulling
can manifest over multiple timescales ranging from a few periods to hours at a time and in the case of intermittent pulsars from
weeks to months. The nulling fractions also show wide variations ranging from less than a few percent to more than seventy
percent of time when the emission nulls. There are studies which show that pulsars with similar nulling fractions sometimes
exhibit very different nulling patterns \citep{gaj14a}.

There are around 75 pulsars where nulling has been reported \citep{rit76,ran86,big92,wan07,gaj12}. The physical processes
responsible for the transitions between the null to the burst states are still unclear. The long duration nulls in the
intermittent pulsar B1931+24 have been identified with large changes in the spin down energy loss \citep{kra06}. This has
motivated the idea of nulling being associated with large scale magnetospheric changes in pulsars \citep{cor13,gaj14a,gaj14b}.
However, it is not clear if the short and the intermediate nulls lasting few periods (seconds) to hundreds of periods (hours)
have the same physical origin as nulling in the intermittent pulsars which last for days at a time. Nulling is largely considered
to be a stochastic process switching between two states. But, in some cases the transition from the null to the burst states have
shown periodicity \citep{her07,her09}.

The sparking discharges in the Inner Acceleration Region (IAR) of pulsars are believed to generate the plasma necessary for
radio emission \citep[][hereafter RS75]{rud75}. RS75 considered a system where the rotation and magnetic axis were aligned in
opposite directions. The sparks, also called subbeams, were postulated to rotate around the common axis in a steady pattern due
to the {\bf E}x{\bf B} drifting. \citep{des01} expanded the RS75 model to the pulsar B0943+10 where the rotation axis is not
aligned with the magnetic axis. They considered the sub-beam system to rotate around the magnetic axis in order to explain the
subpulse drifting seen in this pulsar. This model has now been used in a large number of pulsars to explain the subpulse drifting
phenomenon. The periodic nulls are associated with the line of sight passing between the empty regions of the sub-beam pattern.
Although, in our recent work \citep{bas16} we found the rotating sub-beam system around the magnetic axis to be physically
inconsistent in non-aligned pulsars. The sparks are instead expected to move around the rotation axis.

The underlying physical processes responsible for nulling are still unresolved. The longer duration nulls allude to unknown
physical phenomena which require constrains from observations as well as detailed modelling. In this work we have looked for
additional observational clues to characterize nulling. We have carried out detailed analysis of nulling properties in the
Meterwavelength Single-pulse Polarimetric Emission Survey \citep[MSPES,][]{mit16}. The paper is organized as follows: in section
\ref{sec:obs} we present the observing details and the analysis schemes used; section \ref{sec:res} gives the results of our
analysis; in sections \ref{sec:per} and \ref{sec:phy} we discuss the implications of our results on the physical processes in
pulsars and finally a summary is presented in section \ref{sec:sum}.

\section{\large Observation and Analysis} \label{sec:obs}
\noindent The observing procedure and initial data processing are detailed in \citet{mit16,bas16}. We have searched for nulling
in 123 pulsars observed at 333 and 618 MHz for roughly 2100 pulses in each case. The pulse energy distributions are represented
as two histograms corresponding to the on and off pulse energies (see figure~\ref{fig1}, bottom left panel). The off pulse
distributions are centered around zero and reflect the noise characteristics of the baseline level. The off pulse distributions
are expected to exhibit noise like characteristics resembling a gaussian function. However, in some cases low level wings are
seen due to the presence of systematics which smear the statistical boundary between null and burst pulses.

\subsection{Determining Nulling behavior}
\begin{figure*}
\begin{center}
\mbox{\includegraphics[angle=0,scale=0.9]{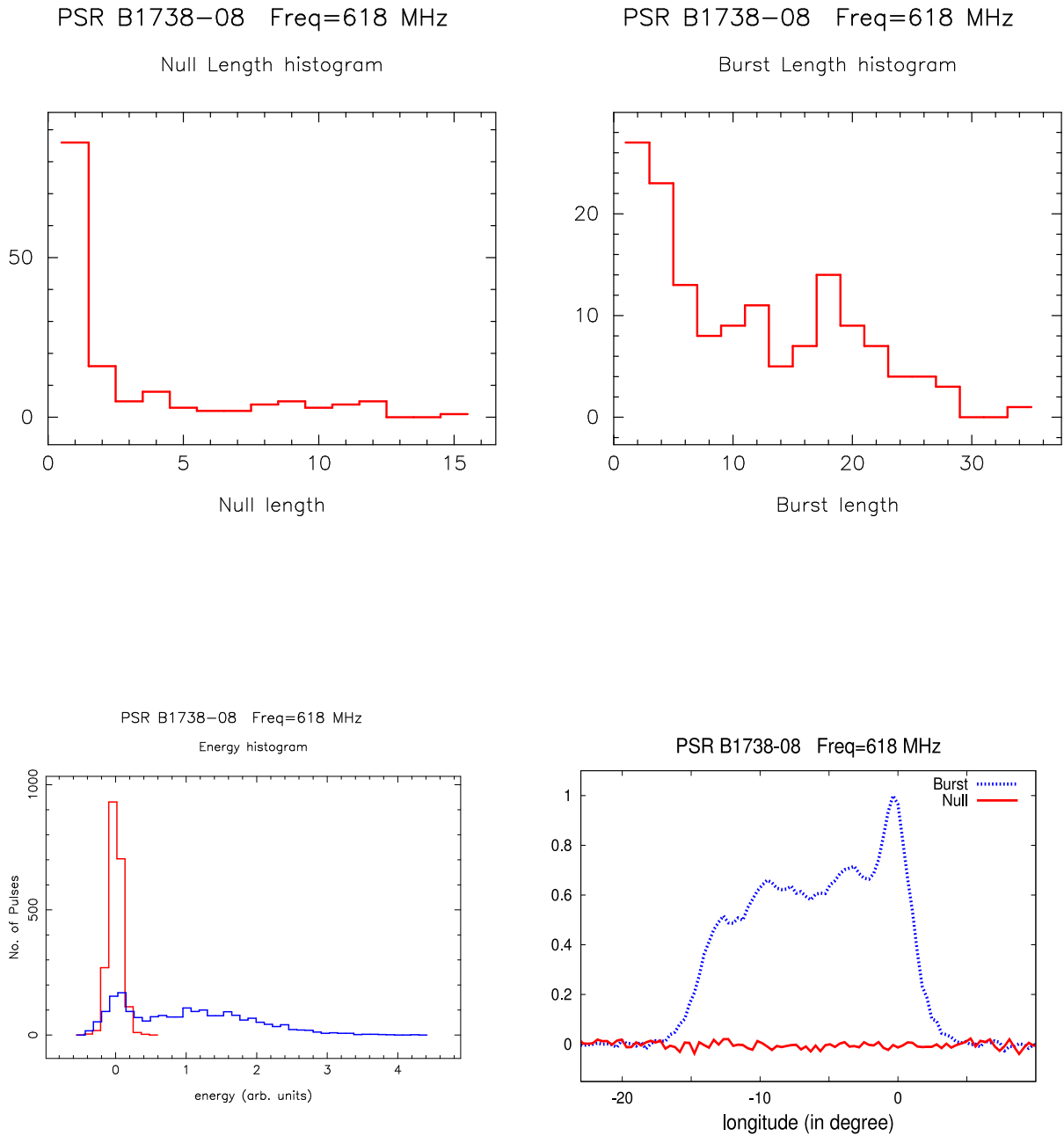}}
\end{center}
\caption{The nulling with well resolved null and burst pulses: Null length
histogram (top left); Burst length histogram (top right); the average energy
distribution (bottom left) for on-pulse (blue histogram) and off-pulse (red
histogram); the folded profile (bottom right) for the null pulses (red line,
noise like characteristics) and burst pulses (blue line).
%(The complete figure set (32 images) is available.)
\label{fig1}}
\end{figure*}

\noindent The nulling was identified when a bimodal shape was seen in the on pulse distribution. The null pulses resembled a
scaled down version of the off pulse distribution and were separated from the burst distribution (see figure~\ref{fig1}, bottom
left panel). In a number of pulsars the detection sensitivities of the single pulses were insufficient to separate out the two.
However, in a few cases averaging 3-5 pulses helped to identify the presence of nulling. The nulling behavior is usually
characterized by the nulling fraction which signifies the fraction of time the pulsar is in the null state. We have determined
the nulling fraction as follows. A gaussian functional fit was estimated for the off pulse distribution which was then scaled
appropriately to the nulling part of the on pulse distribution. The ratio of the two gaussian peaks gave the measured value of
the nulling fraction. In a few cases the entire nulling durations were too small to form well constrained distributions. The
nulling fraction in these cases were estimated by counting the individual null pulses.

\subsection{Resolved Nulls: Estimating Periodicity}
\noindent We carried out additional analysis when the null and the burst pulses were well separated. The 3$\sigma$ noise level of
the off-pulse window was initially considered the boundary between the two states. Each single pulse was identified either as the
null or the burst pulse based on the above boundary. In the next phase all single pulses around the boundary were inspected to
the correct any weak emission identified as nulls. At this stage the null and the burst pulses were separately averaged to form
respective profiles (see figure~\ref{fig1}, bottom right panel). Further analysis were carried out if no significant emission was
detected in the null profile. The null and burst lengths were determined during each transition along with the total number of
transitions between the states. Finally, the null and burst length histograms were estimated (see figure~\ref{fig1}, top panels).

\begin{figure*}
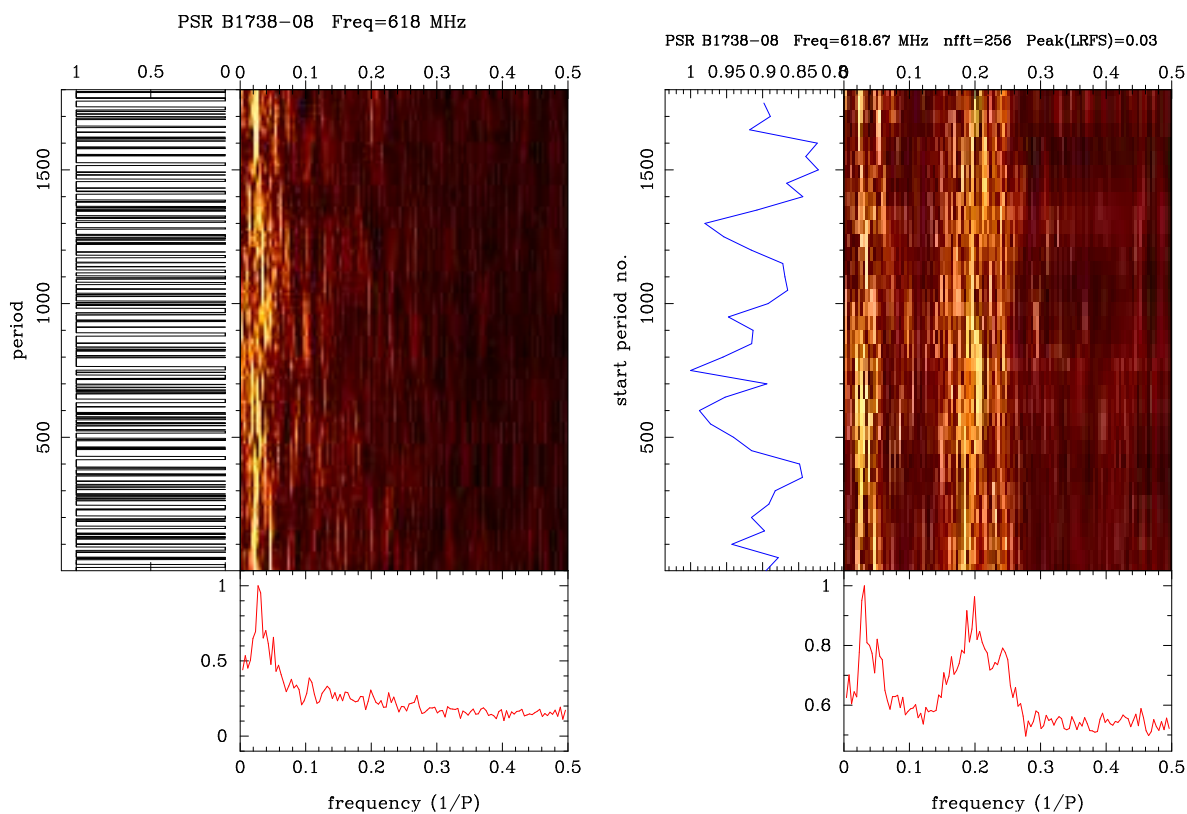

\begin{center}
\begin{tabular}{@{}lr@{}}
{\mbox{\includegraphics[angle=0,scale=0.4]{figure2a.ps}}}&
{\mbox{\includegraphics[angle=0,scale=0.4]{figure2b.ps}}}\\
\end{tabular}
\end{center}
\caption{The figure shows the time varying Fourier transform of the null/burst
(0/1) time series data for the pulsar B1738$-$08 observed at 618 MHz (left).
The longitude resolved fluctuation spectra (LRFS, right panel) from
\citet{bas16} is also shown for comparison. The LRFS shows a wide drifting
feature at higher frequency and a narrower feature at low frequency. The low
frequency feature is identically reflected in the null/burst spectra where the
drifting feature is absent.
%(The complete figure set (30 images) is available.)
\label{fig_nullfft}}
\end{figure*}

Additionally, given the sequence of null and burst pulses we examined the presence of periodicity in the transitions between the
states. The time series data of 0's and 1's were setup corresponding to the null and the burst pulses respectively. This ensured
that all subpulse information was washed away and the only possible periodicity was contained in the transition between the two
states. One dimensional discrete Fourier transform (DFT) was carried out on the time series data. In general we used 256
consecutive points for carrying out the DFT. If the peak frequency was too close to the edge the number of points used for the
DFT were increased accordingly to resolve the periodicity. The starting position was shifted by 10 pulses and the process was
repeated till the end. Finally, all the individual DFTs were averaged to produce a more sensitive spectra. This also enabled us
to examine any time variations in the periodicity (see figure~\ref{fig_nullfft}). In \citet{her07,her09} the pulse
modulation quelling (PMQ) method was used to estimate the periodicity in nulling. One primary difference in this method was that
these authors used the pulses that exhibited the `partial nulls'. For example, PSR B1133+16 has a two component profile where at
certain times the emission corresponding to one of the components is missing. This case is called the partial null. The PMQ used
scaled down average profiles for every burst pulse and a similarly scaled half-average profile for the partial nulls
\citep{her07}. In our work we only considered the complete nulls and the partial nulls were identified as burst emission.

\section{\large Results} \label{sec:res}
\input{table1.tex}

\noindent We have detected nulling in thirty six pulsars including seventeen pulsars where it was reported for the first time.
Table~\ref{tab1} lists the measured nulling fraction at the two observing frequencies. In some cases nulling was found to be
broadband in nature \citep{smi05,gaj14b} while other studies have reported excess nulls at certain frequencies \citep{bha07}. Our
observations were not simultaneous at the two frequencies to investigate this effect but we found the nulling behavior to be
similar at both frequencies. In three pulsars B1114$-$41, B1524$-$39 and B1918+19 nulling could only be measured at 333 MHz but
not at 618 MHz due to lower sensitivity.

In nineteen pulsars the null and burst pulses were well separated and we determined the null length as well as the burst length
histograms. The histograms were dominated by the short nulls with duration of a few periods. In five pulsars B0031$-$07,
B0301+19, B1706$-$16, J1727$-$2739 and B1944+17 the average null lengths exceeded five periods. In eleven pulsars we detected
periodicity in the transition between the two states. The average null lengths of the pulsars with periodic nulling were higher
than the non periodic cases.

\subsection{Statistical Study}
\noindent A number of statistical studies exist in the literature to examine randomness in the null/burst sequence. One of the
more widely used techniques is the `Wald-Wolfowitz runs test' applied to a two state system \citep{red09,gaj12}. It is assumed
that the transition from a null to a burst state is random resembling a coin toss experiment. The run (R) correspond to number of
transitions between the two states with a total of $n_1$ nulls and $n_2$ bursts. The statistics pertaining to R are defined by
the the mean ($\mu$), standard deviation ($\sigma$) and the factor $Z$ which are calculated as:
\begin{eqnarray}
\mu & = & \frac{2n_1n_2}{n_1+n_2} + 1, \nonumber \\
\sigma & = & \sqrt{\frac{2n_1n_2(2n_1n_2-n_1-n_2)}{(n_1+n_2)^2(n_1+n_2-1)}}, \nonumber \\
Z & = & \frac{R - \mu}{\sigma}.
\label{eq1}
\end{eqnarray}
Here, $Z$ is Gaussian distribution with zero mean and standard deviation unity. The randomness of the sequence with 95\%
confidence level correspond to $-$1.96 $< Z <$ 1.96. Table \ref{tab2} shows the $Z$ factor calculated assuming that the two
frequencies represent different segments of a larger time series configuration. The runs test indicates the null/burst sequence
to be non-random in the majority of pulsars, with B0834+06 and B1747$-$46 being the notable exceptions.

The inherent assumption in the runs test is that the transition probabilities from one state to another are identical. In the
cases where the transition probabilities differ the 2-state Markov process provides a more appropriate description. The
transition probabilities are expressed in a 2$\times$2
transition matrix ${\bf Q}$ as : \\
${\bf Q} = \left( \begin{matrix} q_{11} & q_{12} \\ q_{21} & q_{22}
\end{matrix} \right)$ \\
Here $q_{11}$ and $q_{22}$ are the probabilities of continuing in the null and
burst states respectively, while $q_{12}$ and $q_{21}$ signify the transition
probabilities from the null to burst states and vice versa. These quantities
are estimated as \citep{cor13}:
\begin{eqnarray}
\langle NL \rangle & = & 1/q_{12}, \nonumber \\
\langle BL \rangle & = & 1/q_{21}
\label{eq2}
\end{eqnarray}
along with the normalization condition $\sum_j q_{ij}$ = 1. Here $\langle NL
\rangle$ and $\langle BL \rangle$ are the average null lengths and burst
lengths respectively. The expected nulling fraction is calculated as:
\begin{equation}
\mbox{NF} = \frac{q_{21}}{q_{12} + q_{21}}
\label{eq3}
\end{equation}
The estimated transition probabilities are shown in Table \ref{tab2} along with
the expected nulling fractions. We have once again combined the sequences from
the two frequencies for these calculations. The expected nulling fractions from
a 2-state Markov process, particularly with periodic nulling, were larger than
the measured values. This implies that the 2-state Markov process is not an
appropriate explanation for the transition from the null to the burst state and
vice versa.

\input{table2.tex}

\subsection{The Nulling Periodicity}
\noindent As mentioned earlier we have measured periodic nulling in eleven pulsars. In six pulsars B0031$-$07, B0301+19,
J1727$-$2739, B1738$-$08, B1944+17 and B2045$-$16 both nulling and subpulse drifting were seen in the same system as evidenced by
the multiple peaks in their fluctuation spectra \citep[see][for a detailed analysis]{bas16}. The nulling periodicities coincided
with the low frequency peaks (see figure \ref{fig_nullfft}) while the high frequency structures were identified with subpulse
drifting. The nulling periodicities, ranging from 20 $P$ to 600 $P$, were larger than typical subpulse drifting periodicities.

\subsection{Individual pulsars}
\noindent In this section we describe the single pulse behavior in individual pulsars. In some cases nulling was affected by
signal to noise issues. In others we noticed peculiar mode changing. The single pulse plots are available in
\citet{mit16}\footnote{The single pulse time series plots can also be accessed
from http://mspes.ia.uz.zgora.pl/}.\\

\noindent {\bf B0628$-$28} was studied in \citet{rit76} who did not find any significant nulling. The single pulse behavior were
different at the two frequencies. Nulling was seen only at 333 MHz with bursts of emission interspersed by longer duration nulls
or very low level emission. The 618 MHz data showed a continuous but steadily decreasing burst state. Nulling appeared to be
periodic or quasi periodic in nature. The periodicity study could not be conducted due to
the presence of weaker burst pulses particularly around the nulls. \\

\noindent {\bf B0942$-$13} exhibited two types of nulling at 333 MHz within our short window of 2000 pulses. In the beginning the
nulls were of short duration interspersed with the burst state. The mean intensity of the emission started decreasing gradually
around the 1000$^{th}$ pulse and completely switched off around the 1500$^{th}$ pulse. A short burst of emission with much lower
intensity was seen between the 1800$^{th}$ and 1900$^{th}$ pulse. The 618 MHz emission had weaker signal and nulling could not be
identified in the energy
histograms. \\

\noindent {\bf B1556$-$44} did not show much variation in intensity for the majority of the observations. At the very end of the
618 MHz data the intensity dropped for 5--10 pulses leading to possible nulls. The pulsar profile is dominated by a
core component which makes the presence of nulling interesting. \\

\noindent
{\bf B1706$-$16} showed periodic nulling as reported in table~\ref{tab2}.
\citet{nai15} also reported longer duration nulls lasting several thousand
periods. The pulsar belongs to a small group which show extreme nulling. \\

\noindent
{\bf J1808$-$0813} showed the likely presence of periodic or quasi periodic
nulling at both frequencies. Due to lower sensitivity of the emission we were
not able to separate the null and the burst pulses and carry out a more
detailed study of the nulling periodicity. \\

\noindent {\bf B1819$-$22} exhibits phase modulated drifting as reported in \citet{bas16}. The nulling appeared to exhibit a
periodic or quasi periodic nature. We were not able to estimate the nulling periodicity due to weaker signal. However, a low
frequency feature was seen in the LRFS in addition to
subpulse drifting which may be indicative of nulling periodicity. \\

\noindent
{\bf B2003$-$08} has a multi component profile with a likely core emission.
The pulsar exhibits amplitude modulated drifting as reported in \citet{bas16}.
Multiple peaks were seen in the LRFS with the possibility of periodicity due to
nulling. The single pulses were once again of lower sensitivity to carry out a
more detailed analysis.\\

\noindent
{\bf J2346$-$0609} showed relatively high nulling fractions with bursts of
emission. The 333 MHz data were affected by radio frequency interference (RFI).
The 618 MHz data had lower sensitivity for carrying out the nulling periodicity
studies.\\

\subsection{Non Detections}
\noindent
We were not able to carry out measurements in ten pulsars where nulling was
previously reported. In seven pulsars B0148$-$06, B0450$-$18, B0656+14,
B0736$-$40, B1907+03, B1929+10 and B2315+21 the energy histogram analysis
were unable to separate the two states due to lower sensitivity of signals.
Three pulsars B0740$-$28, B1727$-$47 and B1818$-$04 did not show any
nulling despite high sensitivity. In all three cases the previously reported
nulling fractions were upper limits with low values \citep[$<$ 0.1
percent,][]{big92}. They are relatively high energetic with spin down energy
loss greater than 10$^{33}$ erg~s$^{-1}$. It is likely that nulling is not
present, however, one cannot rule out the possibility of shorter duration less
frequent nulls.

\section{\large Comparing Nulling Periodicity and Subpulse Drifting}\label{sec:per}

\input{table3.tex}

\begin{figure*}
\begin{center}
\includegraphics[angle=0.0,scale=0.72]{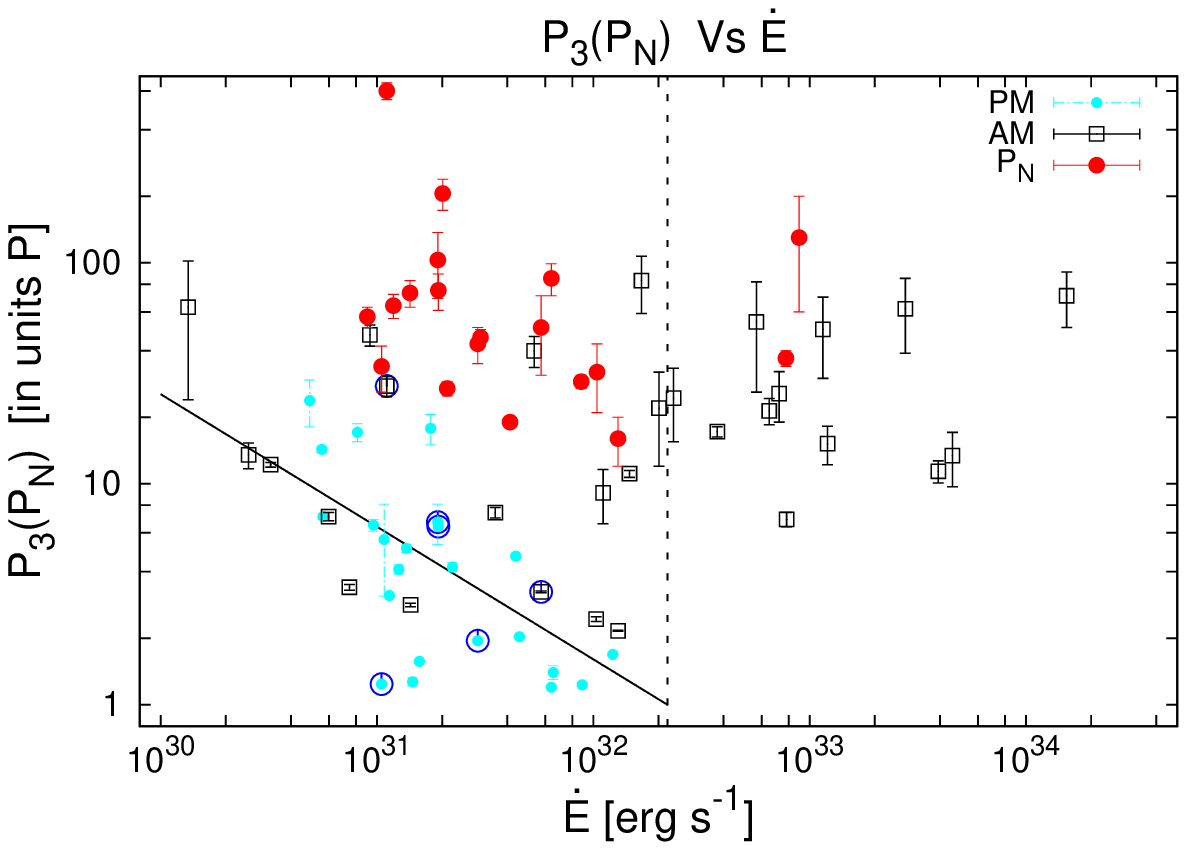}
\caption{The figure shows the Nulling periodicity ($P_N$ as well as the
drifting periodicity ($P_3$) as a function of spin-down energy loss
($\dot{E}$). The drifting phenomenon can be categorized into the phase
modulated drifting (PM, light blue dots) and the periodic amplitude modulation
(AM, black open squares). The PM periodicities are anti-correlated with
$\dot{E}$ (black line) and restricted to a narrow window (black dashed line).
The majority of AM periodicities are spread out over a large $\dot{E}$
range and generally have large periodicities. The nulling periodicities (red
dots) have properties similar to the AM cases. In some pulsars the $P_N$
coexist with $P_3$ (open blue circles identify drifting pulsars with $P_N$) and
are larger than the corresponding drifting periodicities.
\label{fig_edotnull}}
\end{center}
\end{figure*}

\noindent 
The nulling periodicities detected in our studies correspond to complete nulls, i.e. no partial nulls have been
included, with periodicities similar to certain drifting cases. However, in contrast to particularly the phase modulated subpulse
drifting the periodicities associated with nulling show relative spread as evidenced by the null length and burst length
histograms. In an attempt to connect these two phenomena the periodic nulls have been termed as `pseudo-nulls' and interpreted by
\citet{her07,her09,ran08,ran13} using the rotating subbeam `carousel' system \citep{des01}. The pulsar emission beam is believed
to consist of a central core component with concentric conal subbeams rotating around the core \citep{ran90,ran93a}. The core
components are phase stationary and do not participate in the carousel rotation. If the line of sight cuts the emission beam
tangentially this model predicts that there should be the phase modulated subpulse drifting. In the other case when the line of
sight traverses the beam,  the amplitude modulation of conal components (without any phase variations) are expected. The periodic
nulling can manifest the carousel model in two possible ways. In the first case nulling occurs for short durations of few periods
when the line of sight passes between two conal subbeams during their rotation cycle. This can also explain the partial nulls
where the sight-line nulls only appear on one side of the subbeam system. A typical example of this phenomenon was reported in the
pulsar B1133+16 by \citet{her07}. In the second scenario parts of the rotating subbeam system are postulated to be missing. The
longer duration periodic nulls lasting several tens of periods are seen when the line of sight passes through these empty
regions. The longer duration nulls in PSR J1819+1305 were reported to exhibit this phenomenon by \citet{ran08}. The periodic
nulling in the carousel model are associated with conal components \citep{her09} and provide an immediate prediction that these
should not be manifested in the stationary core components. To test this prediction we have collected the nineteen pulsars with
reported nulling periodicity in table \ref{tab3} along with their profile classification. The table shows that in addition to
conal components (S$_d$, D, $_c$T and $_c$Q profile classes) the pulsars with prominent core emission (S$_t$, T and M classes)
also show periodic nulling. The nulling periodicity in the core components is difficult to understand using the carousel model.
It seems more likely that the complete nulls correspond to the conditions when the emission is completely ceased. Additionally, \citet{bas16} reports the presence of periodic amplitude modulation in many pulsars which are core dominated.
In a recent work by \citet{mit17} highly periodic amplitude modulation was seen in the core component of the pulsar B1946+35. This also could
not be explained using the carousel model and was considered as a newly emergent phenomenon.

\citet{bas16} showed from theoretical perspectives that there are difficulties
in explaining the subpulse drifting phenomenon using a carousel system rotating
around the magnetic axis. However, an interesting correlation was seen between
the spin-down energy loss ($\dot{E}$) and the drifting periodicity ($P_3$). The
drifting population can be broadly separated into two groups. The first
corresponding to phase modulated drifting showed significant subpulse motion
across the pulse window. The periodicities were anti-correlated with $\dot{E}$
($P_3 \propto \dot{E}^{-0.6}$) and restricted to a narrow $\dot{E}$ range
($<$ 2$\times$10$^{32}$ erg~s$^{-1}$). The second group showed periodic
amplitude modulation with very little subpulse motion across the pulse window.
They were spread out over a wider $\dot{E}$ range and generally had large $P_3$
values. In a few cases the periodic amplitude modulation had smaller
periodicities which could be anti-correlated with $\dot{E}$. They were possibly
phase modulated drifters with specialized geometries resulting in small phase
variations. At the lower $\dot{E}$ range ($\sim$ 10$^{30}$ erg~s$^{-1}$) the
two classes had comparable $P_3$ values.

We have explored the dependence of the nulling periodicities ($P_N$) on $\dot{E}$ in order to explore if a connection exists with
subpulse drifting. The $P_3$ versus $\dot{E}$ plot in \citet{bas16} is reproduced in figure \ref{fig_edotnull} where the $P_N$
values have also been added. The periodic nulling seems to form a homologous group with the periodic amplitude modulation class.
As reported earlier, in many pulsars both periodic nulling and phase modulated drifting appear. We propose the idea that there
are two distinct periodic phenomenon connected with the radio emission in pulsars. The {\bf E}$\times${\bf B} drift in the IAR
gives rise to the phase modulated drifting. This phenomenon is correlated with $\dot{E}$ and discussed in more details by
\citet{bas16}. On the other hand the periodic amplitude modulation and periodic nulling, which can be considered as an extreme
form of amplitude modulation, are related to an unexplored periodic phenomenon in pulsars.

\section{\large Physical implications of Nulling} \label{sec:phy}
\noindent The duration of the most widespread nulling is short, just one or two pulse periods. This is evidenced by the null
length histograms which generally peak at the lowest bin. In some cases the nulling also lasts for hundreds of periods at a time
with a small but significant population of such pulsars now known \citep{lew04,wan07,gaj12,gaj14a,ker14,you15}. But, even in
these cases the distribution is dominated by the short duration nulls. It has been argued that the processes that cause nulling
at different timescales have a different physical origin. The short duration nulls are associated with stochasticity in the
emission process while the longer duration nulls are large scale changes in the magnetosphere \citep[see][and references
therein]{cor13}. One obvious way for nulling to occur is the pulsar emission beam moving in and out of the observers' line of
sight. Effects like precession or the emission being blocked by an external object orbiting around the neutron star are some
likely scenarios for this to happen. This necessitates nulling to be highly periodic with exact null and burst length intervals.
However, as evident in the single pulses there is no repetitive pattern in the nulling phenomenon even in the cases that show an
underlying periodicity. This is also represented in the null length and the burst length histograms which show a distribution and
rules out these mechanisms as the only possible source of  nulling\footnote{incidentally the carousel model of rotating sub-beams
discussed in the previous section is also a form of geometric nulling}. It is more likely that nulling is related to disruptions
in the radio emission process.

The radio emission can be disrupted if the distribution of particles flowing along the dipolar magnetic field lines changes.
\citet{gol69} demonstrated that the region around a fast spinning neutron star of surface magnetic field $10^{12}$ G cannot
remain as vacuum. They determined a charge separated force-free magnetosphere with density $n_{GJ} = \Omega \cdot B/2\pi c e$. It
has been reported \citep[e.g. most recently by][and references therein]{phi15,pet16} that a sufficient supply of charges from the
neutron star surface as well as from the pair production results in transition from a charge separated magnetosphere to a
force-free magnetosphere. The charge separated magnetosphere corresponds to the case where charged particles are trapped in
charge separated zones which co-rotate with the star. On the other hand the force-free magnetosphere is associated with the
electromagnetically dominated pulsar wind and separated into the closed field line region in which charges corotate with the star
and the open field line region in which charges can flow radially outwards. The coherent radio emission is excited due to some
instabilities in the relativistic plasma outflowing along the open dipolar magnetic field lines. If the fluctuations in the pair
production process are such that the pulsar exists in the transition region between an electrosphere and a force-free
magnetosphere, nulling is expected in its electrosphere state when the particle flow is halted. The timescale for which the
particle flow can stop is comparable to the return current generation time, which is a few hundred milliseconds for a typical one
second pulsar. This provides an explanation for the single period nulls. However, it requires detailed investigation to see if
the electrosphere can survive for longer durations and the transition between the magnetospheric states can support long period
nulls.

Next we examine the conditions which may lead to disruption of the radio emission in the presence of particle flow. Several
observational evidence suggest that the radio emission arises around 500 kilometers above the neutron star due to coherent
curvature radiation \citep{mit02,kij03,mit04,krz09,mit09}. These basic observational features can be best explained by the so
called steady state polar cap models \citep[e.g.,][]{stu71,rud75}. In pulsars where ${\bf\Omega\cdot\mbox{B}}< 0$, the positive
ions cannot be extracted from the stellar surface due to high binding energy. This results in the formation of a vacuum gap or
inner acceleration region (IAR) above the magnetic poles. Once the IAR is formed the radio emission process can be broadly
divided into three stages occurring at distinct locations. The first stage consists of breakdown of the IAR by isolated
discharges on the polar cap due to magnetic e$^-$e$^+$ pair production. High energy gamma-rays ($>$ 1.02 MeV) can form pairs in
the presence of strong magnetic fields \citep{shu82}. The large electric fields separate out the pairs with the back streaming
electrons heating the surface to very high temperatures (upto million degree kelvin). They further emit soft X-rays due to
thermal radiation. The positrons, also called primary particles, accelerate to high Lorentz factors and stream away from the
stellar surface. The second stage consists of the primary particles producing further higher energy photons via curvature
radiation or inverse Compton scattering leading to more pairs. This eventually results in a cascade giving rise to a more dense
but less energetic secondary pair plasma outside the IAR. The outflowing streams of secondary plasma clouds with the specific
energy distribution function are formed at the end of this stage. At the final stage, around heights of $\sim$ 500 kilometers,
the two stream instability develops due to the overlapping of fast and slow moving particles of adjacent secondary plasma clouds
\citep{ass98} along a particular set of field lines. This leads to the generation of strong electrostatic Langmuir wave
turbulence. The Langmuir waves are modulationally unstable and their nonlinear evolution produces the charged solitons that can
excite the radio waves via the curvature radiation mechanism \citep{mel00,gil04}. The radiation finally emerges as ordinary and
extraordinary modes from the secondary plasma \citep{mel14,mit16b}.

The emission mechanism is extremely sensitive to the plasma parameters. Any random fluctuations in these parameters are likely to
appear between timescales of 100 nanoseconds (time to empty the gap after sparking has stopped) to 5 microseconds (spark
formation time) which are much shorter than the typical nulling timescales. The only possibilities that can change the plasma
parameters are additional physical processes which alter the nature of the IAR. There are two different mechanisms in the
literature which can be used in this context. The first proposed by \citet{tim10} suggests changes in the magnetospheric state
leading to variations in the spin-down rates. This is brought about by the changes in the current flow and the size of the open
field line region which might be effective in altering the parameters of the outflowing plasma. The second method leads to
modifications in the potential drop across the IAR by introducing the concept of a partially screened gap
\citep[PSG,][]{gil03,sza15}. The surface temperature is considered to be at the critical level for the ions to be emitted and in
the process reducing the gap potential. This results in altering the energy of the primary particle which can further change the
process of secondary particle production. The secondary pair production process can switch from being dominated by curvature
radiation mechanism to inverse Compton scattering resulting in a modified energy distribution function. However, the transition
between different states would still require an additional triggering mechanism which are not addressed at present. This
triggering mechanism would likely involve an additional source of high energy photons from outside the IAR. Additionally, the
processes generating these photons should be self regulating to switch from the null to the burst states and vice versa. The
underlying mechanisms which can lead to such high energy photons outside the IAR need to be developed with detailed modelling.
However, one of our results provide some constraints on this process. The presence of nulling periodicity seen in a number of
pulsars suggest that the additional mechanism should be periodic in certain cases.

\section{\large Summary} \label{sec:sum}
\noindent
A detailed study of the nulling was conducted for the 123 pulsars observed in
the Meterwavelength Single-pulse Polarimetric Emission Survey. Nulling was
observed in thirty six pulsars including eleven pulsars showing the presence of
nulling periodicity. The nulling periodicity was demonstrated to be different
and of longer duration than the phase modulated subpulse drifting. The presence
of nulling require a triggering mechanism to change the pair production process
within the magnetosphere. The triggering mechanism is not yet known but should
be periodic in some pulsars with periodic nulling.\\\\

{\bf Acknowledgments}: We thank the referee for the valuable comments which helped to improve this paper. This work was supported by grant DEC-2013/09/B/ST9/02177 of the Polish National Science Centre. We would like to thank
the staff of Giant Meterwave Radio Telescope and National Center for Radio Astrophysics for providing valuable support in
carrying out this project. GMRT is run by the National Centre for Radio Astrophysics of the Tata Institute of Fundamental
Research.

\appendix
\section{\large Individual Pulsars in our Sample}
We discuss the previously reported nulling studies for the pulsars, particularly the cases where nulling periodicities are
reported, and compare them with our results. \\

{\bf B0031$-$07} is a conal single profile which exhibits the phase modulated subpulse drifting as well as nulling. The drifting
is seen in three different modes A, B and C with $P_3$ values of roughly 12 $P$, 7 $P$ and 4 $P$ respectively \citep{hug70}.
\citet{smi05,gaj14b} showed the nulling to be broadband in this pulsar. We find the nulling to exhibit a longer periodicity of
75$\pm$14 $P$ which cannot be connected harmonically with any of the three modes. \\

{\bf B0301+19} is a conal double profile which also shows both nulling and subpulse drifting. The LRFS of this pulsar show two
feature, one associated with drifting with $P_3$ roughly 6 $P$ and another corresponding to nulling at 103$\pm$34 $P$.
\citet{her09} also identified the low frequency feature in their PMQ analysis. Interestingly \citet{you12} showed that the bridge
emission between the conal components has a hidden core emission. We find that both the bridge and the conal components vanish
during the periodic nulls and it is unlikely to be a result of carousel rotation. \\

{\bf B0525+21} is a conal double profile where the PMQ analysis of \citet{her09} showed the presence of a low frequency double
peaked structure which they identified with nulling. \citet{you12} reports the bridge emission to contain a hidden core
component. We recorded only 400 pulses for this pulsar and the periodicity was not clear in our analysis. However, we found
nulling to extend to all the components which make the carousel interpretation unlikely.\\

{\bf B0818$-$13} is a conal single profile which shows the presence of two distinct emission states \citep{lyn83,jan04}. In the
brighter phase the pulsar shows the presence of prominent drift bands with occasional single period nulls. The weaker emission
state shows longer nulling while the drifting continues to exist during the weaker burst states. There is no detectable
periodicity in the null to burst transition during either of the two states. \\

{\bf B0834+06} is a conal double profile which shows the presence of drifting as well as nulling. The LRFS in this pulsar shows
the presence of a periodicity corresponding to 2.2 $P$ \citep{wel06,bas16}. This feature corresponds to the amplitude modulation.
The PMQ analysis of \citet{her09} found the same feature in a weaker way as well as a low frequency feature. The periodicity was
interpreted as corresponding to non random partial nulls \citep{ran07}. In our analysis with only the complete nulls we did not
see any periodicity for the null to burst transitions. Nulling is seen for short durations in this pulsar lasting for few
periods. We have argued that complete nulls are not geometrical in nature but arise due to changes in the pulsar magnetosphere
over short timescales. In section \ref{sec:phy} we discuss possible conditions, such as transition from a charge separated
magnetosphere to a force free magnetosphere, etc., where physical mechanism can change over such short timescales. \\

{\bf B1133+16} is a conal double profile similar to B0301+19 and B0525+21 with the bridge likely once again exhibiting core
emission \citep{you12}. The LRFS show the presence of a low frequency feature as reported in \citep{bac70,wel06}. The feature
becomes more prominent in both the PMQ and our analysis and is clearly related to periodic nulls. \citet{her07} identified the
periodicity as the longer circulation time of the rotating carousel model with sparse and occasional beamlets. They interpreted
the shorter nulls as empty sight-line traverse between subbeams. Our analysis uses only the complete nulls and the presence of a
hidden core component makes it unlikely to be a result of carousel circulation. Nulling is supposed to be a broadband phenomenon
\citep{smi05,gaj14b} but simultaneous observations by \citet{bha07} report the presence of excess nulls below 1.4 GHz for this
pulsar. However, more observations are needed to show if the periodicity persists across the frequencies.\\

{\bf J1727$-$2739} is a conal double profile with both nulling and drifting. \citet{wan07} reported frequent short bursts
separated by null intervals in this pulsar. A more detailed study by \citet{wen16} showed the presence of three distinct modes
during the burst state with modes A and B showing different drifting properties while in mode C no subpulse drifting is
detected. We have additionally found nulling to exhibit periodicity. \\

{\bf B1918+19} shows a three component conal profile with nulling as well as subpulse drifting. \citet{han87} showed the presence
of three different drift modes A, B, C with periodicities $\sim$6 $P$, $\sim$4 $P$ and $\sim$3 $P$ respectively, along with a
disordered mode (N) without any drifting. \citet{ran13} found the nulls to be confined mainly in the B and N modes with the modes
exhibiting specific modal sequences. The three different drift modes have been interpreted in the framework of the carousel model
as the number of sparks or subbeams changing in each mode. The PMQ analysis of \citet{her09} show the presence of two
harmonically connected peaks at 85$\pm$14$P$ and 43$\pm$4$P$ in addition to a wide structure at 12$P$. The 12$P$ periodicity was
taken as a confirmation of the carousal circulation time, the longer periodicities has been associated with the mode sequences.
The single pulse data were not sensitive for us to carry out a detailed null periodicity study.\\

{\bf B1944+17} is a conal profile with long null intervals as well as subpulse drifting. \citet{dei86} identified 4 distinct
modes in the pulsar with modes A and B associated with subpulse drifting having periodicities $\sim$14 $P$ and $\sim$6 $P$
respectively while mode C and D do not show any drifting periodicity. \citet{klo10} divided the nulling behavior into two groups,
the short duration nulls were assumed to be pseudo-nulls resulting from carousel rotation while the longer duration nulls were
expected to be true nulls without any periodicity. But the results of our periodicity studies show contrasting nulling behavior
for this pulsar. The null periodicity 600$\pm$52 is the longest known case and correspond to the longer duration nulls. No
corresponding periodicity was detected for the shorter null lengths.\\

{\bf B2303+30} shows the presence of nulling as well as subpulse drifting. The pulsar shows two emission modes with the B mode
having a near 2$P$ periodicity while the Q mode show a 3$P$ periodicity \citep{red05}. Nulling is reported to be mainly seen in
the Q mode. The PMQ analysis of \citet{her09} showed the presence of low frequency structure which is verified in our nulling
periodicity studies. \\

\section{\large The plots detailing the nulling phenomenon with well resolved
null and burst pulses.}
\input{appendix1.tex}

\clearpage

\section{\large The plots detailing the Fourier transform analysis to estimate
the nulling periodicity.}
\input{appendix2.tex}

\end{document}

%% file: table1.tex
\begin{table*}
\begin{center}
\caption{The list of nulling pulsars is presented with the pulsar name in 
column 1 along with the period from the ATNF database in column 2. Column 3 and
4 quotes the number of single pulses and the nulling fraction at 333 MHz while 
column 5 and 6 presents the corresponding values at 610 MHz. In some cases our 
criterion for estimating nulling (see text) implied that nulling could be 
measured at only one frequency and the non detections are represented as `N'. 
In some cases nulling fraction could only be estimated by counting the number 
of pulses below a statistical threshold and no error is calculated for these 
cases with the numbers serving as an upper limit. Finally, column 7 lists the 
references for previous nulling studies.
\label{tab1}}
{\begin{tabular}{ccc|cc|cc|c}
\multicolumn{3}{c}{} & \multicolumn{2}{c}{\underline{333 MHz}} & \multicolumn{2}{c}{\underline{618 MHz}}&  \\
     &   PSR  & Period & N$_p$ & NF & N$_p$ & NF & References \\
     &        &  (sec) &       & \% &       & \% &   \\
\tableline
  1. & B0031$-$07 & 0.943 & 2096 & 31.3$\pm$2.3 & 1119 & 22.8$\pm$1.8 & 1,2,3,4,5,6 \\
  2. &  B0301+19  & 1.388 & 2115 & ~8.7$\pm$1.2 & 2115 & ~6.1$\pm$0.6 &  7  \\
  3. &  B0525+21  & 3.746 & ~400 & 14.4$\pm$1.2 &  --- &      ---     &  8  \\
  4. & B0628$-$28 & 1.244 & 2112 & 13.6$\pm$1.9 & 2111 &       N      &  8  \\
  5. & B0818$-$13 & 1.238 & ~969 & ~9.8$\pm$1.2 & 2122 &      0.8     & 9,10,11\\
  6. &  B0834+06  & 1.274 & 1052 & ~3.9$\pm$0.3 & 2116 & ~4.7$\pm$0.7 & 8,12\\
  7. & B0906$-$17 & 0.402 & 2078 & 26.8$\pm$1.7 & 2234 & 25.7$\pm$1.3 & --- \\
  8. & B0942$-$13 & 0.570 & 2100 & 14.4$\pm$0.9 & 2196 &   N      & --- \\
  9. & B1114$-$41 & 0.943 & 2095 & ~3.3$\pm$0.5 & 1918 &   N      & --- \\
 10. &  B1133+16  & 1.188 & ~753 & 13.7$\pm$2.1 & ~494 & 11.9$\pm$2.3 & 8,13,14\\
 11. &  B1237+25  & 1.382 & 1081 & ~2.0$\pm$0.1 & ~864 & ~3.1$\pm$0.4 &  8  \\
 12. & B1325$-$49 & 1.479 & 2100 &    4.0   & 2100 &      4.4     & --- \\
 13. & B1524$-$39 & 2.418 & ~862 & ~5.1$\pm$1.3 & 1233 &   N      & --- \\
 14. & B1556$-$44 & 0.257 & 2293 &       N      & 2484 &  0.24    & --- \\
 15. & B1700$-$32 & 1.212 & 1716 &      1.6     & 2125 &  0.4     & --- \\
 16. & B1706$-$16 & 0.653 & 2106 & ~3.7$\pm$1.3 & 2106 & ~4.9$\pm$0.3 & 15 \\
 17. & J1727$-$2739 & 1.293 & 2126 & 57.0$\pm$2.3 & 2127 & 48.3$\pm$1.8 & 16,17 \\
 18. & B1730$-$37 & 0.338 &  --- &      ---     & 2110 & 52.4$\pm$3.5 & --- \\
 19. & B1738$-$08 & 2.043 & 1753 & 15.7$\pm$1.7 & 2052 & 15.8$\pm$1.4 & --- \\
 20. & B1742$-$30 & 0.367 & 2120 & 40.2$\pm$1.9 & 2119 & 24.8$\pm$1.0 & 18  \\
 21. & B1747$-$46 & 0.742 & 2017 & ~2.4$\pm$0.5 & 2084 &  2.4     & --- \\
 22. & B1749$-$28 & 0.563 & 2129 &      0.2     & 2129 & ~1.7$\pm$0.4 & 18  \\
 23. & B1758$-$03 & 0.921 & 2145 & 27.7$\pm$1.3 & 1944 & 26.1$\pm$2.6 & --- \\
 24. & J1808$-$0813 & 0.876 & 2110 & 12.8$\pm$1.3 & 1557 & ~8.2$\pm$1.0 & --- \\
 25. & B1813$-$36 & 0.387 & 2158 & 16.7$\pm$0.7 &  --- &  ---     & --- \\
 26. & B1819$-$22 & 1.874 & 2106 & ~4.7$\pm$0.9 & 1309 & ~5.5$\pm$0.7 & --- \\
 27. & B1857$-$26 & 0.612 & 2152 & ~5.1$\pm$0.8 & 2152 & ~8.1$\pm$0.5 & 8,19 \\
 28. & J1901$-$0906 & 1.782 & 2076 &      2.9     & 1068 & ~5.6$\pm$0.7 & --- \\
 29. &  B1918+19  & 0.821 & 2105 &      2.0     & 2180 &       N      & 20 \\
 30. &  B1944+17  & 0.441 & 5566 & 29.7$\pm$1.4 & 2174 & 37.9$\pm$2.3 & 8,21,22 \\
 31. & B2003$-$08 & 0.581 & 2157 & 15.6$\pm$1.0 & 2157 & 24.2$\pm$1.5 & --- \\
 32. & B2045$-$16 & 1.962 & ~924 & ~8.3$\pm$1.4 & 1828 & ~9.0$\pm$0.5 & 8   \\
 33. &  B2303+30  & 1.576 & 1698 & ~5.3$\pm$0.5 &  --- &      ---     & 23  \\
 34. &  B2310+42  & 0.349 & 2572 & ~3.7$\pm$0.5 &  --- &  ---     & 24 \\
 35. & B2327$-$20 & 1.644 & 1284 & ~9.6$\pm$0.9 & 1500 & 13.1$\pm$1.5 & 18  \\
 36. & J2346$-$0609 & 1.181 & 2130 & 42.5$\pm$3.8 & 2128 & 28.7$\pm$1.8 & --- \\
\tableline
\end{tabular}}
\tablenotetext{~}{{\bf References :} 1-\citet{hug70}, 2-\citet{viv95},
3-\citet{viv97}, 4-\citet{jos00}, 5-\citet{smi05}, 6-\citet{gaj14b},
7-\citet{ran86}, 8-\citet{rit76}, 9-\citet{lyn83}, 10-\citet{jan04},
11-\citet{gaj12}, 12-\citet{ran07}, 13-\citet{bha07}, 14-\citet{her07},
15-\citet{nai15}, 16-\citet{wan07}, 17-\citet{wen16}, 18-\citet{big92},
19-\citet{mit08}, 20-\citet{ran13}, 21-\citet{klo10}, 22-\citet{dei86}, 
23-\citet{red05}, 24-\citet{wri12}}
\end{center}
\end{table*}

%% file: table2.tex
\begin{table*}
\begin{center}
\caption{The nulling statistics of pulsars where the null and burst periods 
were resolved. Columns 2, 3 and 4 represents the number of transitions between 
the null and burst states, N$_{T}$, the average burst length, 
$\langle$BL$\rangle$, and the average null length, $\langle$NL$\rangle$, 
respectively at 333 MHz, while column 5, 6 and 7 present the corresponding 
values at 618 MHz. Column 8 presents the peridicity (P$_N$) in the transition 
from the null to the burst states, while column 9 presents the $Z$ statistics 
for the runs test. Columns 10 and 11 show the estimated transition probability
from the null to burst state and vice versa for a 2-state Markov process with
the estimated nulling fraction from these transition probabilities, NF(E), is 
shown in column 12.
\label{tab2}}
{\begin{tabular}{cc|ccc|ccc|c|c|ccc}
\multicolumn{2}{c}{} & \multicolumn{3}{c}{\underline{333 MHz}} & \multicolumn{3}{c}{\underline{618 MHz}} & \multicolumn{5}{c}{}  \\
     &   PSR  & N$_{T}$ & $\langle$BL$\rangle$ & $\langle$NL$\rangle$ & N$_{T}$ & $\langle$BL$\rangle$ & $\langle$NL$\rangle$ & P$_N$ & $Z$ & q$_{12}$ & q$_{21}$ & NF(E)\\
     &    &   &  (P) & (P) &  & (P) & (P) & (P) &   &   &   & (\%) \\
\tableline
     &              &     &       &      &     &       &      &           &         &  &  &    \\
  1. & B0031$-$07 & ~45 & ~26.0 & 19.3 & ~14 & ~42.3 & 34.1 &~75$\pm$14~& $-$52.0 & 0.044 & 0.034 & 43.4 \\
  2. &  B0301+19  & ~43 & ~41.9 & ~6.3 & ~30 & ~61.2 & ~6.6 &103$\pm$34~& $-$52.9 & 0.155 & 0.020 & 11.4 \\
  3. &  B0525+21  & ~17 & ~16.8 & ~4.9 & --- &  ---  &  --- &    ---    & $-$14.4 & 0.202 & 0.059 & 22.7 \\
  4. & B0818$-$13 & ~58 & ~14.6 & ~2.1 & ~15 & 111.5 & ~1.0 &    ---    & $-$22.6 & 0.537 & 0.029 & ~5.1 \\
  5. &  B0834+06  & ~64 & ~15.5 & ~1.0 & 156 & ~12.4 & ~1.0 &    ---    & $-$1.3  & 0.905 & 0.075 & ~7.7 \\
  6. &  B1133+16  & --- &  ---  &  --- & ~57 & ~~6.9 & ~1.5 &~29$\pm$2~~& $-$4.0  & 0.679 & 0.145 & 17.6 \\
  7. &  B1237+25  & ~41 & ~24.0 & ~1.6 & ~24 & ~33.1 & ~1.5 &    ---    & $-$13.5 & 0.657 & 0.036 & ~5.3 \\
  8. & B1524$-$39 & ~36 & ~22.3 & ~1.3 & --- &  ---  &  --- &    ---    & $-$5.9~ & 0.766 & 0.045 & ~5.5 \\
  9. & B1700$-$32 & ~13 & 127.5 & ~2.1 & ~~5 & 348.4 & ~1.0 &    ---    & $-$29.2 & 0.514 & 0.005 & ~1.0 \\
 10. & B1706$-$16 & ~52 & ~36.9 & ~3.3 & ~48 & ~37.8 & ~5.9 &130$\pm$70~& $-$48.9 & 0.220 & 0.027 & 10.9 \\
 11. & J1727$-$2739 & --- &  ---  &  --- & ~43 & ~12.5 & 35.9 &206$\pm$33~& $-$41.1 & 0.028 & 0.080 & 74.1 \\
 12. & B1738$-$08 & 152 & ~~8.1 & ~3.4 & 144 & ~10.5 & ~3.6 &~34$\pm$8~~& $-$37.4 & 0.287 & 0.108 & 27.3 \\
 13. & B1747$-$46 & ~75 & ~25.2 & ~1.0 & ~46 & ~42.2 & ~1.0 &    ---    & $-$1.2  & 0.953 & 0.032 & ~3.2 \\
 14. & B1749$-$28 & ~~2 & 820.0 & ~2.0 & ~14 & 147.6 & ~3.6 &    ---    & $-$44.0 & 0.291 & 0.004 & ~1.5 \\
 15. &  B1944+17  & 255 & ~~7.2 & 14.5 & ~78 & ~10.4 & 17.5 &600$\pm$52~& $-$71.0 & 0.066 & 0.126 & 65.7 \\
 16. & B2045$-$16 & ~57 & ~13.2 & ~2.3 & 109 & ~14.3 & ~2.4 &~51$\pm$20~& $-$26.8 & 0.417 & 0.072 & 14.7 \\
 17. &  B2303+30  & ~88 & ~17.3 & ~1.9 & --- &  ---  &  --- &~43$\pm$8~~& $-$17.6 & 0.518 & 0.058 & 10.0 \\
 18. &  B2310+42  & ~44 & ~51.6 & ~3.6 & --- &  ---  &  --- &~32$\pm$11~& $-$35.2 & 0.274 & 0.019 & ~6.6 \\
 19. & B2327$-$20 & ~87 & ~12.0 & ~2.5 & 117 & ~10.3 & ~2.4 &~19$\pm$1~~& $-$26.1 & 0.412 & 0.091 & 18.0 \\
     &              &     &       &      &     &       &      &           &         &  &  &    \\
\tableline
\end{tabular}}
%\tablenotetext{~}{}
\end{center}
\end{table*}

%% file: table3.tex
\begin{table*}
\begin{center}
\caption{The details of Pulsars with Periodic Nulling. The period and spin down
energy loss is listed in column 3 and 4 respectively. Column 5 lists the 
nulling periodicity while column 6 describes the classification of each pulsar.
\label{tab3}}
{\begin{tabular}{cccccc}
     & PSR & Period &          \.{E}         & P$_N$ & Class. \\
     &     & (sec)  &(10$^{30}$ erg s$^{-1}$)&  (P)  &        \\
\tableline
     &            &       &      &           &                \\
  1. & B0031$-$07 & 0.943 & 19.2 &~75$\pm$14 & S$_{d}$  \\
  2. &  B0301+19  & 1.388 & 19.1 &103$\pm$34 &    D     \\
  3. &  B0525+21  & 3.746 & 30.1 &~46$\pm$4~ &    D     \\
  4. &  B0751+32  & 1.442 & 14.2 &~73$\pm$10 &    D     \\
  5. &  B0834+06  & 1.274 & 130  &~16$\pm$4~ &    D     \\
  6. &  B1133+16  & 1.188 & 87.9 &~29$\pm$2~ &    D     \\
  7. & J1649+2533 & 1.015 & 21.1 &~27$\pm$2~ &   ---    \\
  8. & B1706$-$16 & 0.653 & 894  &130$\pm$70 & S$_{t}$  \\
  9. &J1727$-$2739& 1.293 & 20.1 &206$\pm$33 &   ---    \\
 10. & B1738$-$08 & 2.043 & 10.5 &~34$\pm$8~ & $_c$Q?   \\
 11. & J1819+1305 & 1.060 & 11.9 &~64$\pm$8~ &   ---    \\
 12. &  B1839+09  & 0.381 & 776  &~37$\pm$3~ & S$_{t}$  \\
 13. &  B1918+19  & 0.821 & 63.9 &~85$\pm$14 &  $_c$T   \\
 14. &  B1944+17  & 0.441 & 11.1 &600$\pm$52 &  $_c$T   \\
 15. &  B2034+19  & 2.074 & 9.02 &~57$\pm$6~ &   ---    \\
 16. & B2045$-$16 & 1.962 & 57.3 &~51$\pm$20 &    T     \\
 17. &  B2303+30  & 1.576 & 29.2 &~43$\pm$8~ & S$_{d}$  \\
 18. &  B2310+42  & 0.349 & 104  &~32$\pm$11 &    M?    \\
 19. & B2327$-$20 & 1.644 & 41.2 &~19$\pm$1~ &    T     \\
     &            &       &      &           &          \\
\tableline
\end{tabular}}
\tablenotetext{~}{Classifications from \citet{ran90,ran93b}.}
\end{center}
\end{table*}

%% file: appendix1.tex
\clearpage

\begin{figure*}
\begin{center}
\mbox{\includegraphics[angle=0,scale=0.9]{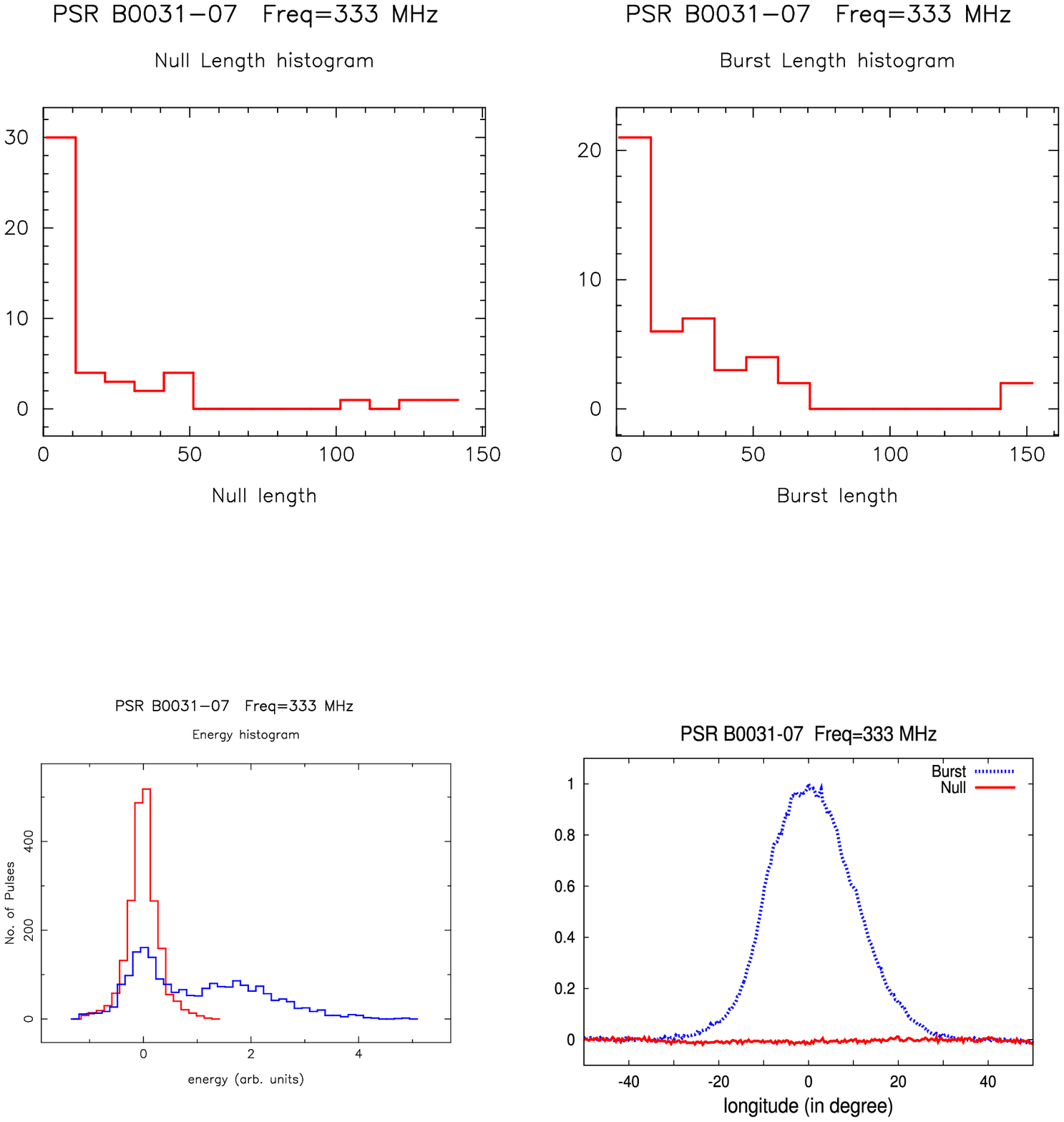}}
\end{center}
\caption{Null length histogram (top left); Burst length histogram (top right); the average energy distribution (bottom left) for on-pulse window (blue hisogram) and off-pulse window (red hisogram); the folded profile (bottom right) for the null pulses (red line, noise like characteristics) and burst pulses (blue line).}
\end{figure*}

\clearpage

\begin{figure*}
\begin{center}
\mbox{\includegraphics[angle=0,scale=0.9]{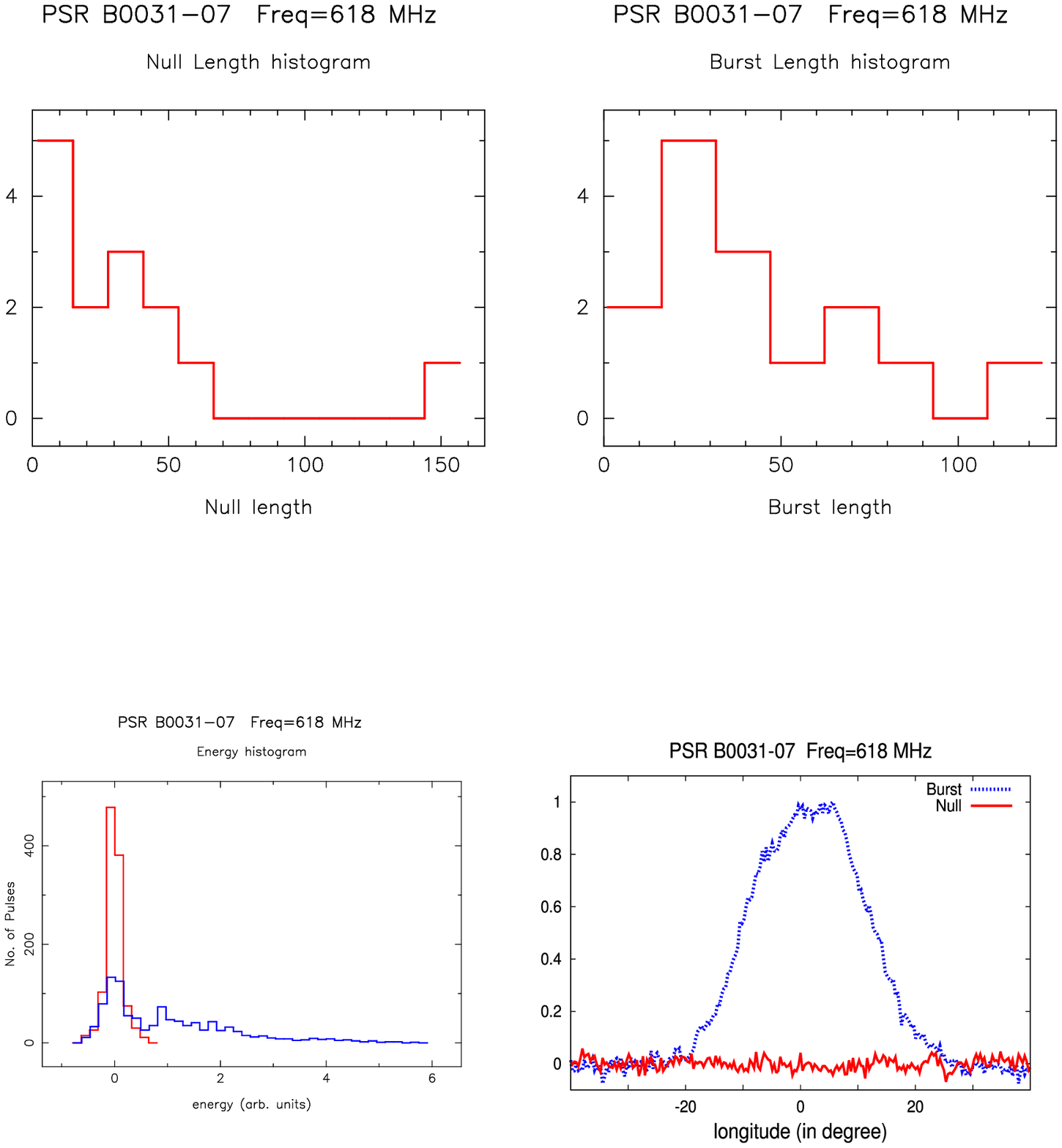}}
\end{center}
\caption{Null length histogram (top left); Burst length histogram (top right); the average energy distribution (bottom left) for on-pulse window (blue hisogram) and off-pulse window (red hisogram); the folded profile (bottom right) for the null pulses (red line, noise like characteristics) and burst pulses (blue line).}
\end{figure*}

\clearpage

\begin{figure*}
\begin{center}
\mbox{\includegraphics[angle=0,scale=0.9]{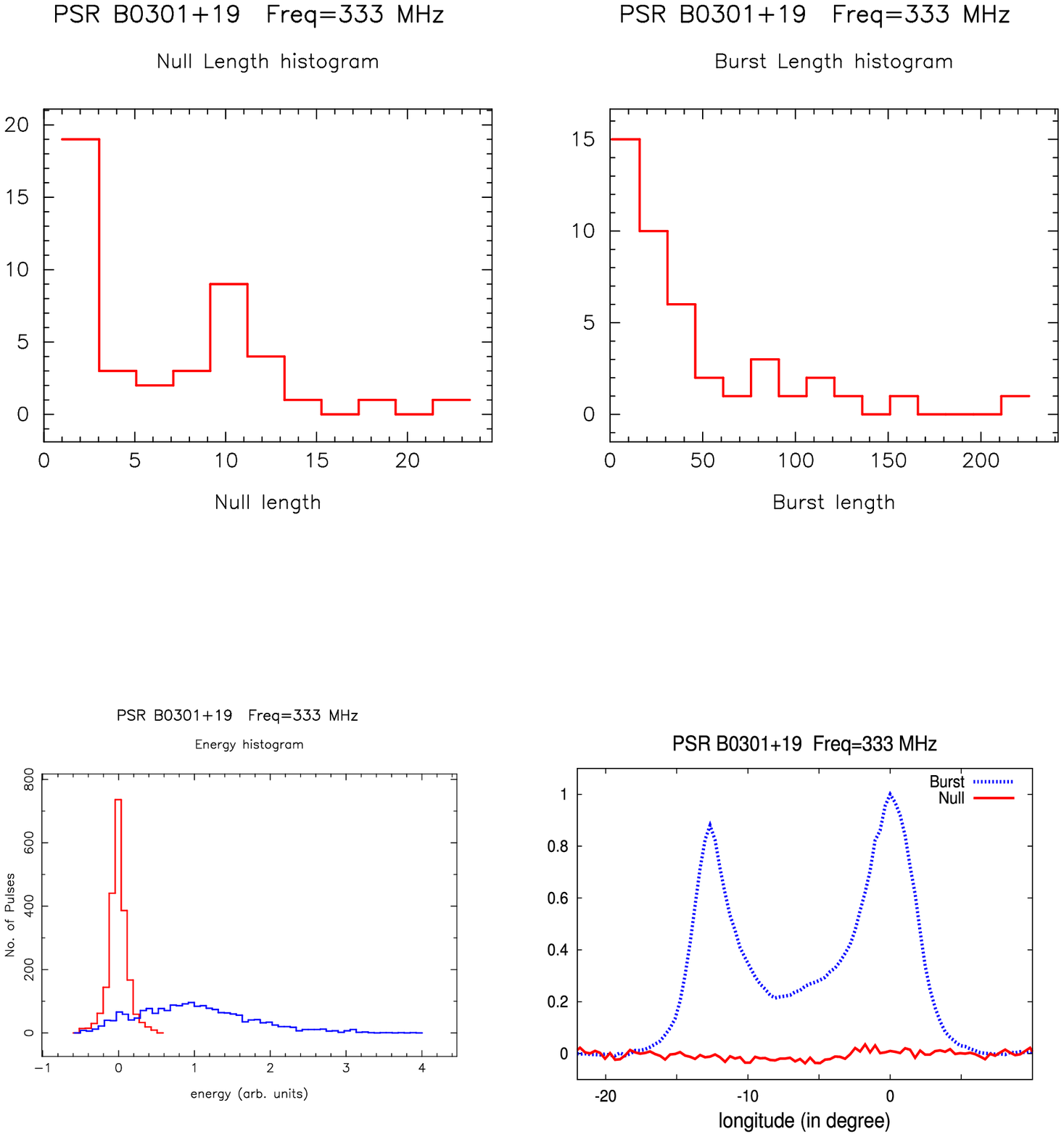}}
\end{center}
\caption{Null length histogram (top left); Burst length histogram (top right); the average energy distribution (bottom left) for on-pulse window (blue hisogram) and off-pulse window (red hisogram); the folded profile (bottom right) for the null pulses (red line, noise like characteristics) and burst pulses (blue line).}
\end{figure*}

\clearpage

\begin{figure*}
\begin{center}
\mbox{\includegraphics[angle=0,scale=0.9]{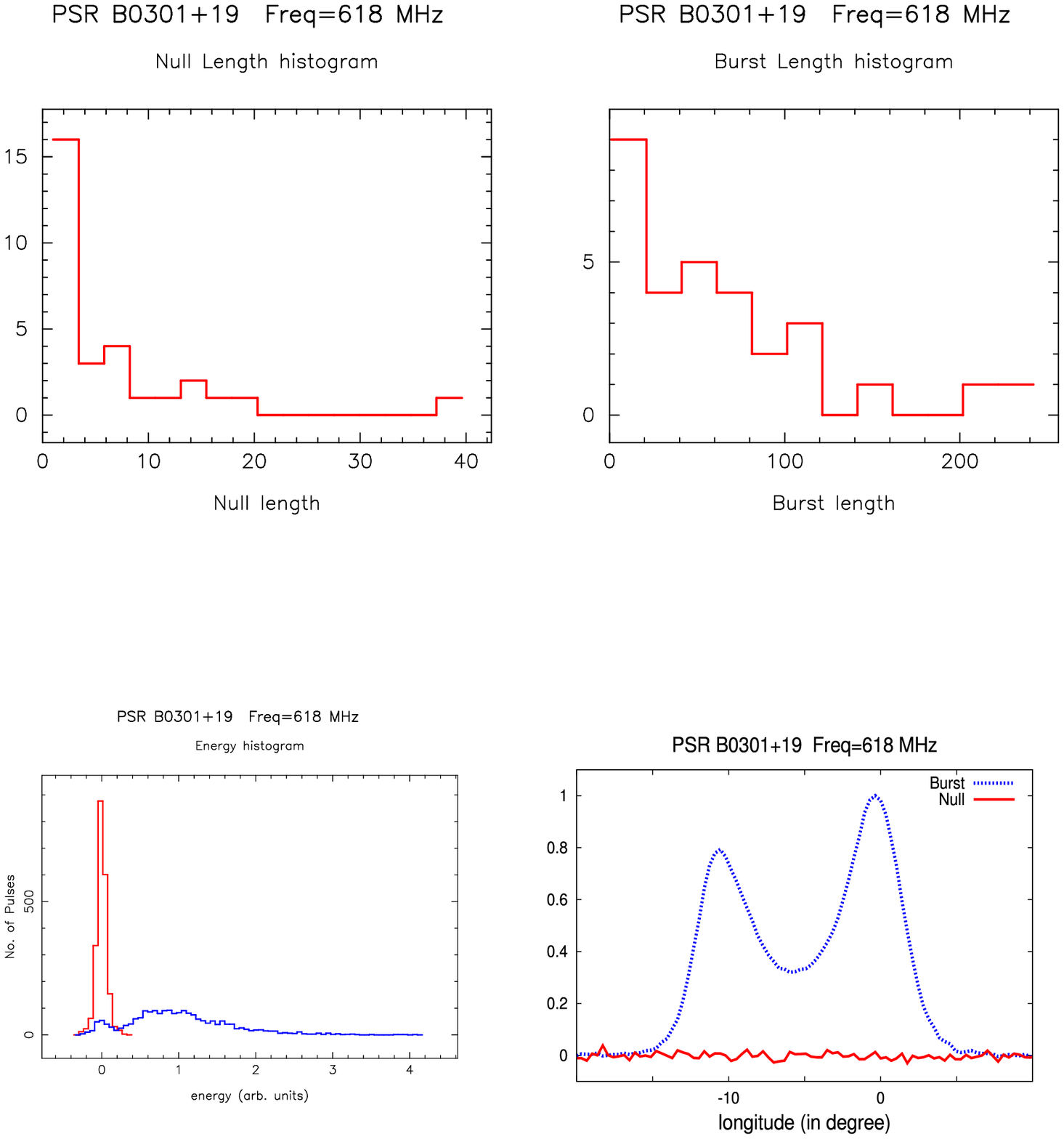}}
\end{center}
\caption{Null length histogram (top left); Burst length histogram (top right); the average energy distribution (bottom left) for on-pulse window (blue hisogram) and off-pulse window (red hisogram); the folded profile (bottom right) for the null pulses (red line, noise like characteristics) and burst pulses (blue line).}
\end{figure*}

\clearpage

\begin{figure*}
\begin{center}
\mbox{\includegraphics[angle=0,scale=0.9]{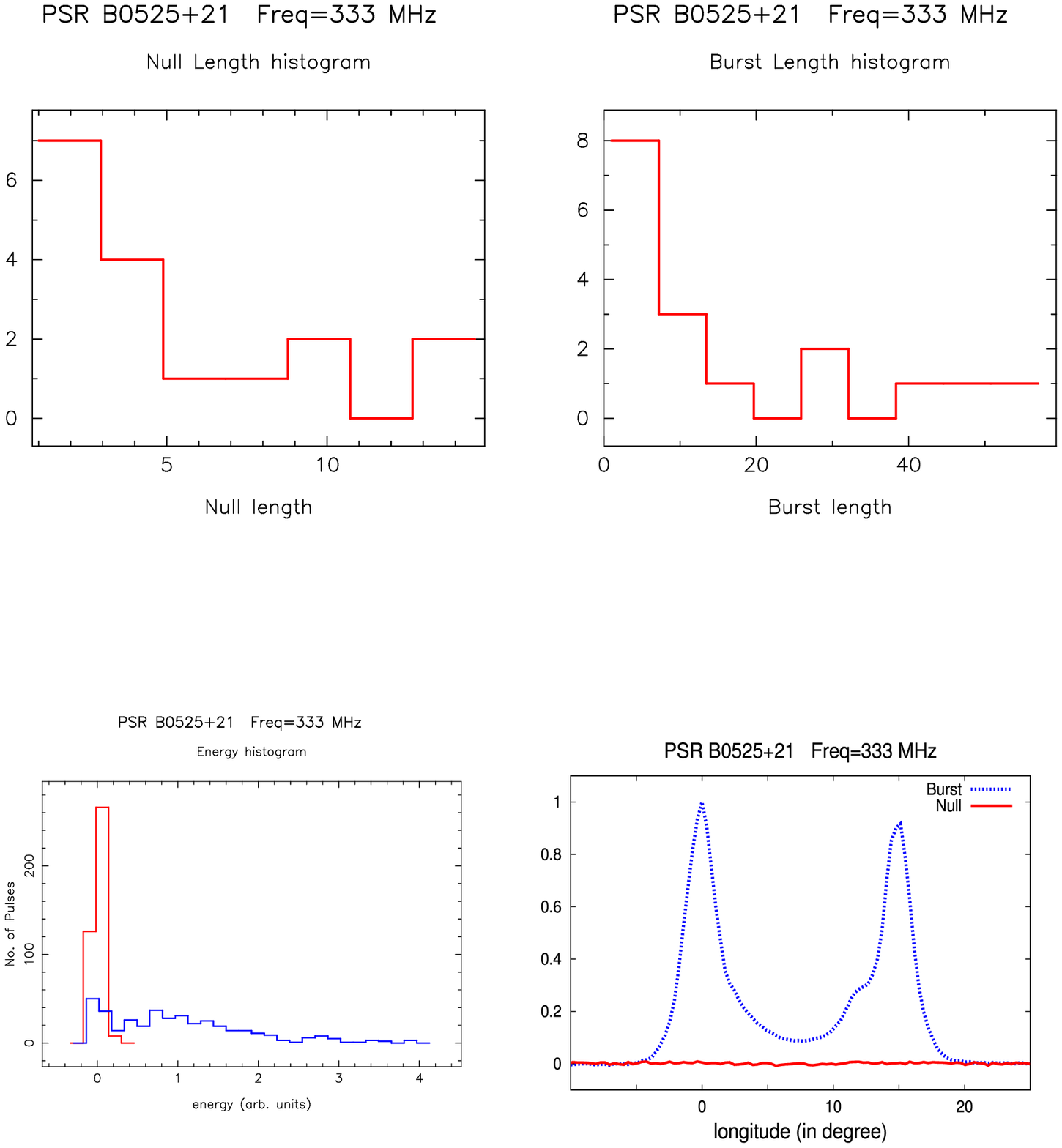}}
\end{center}
\caption{Null length histogram (top left); Burst length histogram (top right); the average energy distribution (bottom left) for on-pulse window (blue hisogram) and off-pulse window (red hisogram); the folded profile (bottom right) for the null pulses (red line, noise like characteristics) and burst pulses (blue line).}
\end{figure*}

\clearpage

\begin{figure*}
\begin{center}
\mbox{\includegraphics[angle=0,scale=0.9]{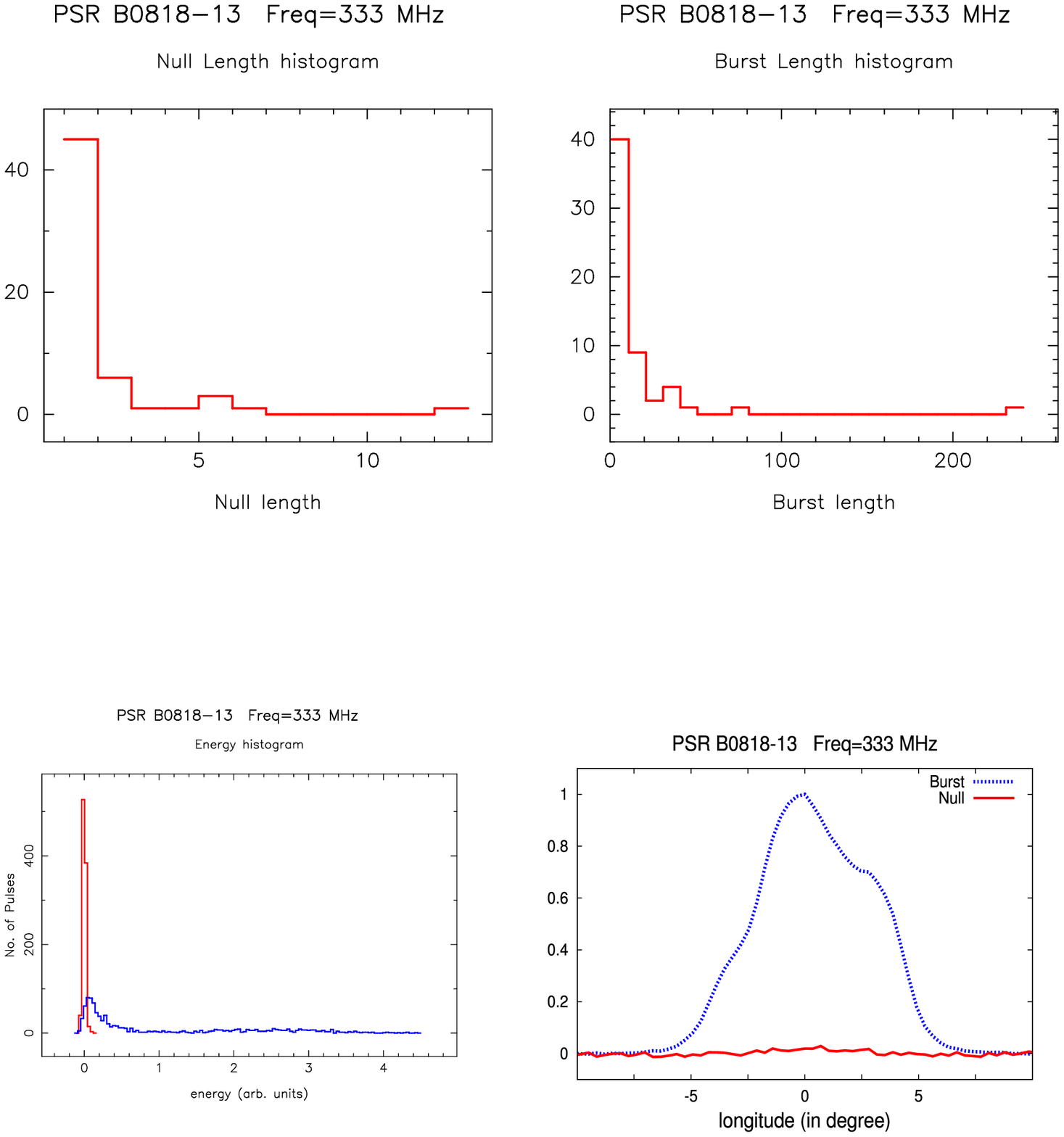}}
\end{center}
\caption{Null length histogram (top left); Burst length histogram (top right); the average energy distribution (bottom left) for on-pulse window (blue hisogram) and off-pulse window (red hisogram); the folded profile (bottom right) for the null pulses (red line, noise like characteristics) and burst pulses (blue line).}
\end{figure*}

\clearpage

\begin{figure*}
\begin{center}
\mbox{\includegraphics[angle=0,scale=0.9]{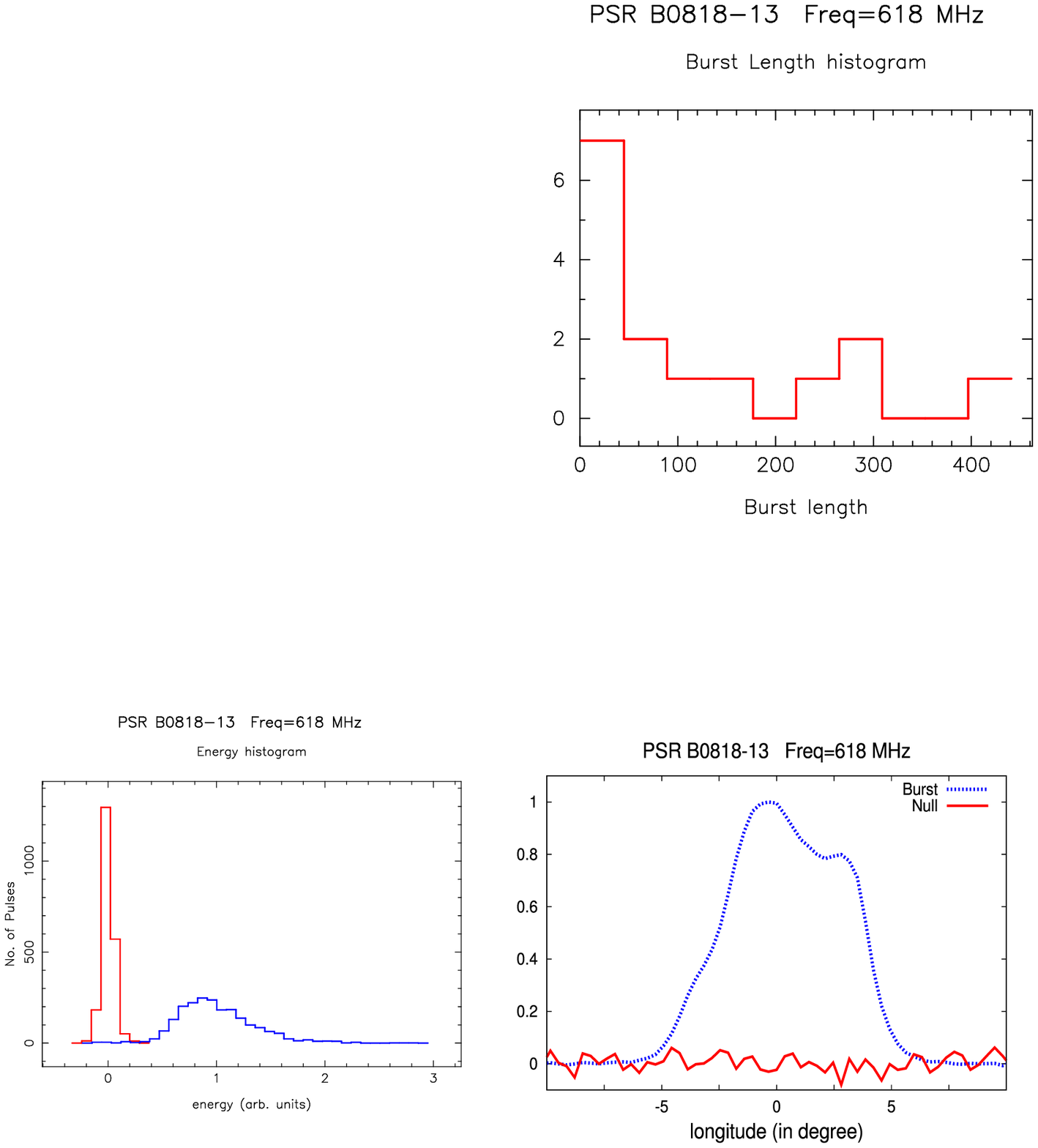}}
\end{center}
\caption{15 single period nulls, no null length histogram; Burst length histogram (top right); the average energy distribution (bottom left) for on-pulse window (blue hisogram) and off-pulse window (red hisogram); the folded profile (bottom right) for the null pulses (red line, noise like characteristics) and burst pulses (blue line).}
\end{figure*}

\clearpage

\begin{figure*}
\begin{center}
\mbox{\includegraphics[angle=0,scale=0.9]{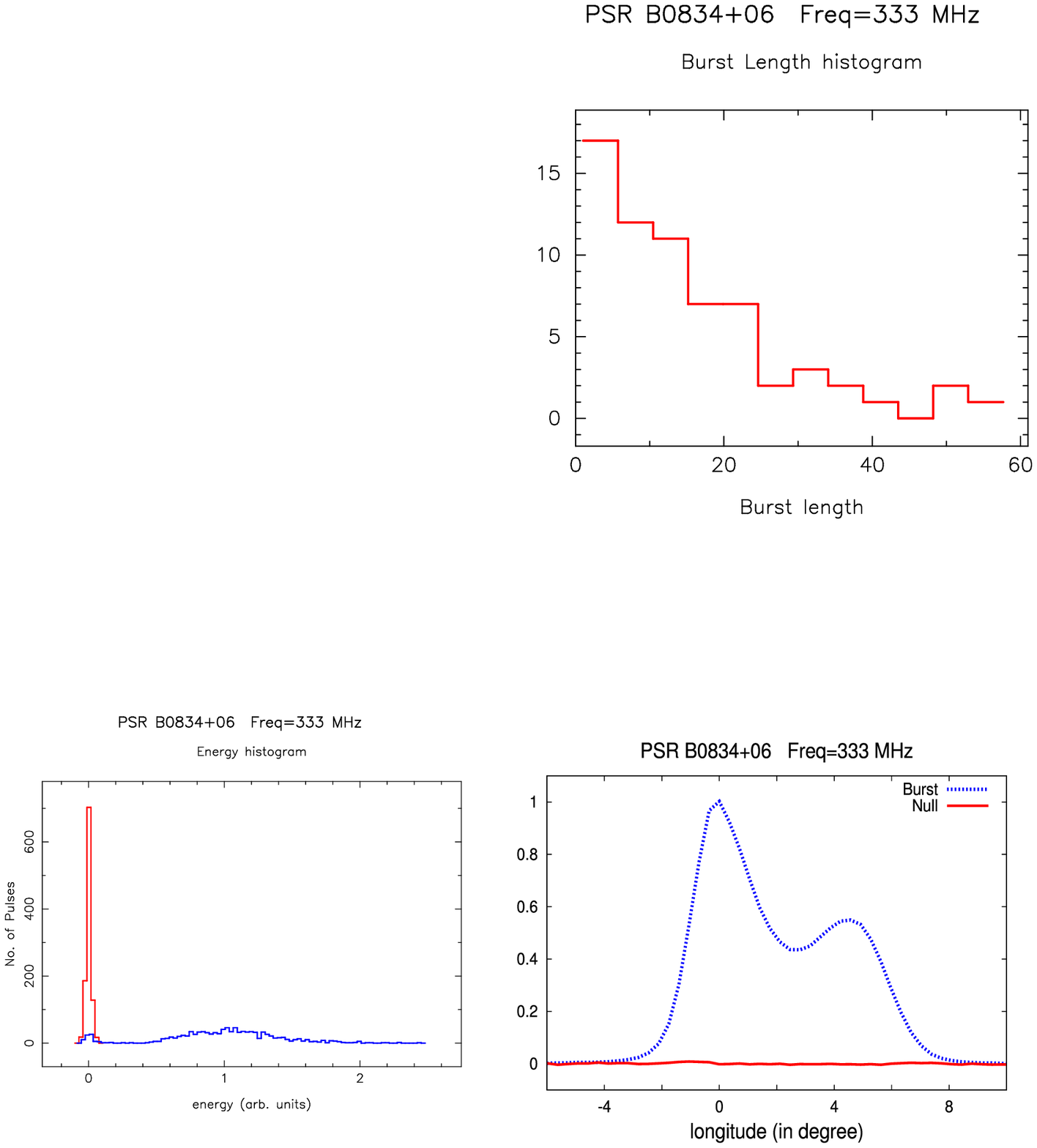}}
\end{center}
\caption{64 single period nulls, no null length histogram; Burst length histogram (top right); the average energy distribution (bottom left) for on-pulse window (blue hisogram) and off-pulse window (red hisogram); the folded profile (bottom right) for the null pulses (red line, noise like characteristics) and burst pulses (blue line).}
\end{figure*}

\clearpage

\begin{figure*}
\begin{center}
\mbox{\includegraphics[angle=0,scale=0.9]{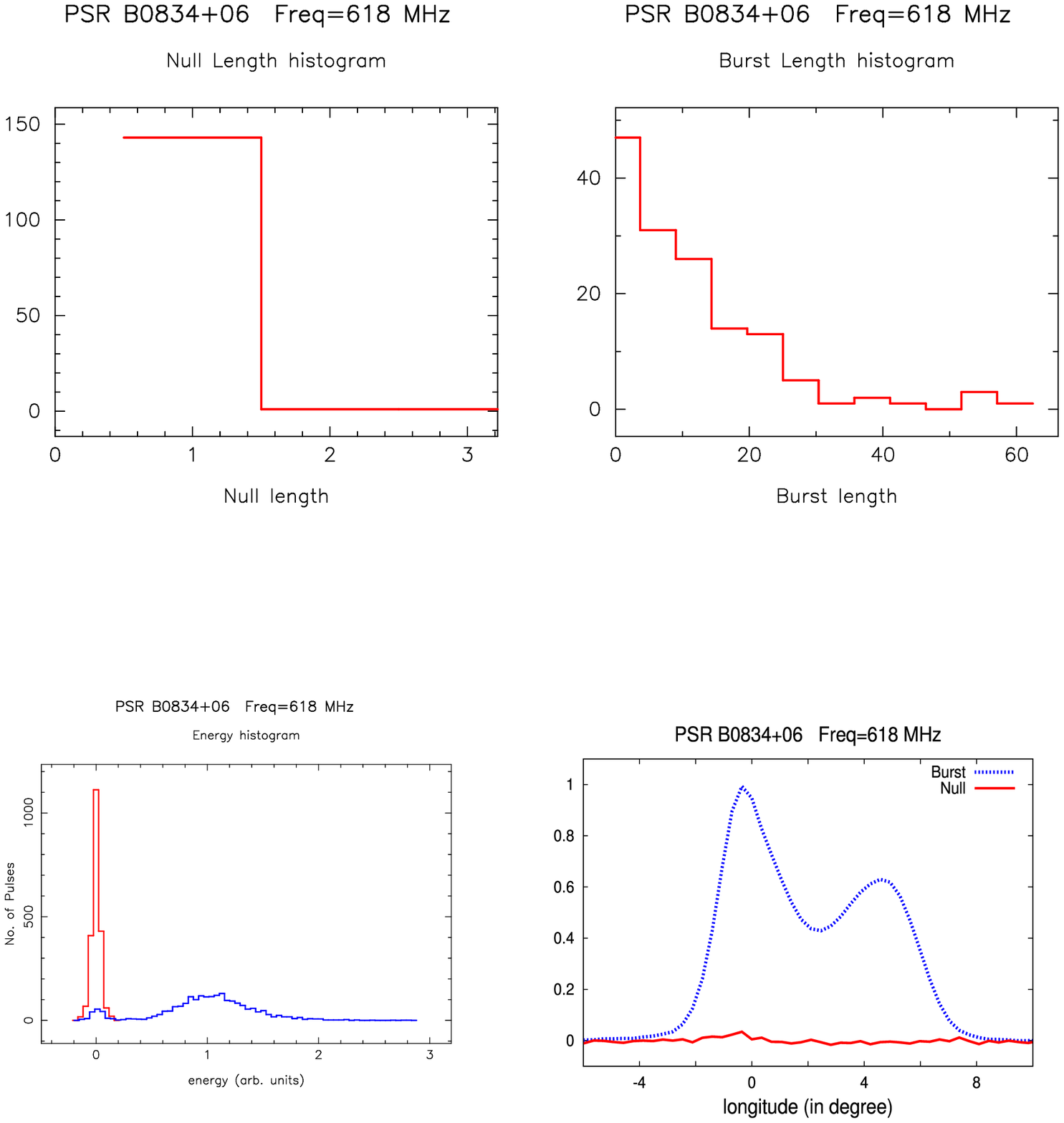}}
\end{center}
\caption{Null length histogram (top left); Burst length histogram (top right); the average energy distribution (bottom left) for on-pulse window (blue hisogram) and off-pulse window (red hisogram); the folded profile (bottom right) for the null pulses (red line, noise like characteristics) and burst pulses (blue line).}
\end{figure*}

\clearpage

\begin{figure*}
\begin{center}
\mbox{\includegraphics[angle=0,scale=0.9]{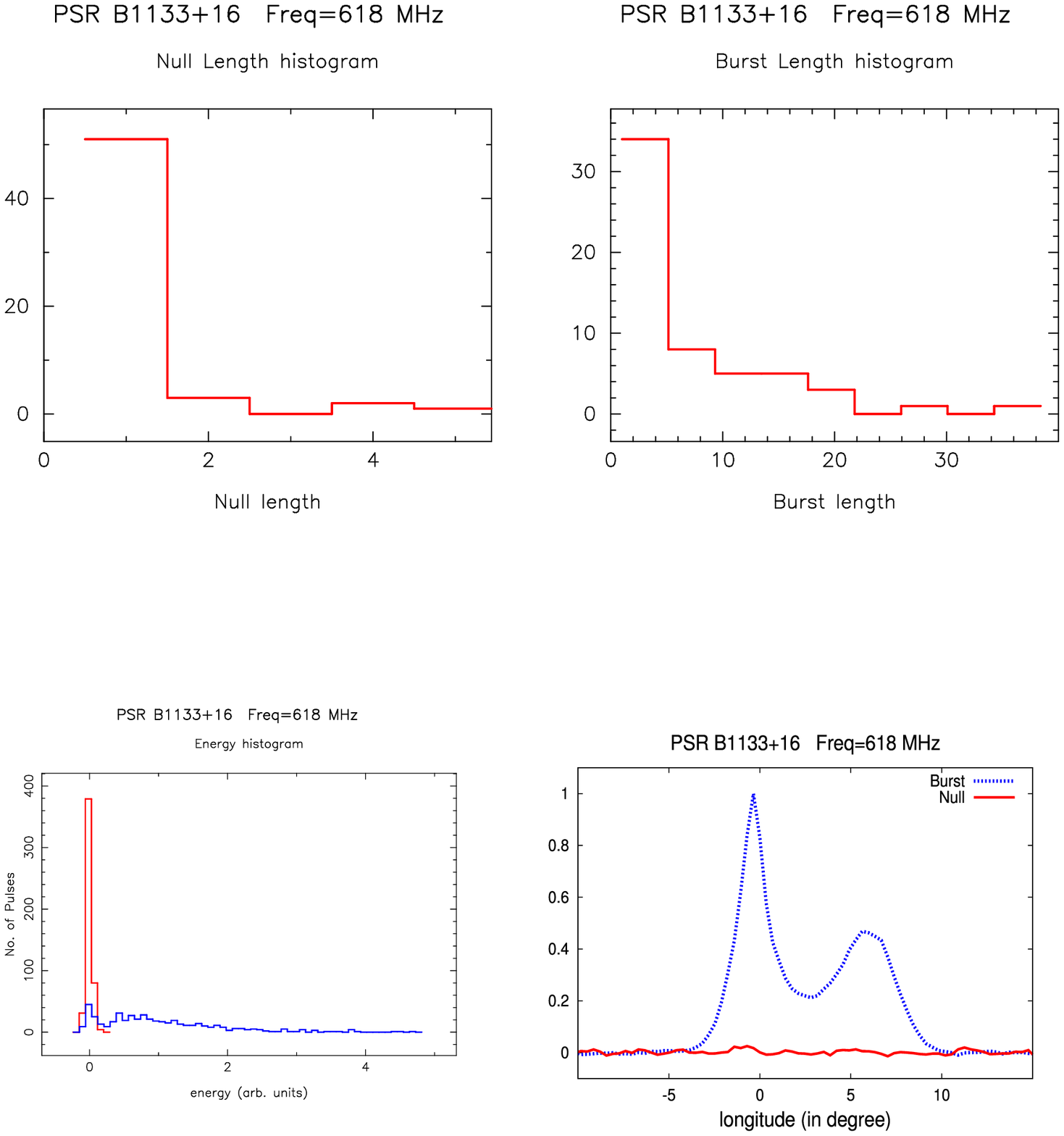}}
\end{center}
\caption{Null length histogram (top left); Burst length histogram (top right); the average energy distribution (bottom left) for on-pulse window (blue hisogram) and off-pulse window (red hisogram); the folded profile (bottom right) for the null pulses (red line, noise like characteristics) and burst pulses (blue line).}
\end{figure*}

\clearpage

\begin{figure*}
\begin{center}
\mbox{\includegraphics[angle=0,scale=0.9]{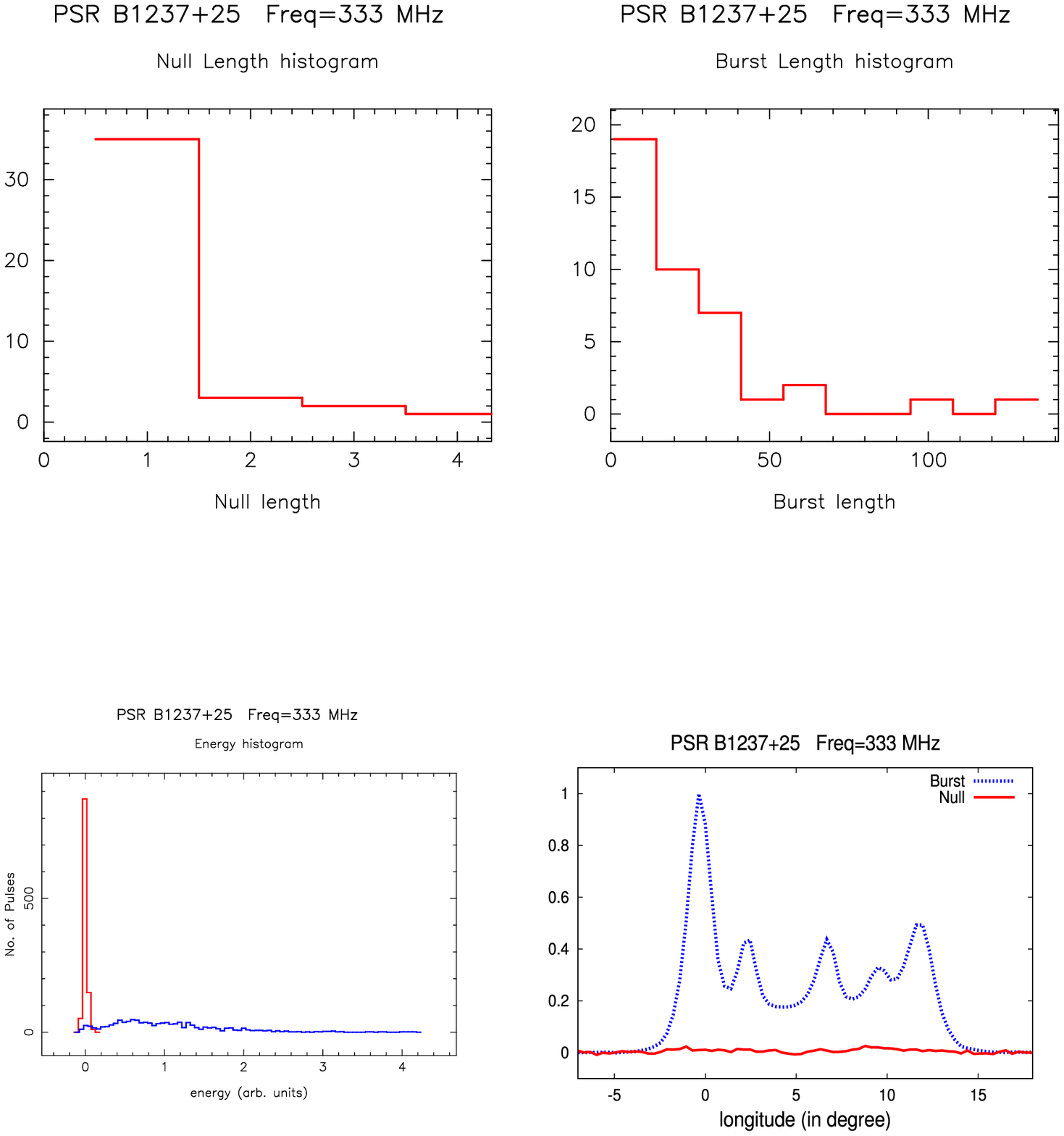}}
\end{center}
\caption{Null length histogram (top left); Burst length histogram (top right); the average energy distribution (bottom left) for on-pulse window (blue hisogram) and off-pulse window (red hisogram); the folded profile (bottom right) for the null pulses (red line, noise like characteristics) and burst pulses (blue line).}
\end{figure*}

\clearpage

\begin{figure*}
\begin{center}
\mbox{\includegraphics[angle=0,scale=0.9]{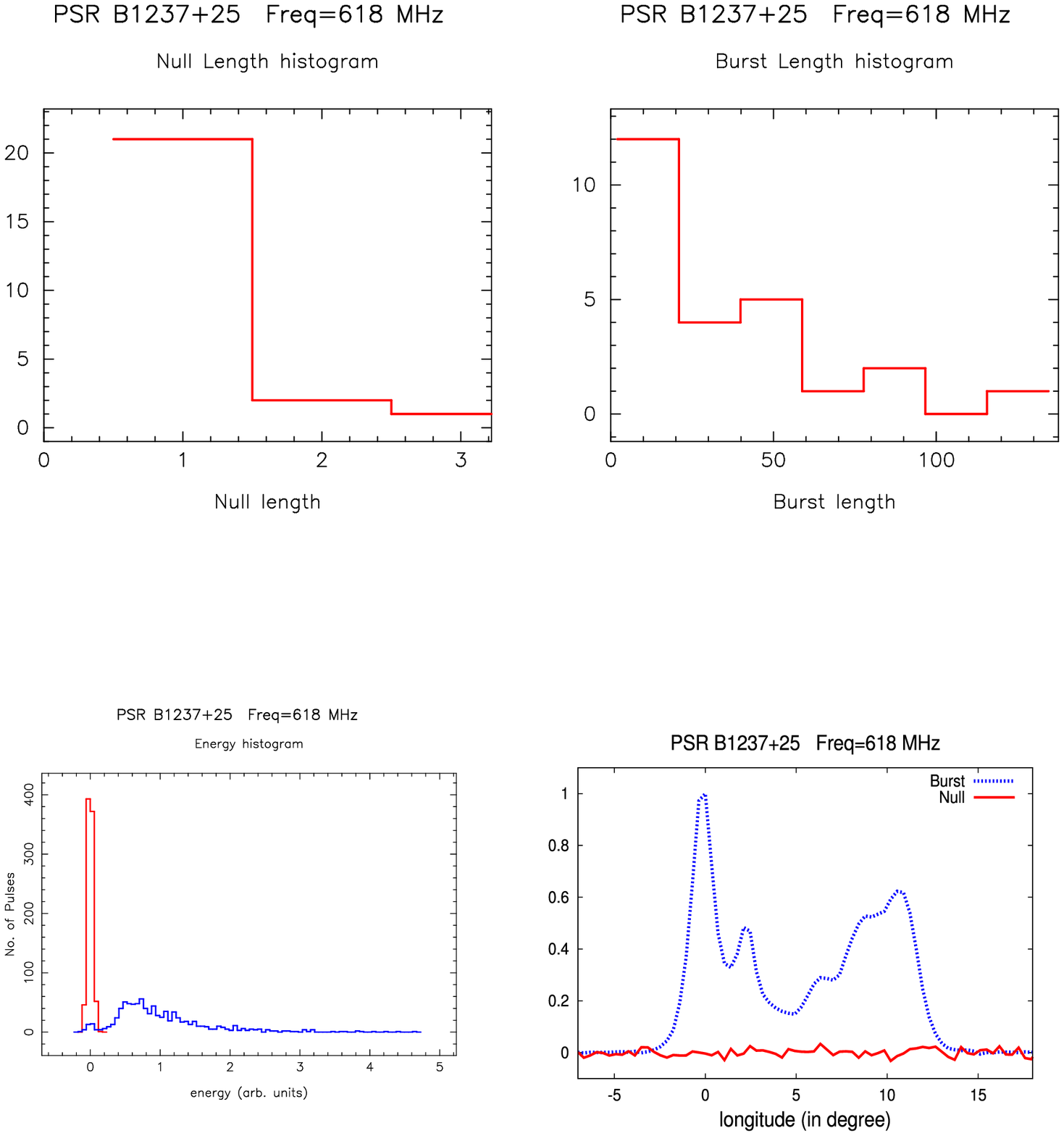}}
\end{center}
\caption{Null length histogram (top left); Burst length histogram (top right); the average energy distribution (bottom left) for on-pulse window (blue hisogram) and off-pulse window (red hisogram); the folded profile (bottom right) for the null pulses (red line, noise like characteristics) and burst pulses (blue line).}
\end{figure*}

\clearpage

\begin{figure*}
\begin{center}
\mbox{\includegraphics[angle=0,scale=0.9]{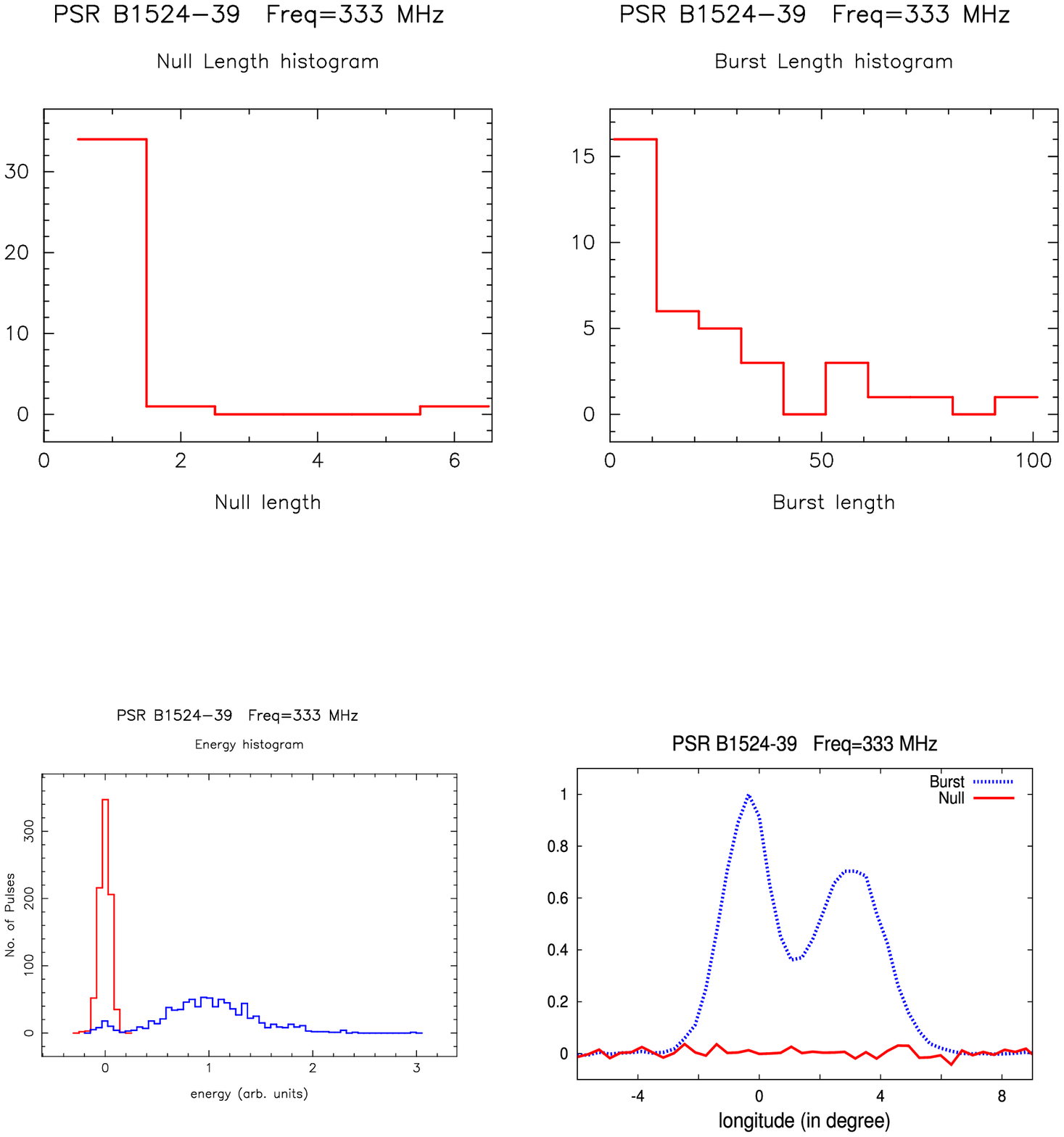}}
\end{center}
\caption{Null length histogram (top left); Burst length histogram (top right); the average energy distribution (bottom left) for on-pulse window (blue hisogram) and off-pulse window (red hisogram); the folded profile (bottom right) for the null pulses (red line, noise like characteristics) and burst pulses (blue line).}
\end{figure*}

\clearpage

\begin{figure*}
\begin{center}
\mbox{\includegraphics[angle=0,scale=0.9]{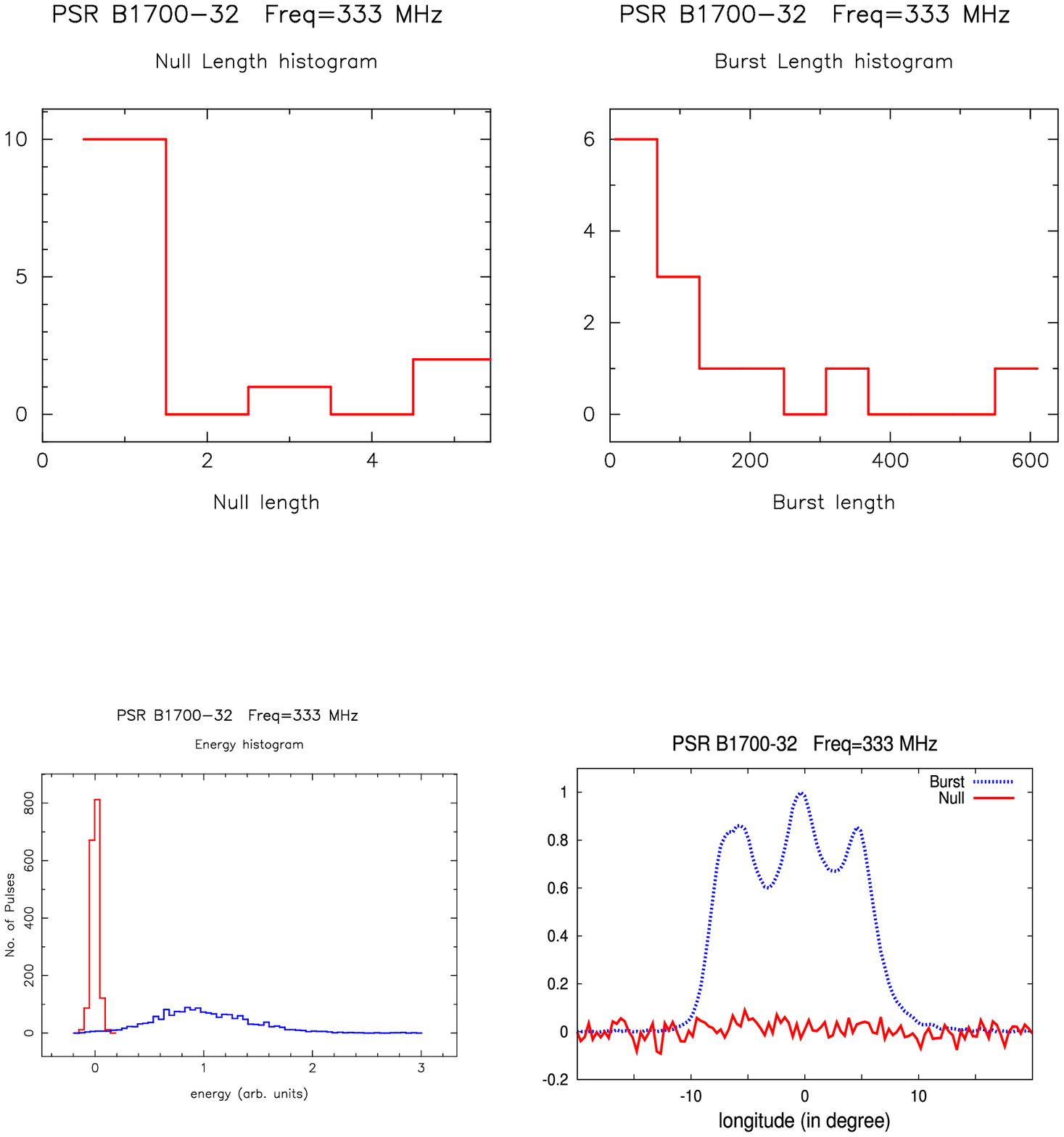}}
\end{center}
\caption{Null length histogram (top left); Burst length histogram (top right); the average energy distribution (bottom left) for on-pulse window (blue hisogram) and off-pulse window (red hisogram); the folded profile (bottom right) for the null pulses (red line, noise like characteristics) and burst pulses (blue line).}
\end{figure*}

\clearpage

\begin{figure*}
\begin{center}
\mbox{\includegraphics[angle=0,scale=0.9]{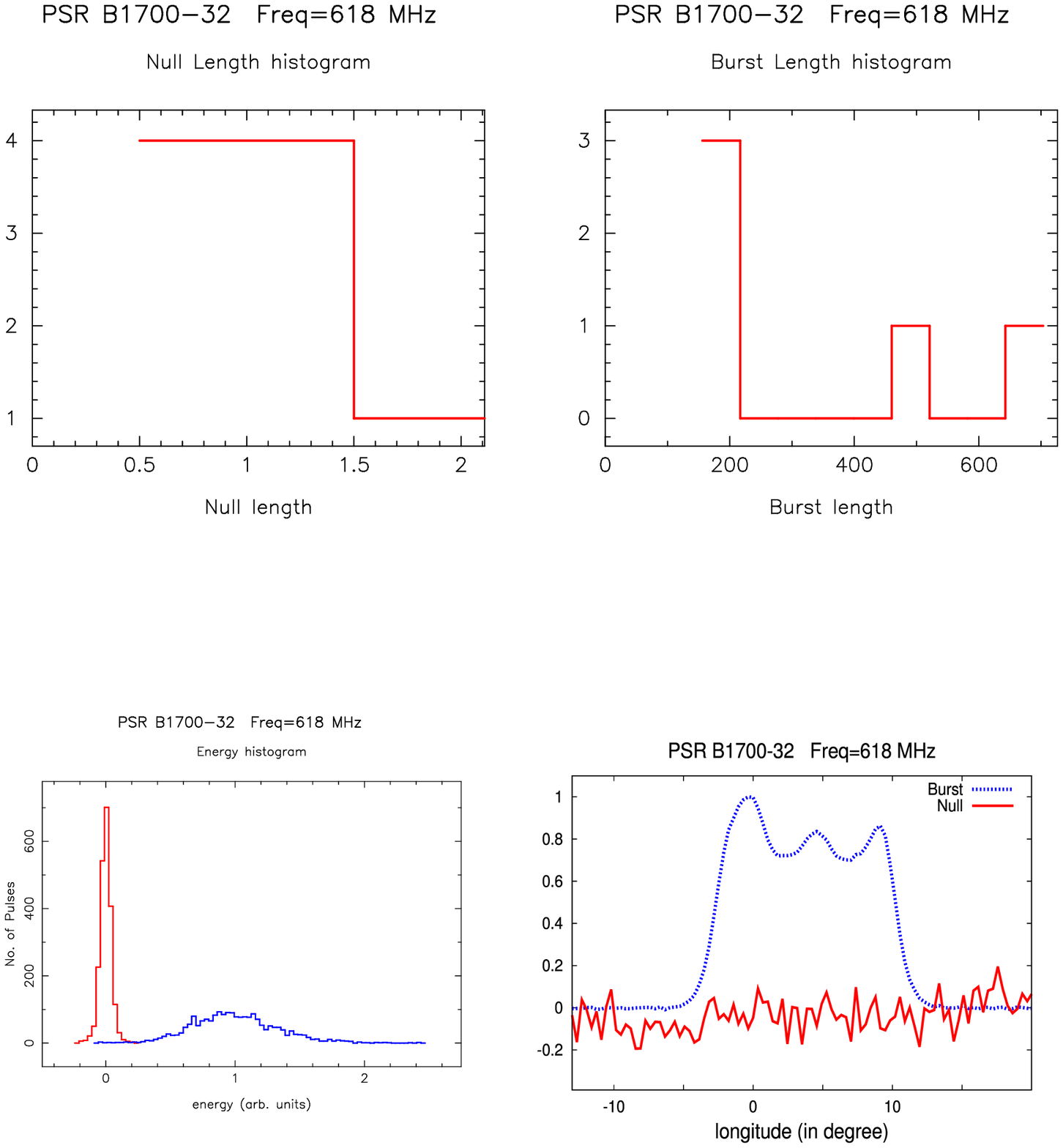}}
\end{center}
\caption{Null length histogram (top left); Burst length histogram (top right); the average energy distribution (bottom left) for on-pulse window (blue hisogram) and off-pulse window (red hisogram); the folded profile (bottom right) for the null pulses (red line, noise like characteristics) and burst pulses (blue line).}
\end{figure*}

\clearpage

\begin{figure*}
\begin{center}
\mbox{\includegraphics[angle=0,scale=0.9]{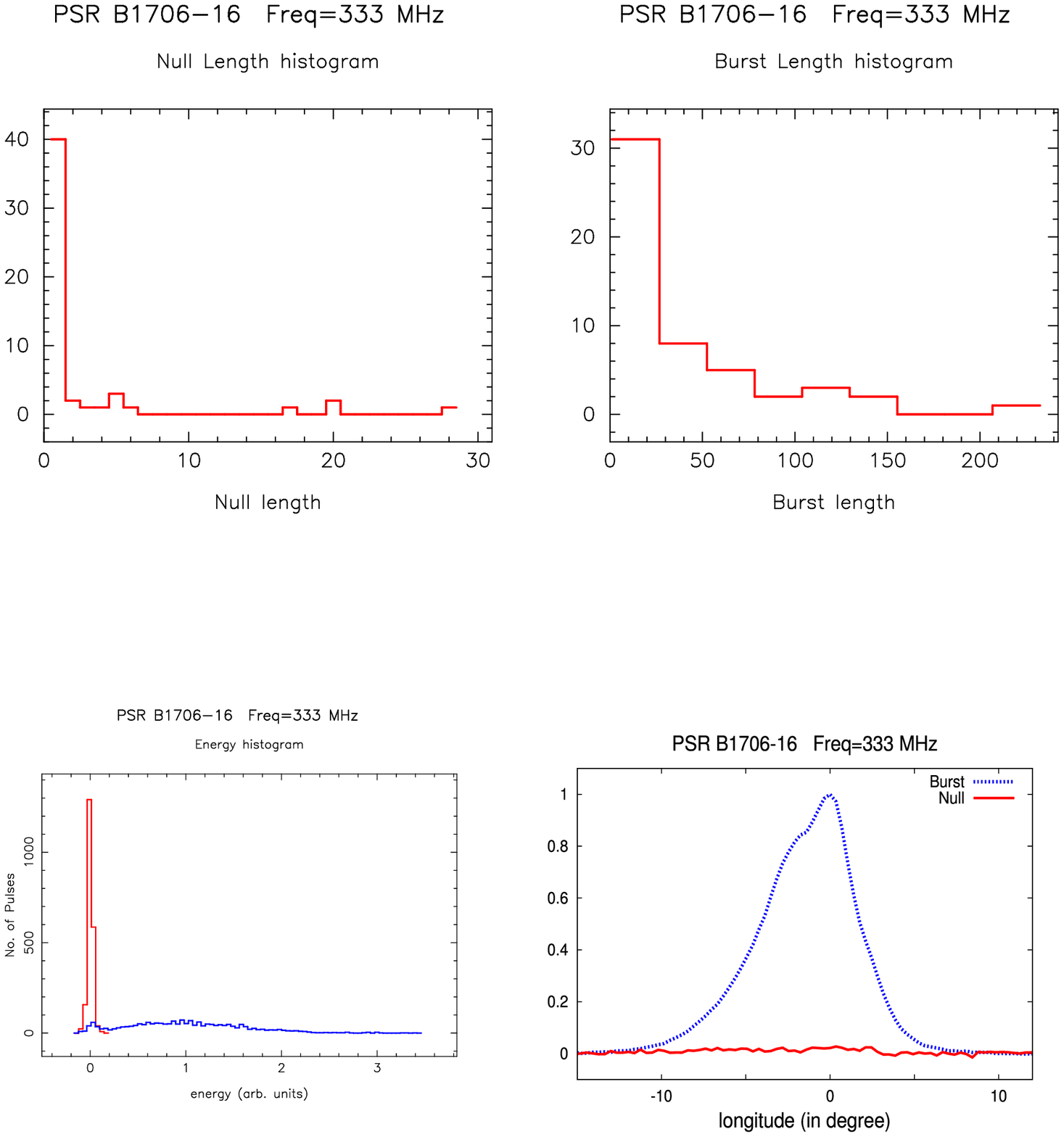}}
\end{center}
\caption{Null length histogram (top left); Burst length histogram (top right); the average energy distribution (bottom left) for on-pulse window (blue hisogram) and off-pulse window (red hisogram); the folded profile (bottom right) for the null pulses (red line, noise like characteristics) and burst pulses (blue line).}
\end{figure*}

\clearpage

\begin{figure*}
\begin{center}
\mbox{\includegraphics[angle=0,scale=0.9]{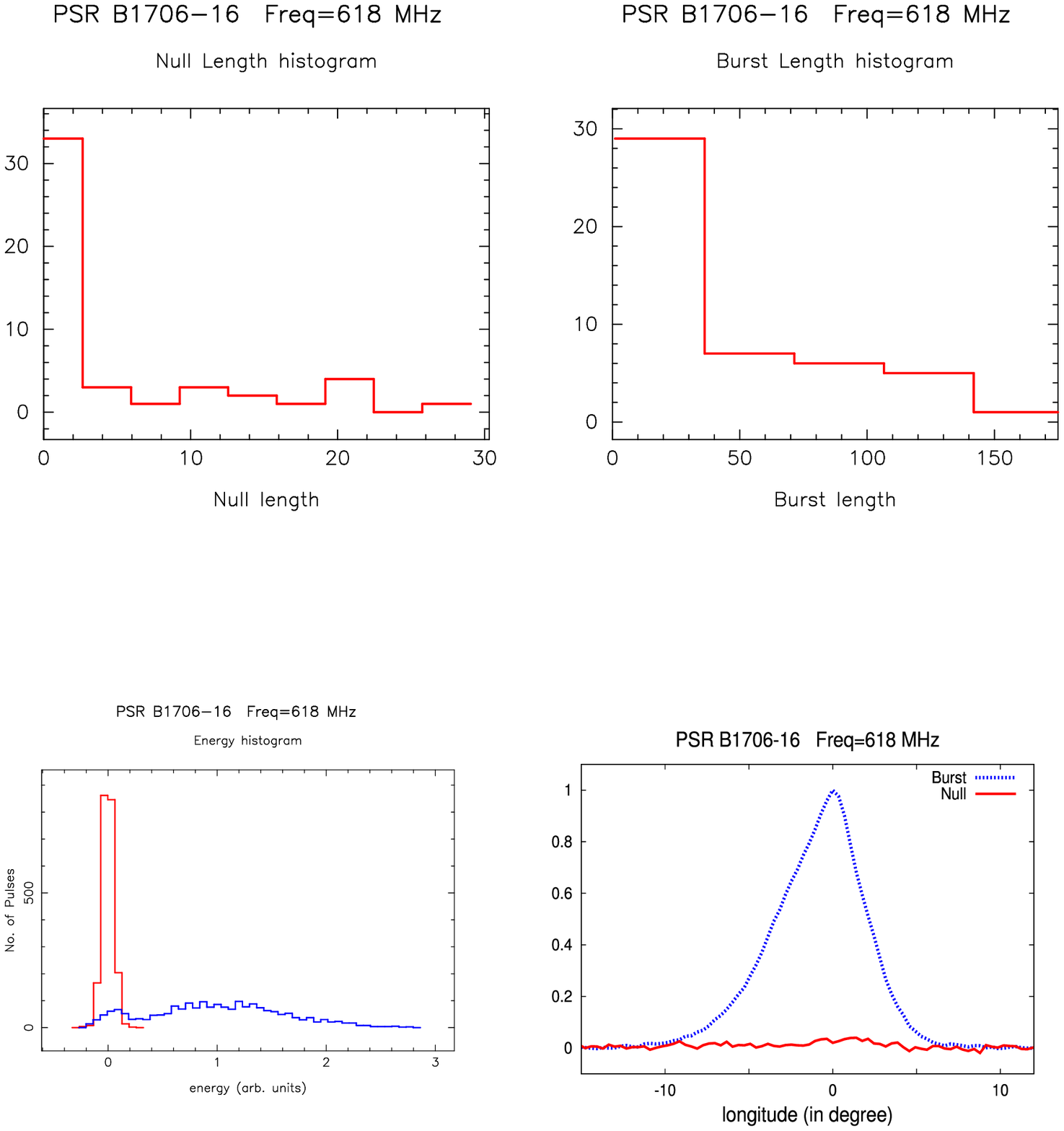}}
\end{center}
\caption{Null length histogram (top left); Burst length histogram (top right); the average energy distribution (bottom left) for on-pulse window (blue hisogram) and off-pulse window (red hisogram); the folded profile (bottom right) for the null pulses (red line, noise like characteristics) and burst pulses (blue line).}
\end{figure*}

\clearpage

\begin{figure*}
\begin{center}
\mbox{\includegraphics[angle=0,scale=0.9]{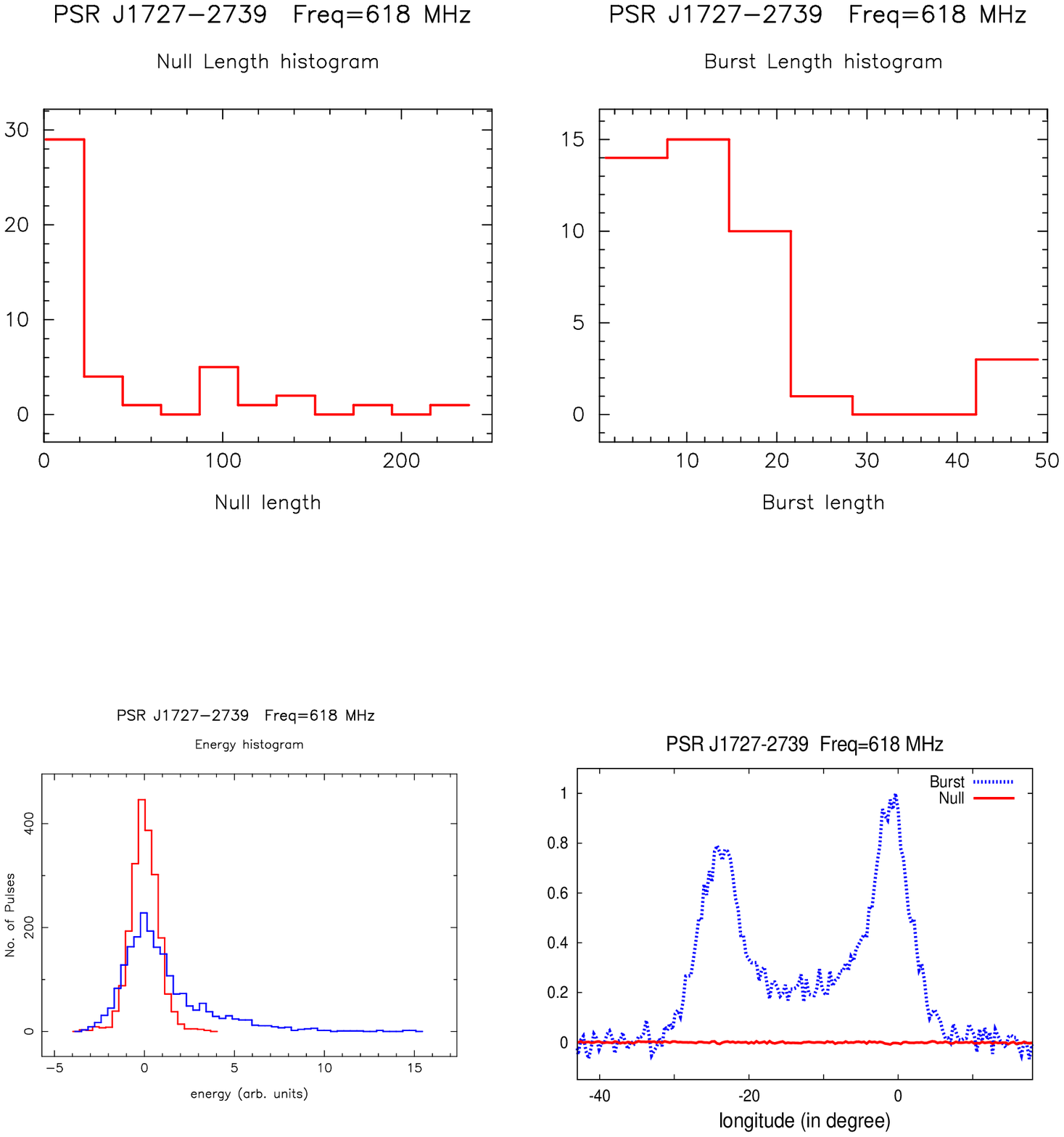}}
\end{center}
\caption{Null length histogram (top left); Burst length histogram (top right); the average energy distribution (bottom left) for on-pulse window (blue hisogram) and off-pulse window (red hisogram); the folded profile (bottom right) for the null pulses (red line, noise like characteristics) and burst pulses (blue line).}
\end{figure*}

\clearpage

\begin{figure*}
\begin{center}
\mbox{\includegraphics[angle=0,scale=0.9]{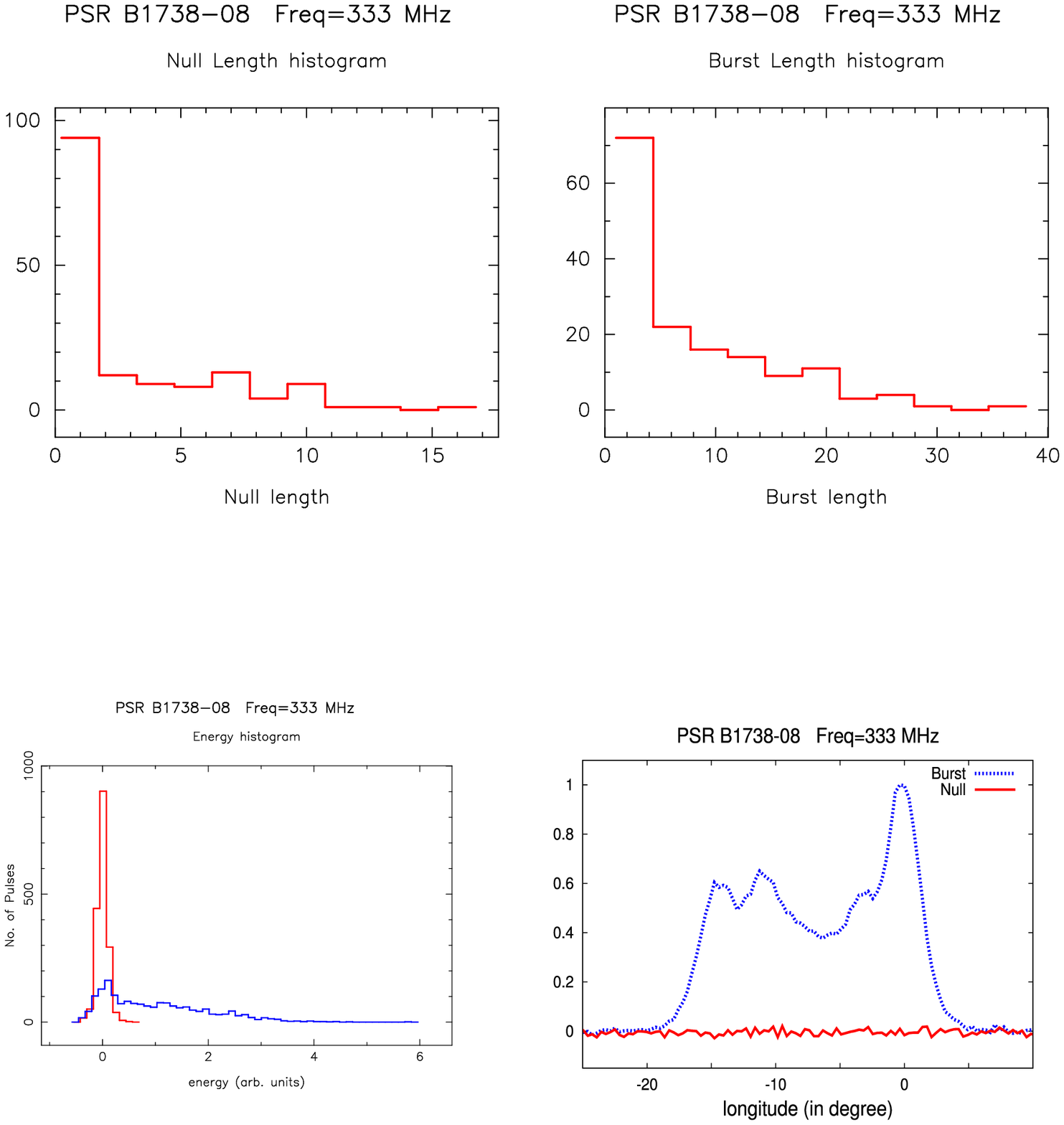}}
\end{center}
\caption{Null length histogram (top left); Burst length histogram (top right); the average energy distribution (bottom left) for on-pulse window (blue hisogram) and off-pulse window (red hisogram); the folded profile (bottom right) for the null pulses (red line, noise like characteristics) and burst pulses (blue line).}
\end{figure*}

\clearpage

\begin{figure*}
\begin{center}
\mbox{\includegraphics[angle=0,scale=0.9]{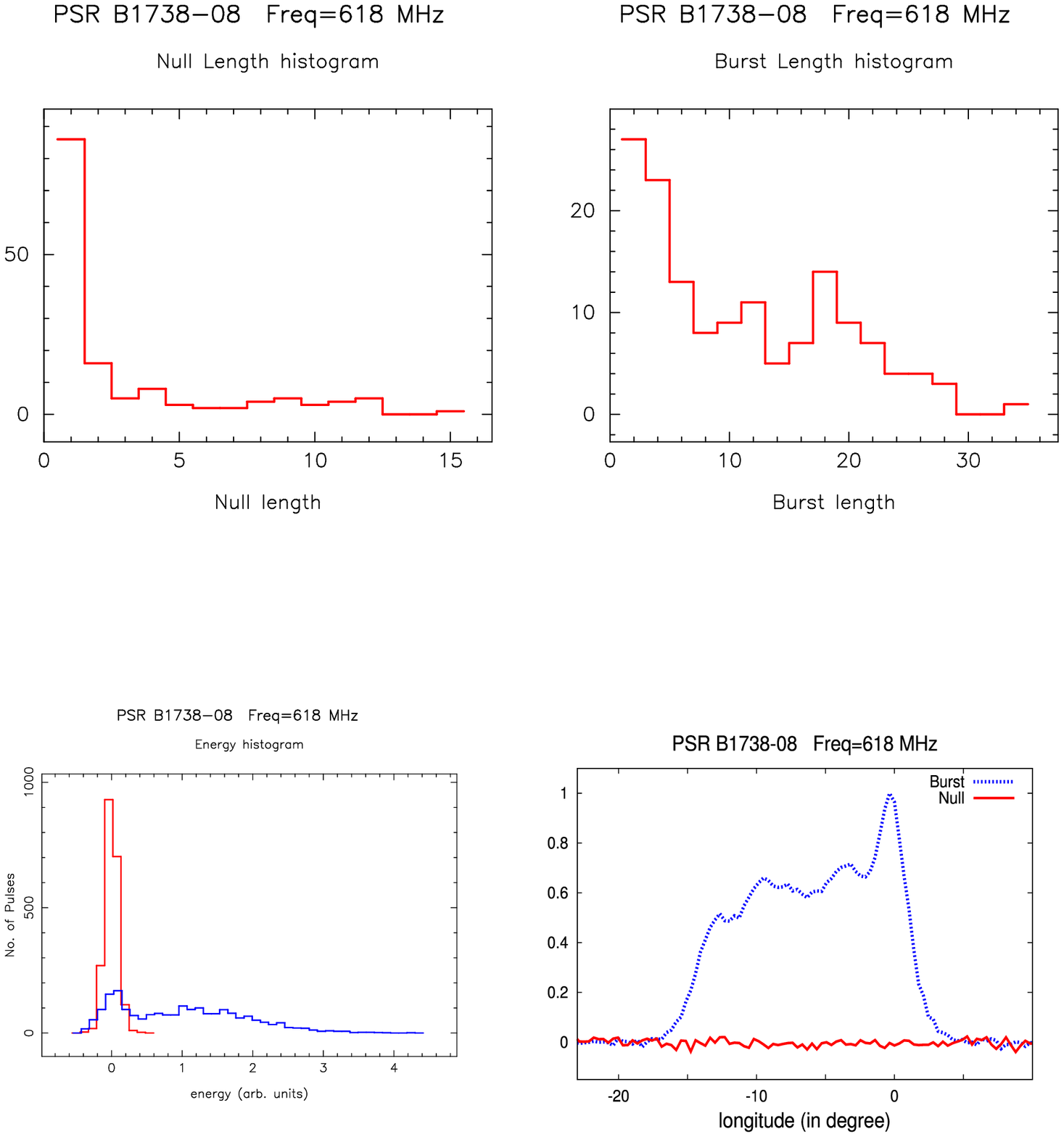}}
\end{center}
\caption{Null length histogram (top left); Burst length histogram (top right); the average energy distribution (bottom left) for on-pulse window (blue hisogram) and off-pulse window (red hisogram); the folded profile (bottom right) for the null pulses (red line, noise like characteristics) and burst pulses (blue line).}
\end{figure*}

\clearpage

\begin{figure*}
\begin{center}
\mbox{\includegraphics[angle=0,scale=0.9]{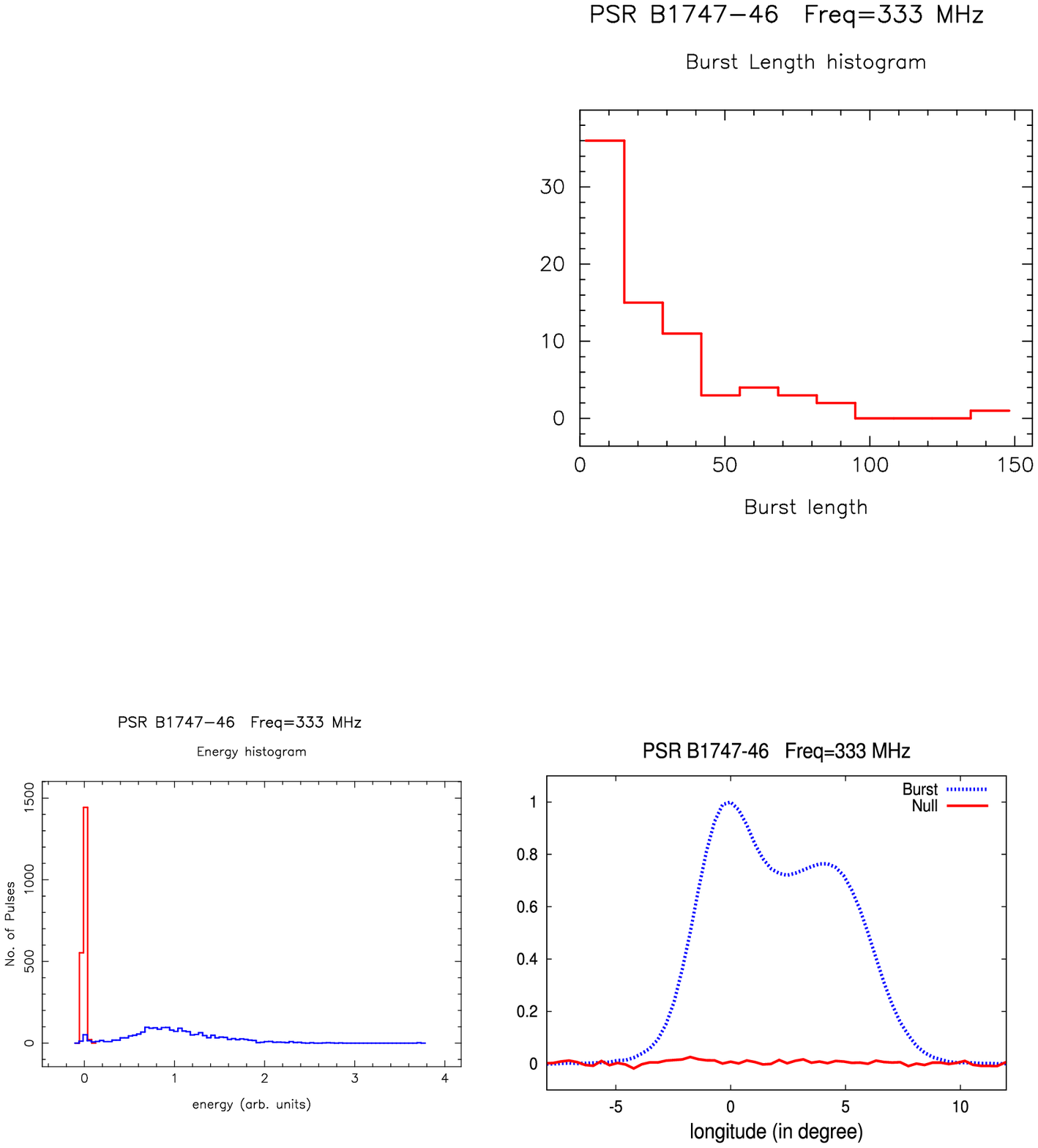}}
\end{center}
\caption{75 single period nulls, no null length histogram; Burst length histogram (top right); the average energy distribution (bottom left) for on-pulse window (blue hisogram) and off-pulse window (red hisogram); the folded profile (bottom right) for the null pulses (red line, noise like characteristics) and burst pulses (blue line).}
\end{figure*}

\clearpage

\begin{figure*}
\begin{center}
\mbox{\includegraphics[angle=0,scale=0.9]{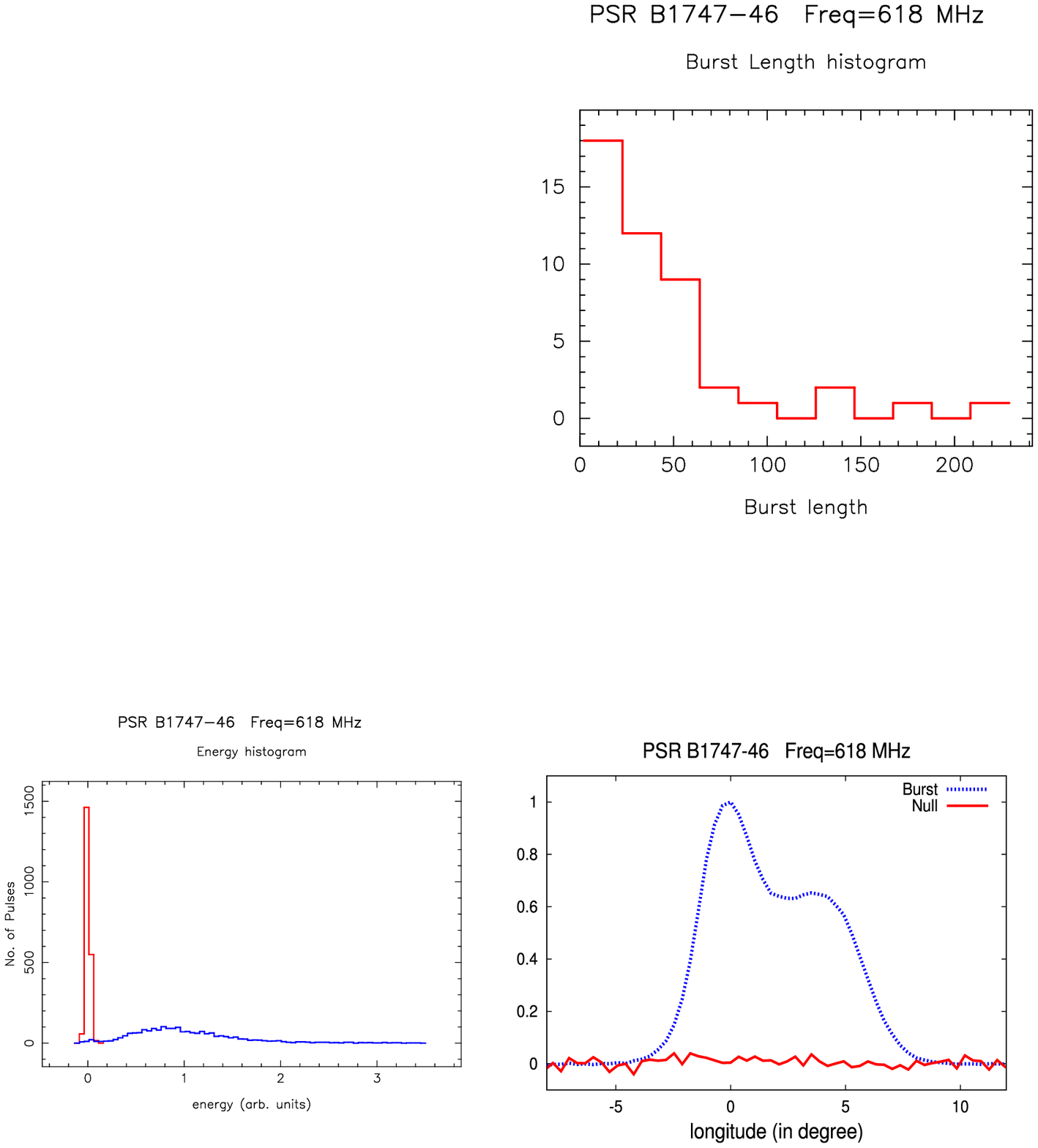}}
\end{center}
\caption{46 single period nulls, no null length histogram; Burst length histogram (top right); the average energy distribution (bottom left) for on-pulse window (blue hisogram) and off-pulse window (red hisogram); the folded profile (bottom right) for the null pulses (red line, noise like characteristics) and burst pulses (blue line).}
\end{figure*}

\clearpage

\begin{figure*}
\begin{center}
\mbox{\includegraphics[angle=0,scale=0.9]{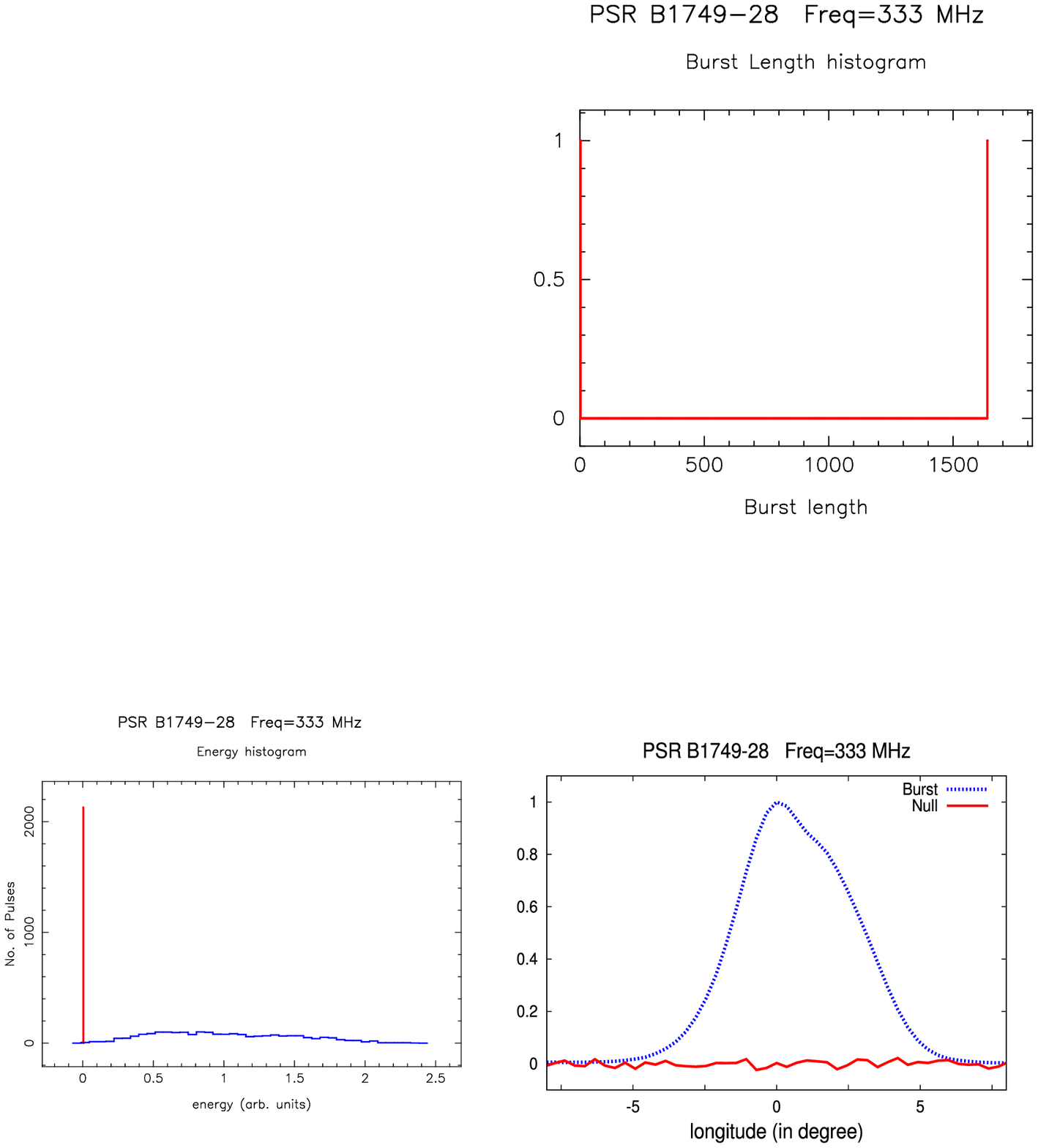}}
\end{center}
\caption{2 two period nulls, no null length histogram; Burst length histogram (top right); the average energy distribution (bottom left) for on-pulse window (blue hisogram) and off-pulse window (red hisogram); the folded profile (bottom right) for the null pulses (red line, noise like characteristics) and burst pulses (blue line).}
\end{figure*}

\clearpage

\begin{figure*}
\begin{center}
\mbox{\includegraphics[angle=0,scale=0.9]{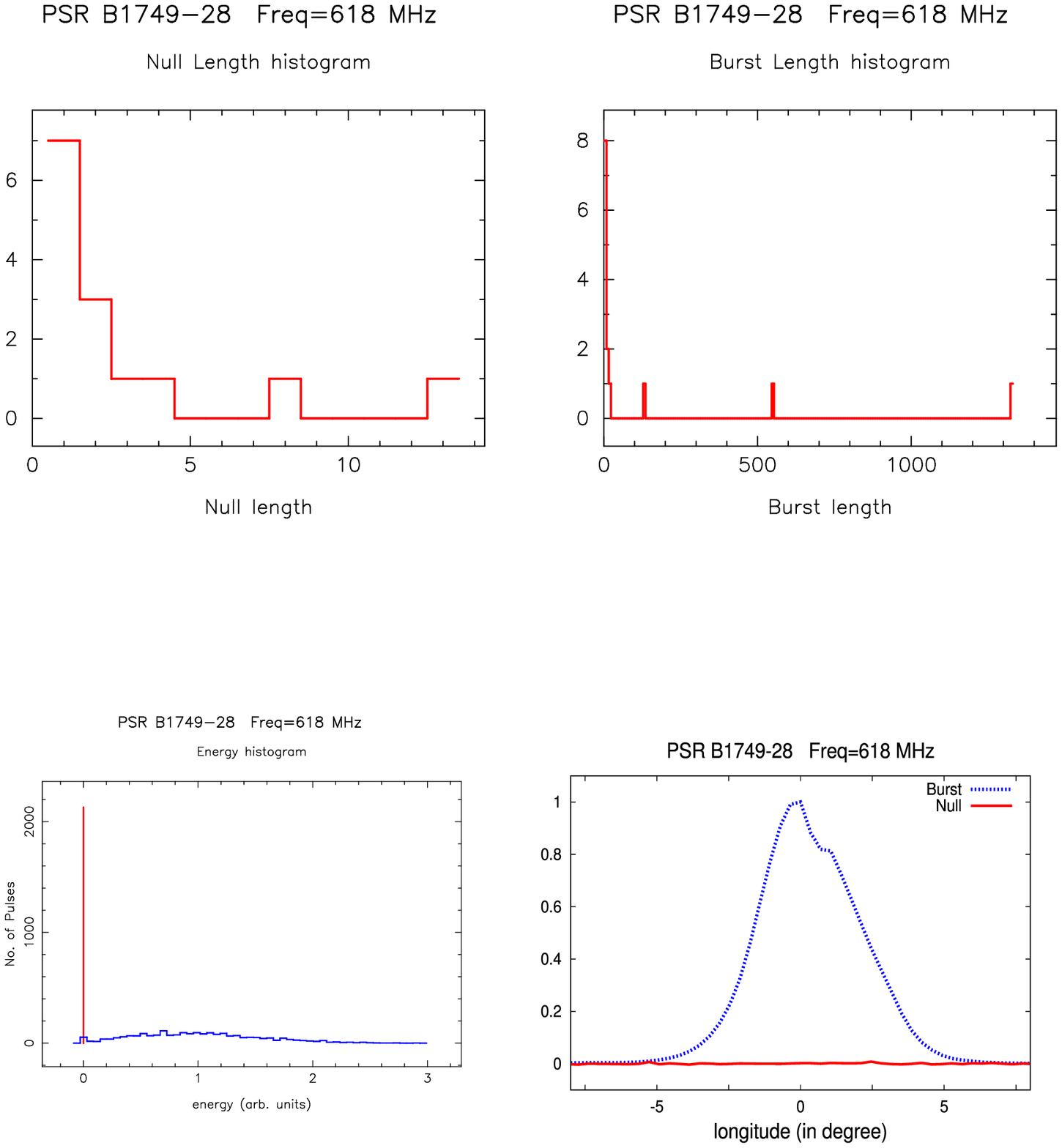}}
\end{center}
\caption{Null length histogram (top left); Burst length histogram (top right); the average energy distribution (bottom left) for on-pulse window (blue hisogram) and off-pulse window (red hisogram); the folded profile (bottom right) for the null pulses (red line, noise like characteristics) and burst pulses (blue line).}
\end{figure*}

\clearpage

\begin{figure*}
\begin{center}
\mbox{\includegraphics[angle=0,scale=0.9]{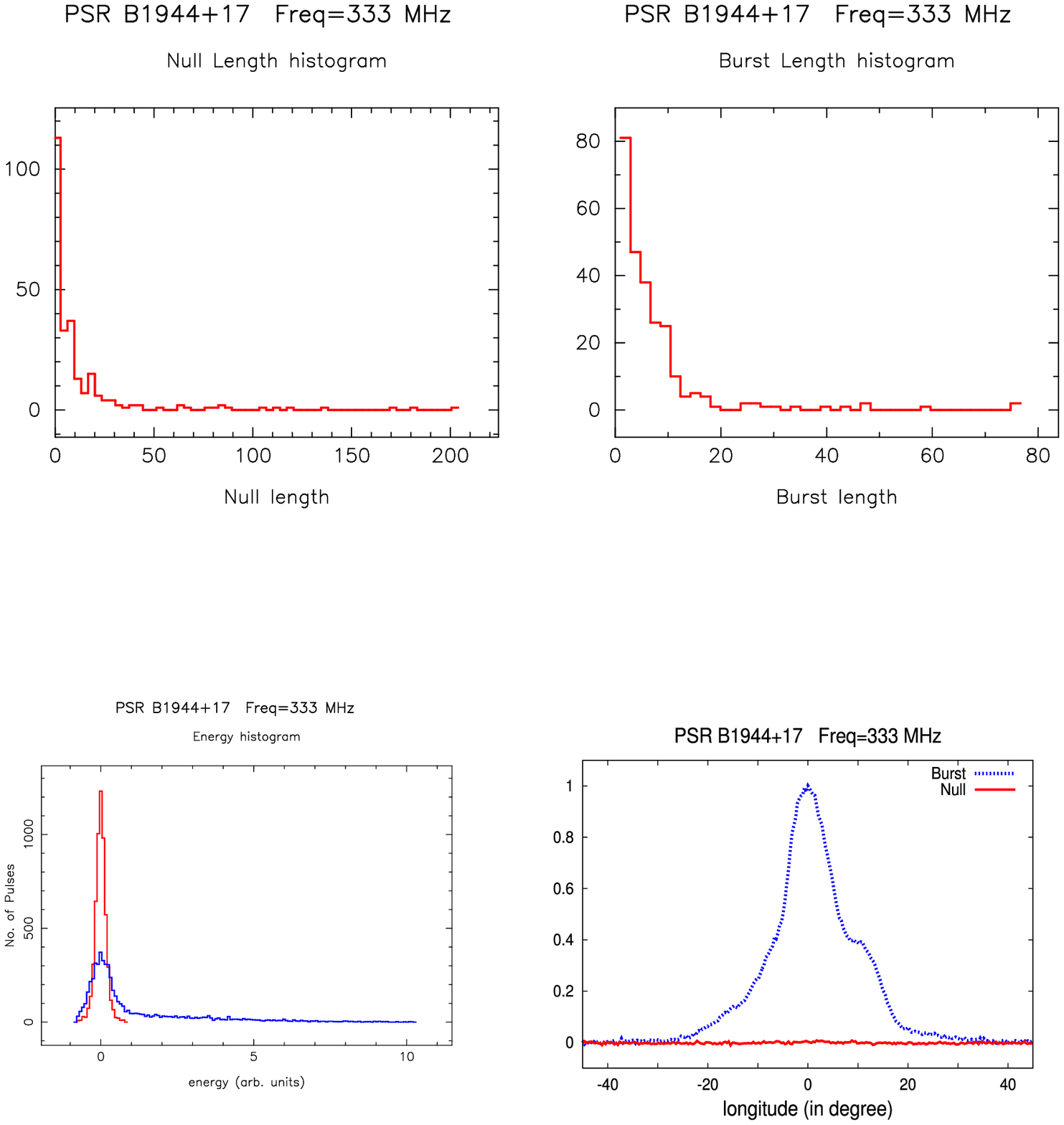}}
\end{center}
\caption{Null length histogram (top left); Burst length histogram (top right); the average energy distribution (bottom left) for on-pulse window (blue hisogram) and off-pulse window (red hisogram); the folded profile (bottom right) for the null pulses (red line, noise like characteristics) and burst pulses (blue line).}
\end{figure*}

\clearpage

\begin{figure*}
\begin{center}
\mbox{\includegraphics[angle=0,scale=0.9]{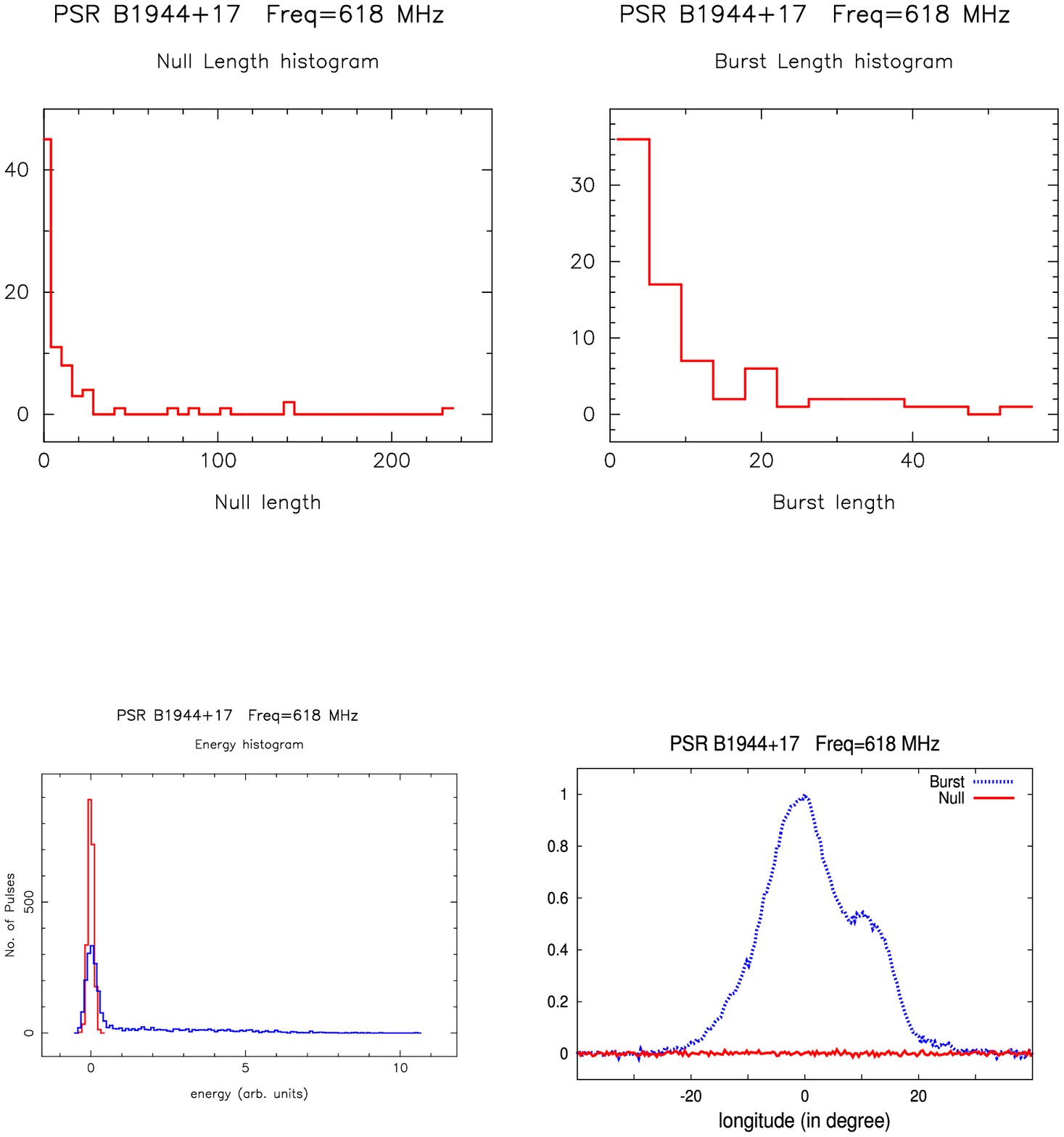}}
\end{center}
\caption{Null length histogram (top left); Burst length histogram (top right); the average energy distribution (bottom left) for on-pulse window (blue hisogram) and off-pulse window (red hisogram); the folded profile (bottom right) for the null pulses (red line, noise like characteristics) and burst pulses (blue line).}
\end{figure*}

\clearpage

\begin{figure*}
\begin{center}
\mbox{\includegraphics[angle=0,scale=0.9]{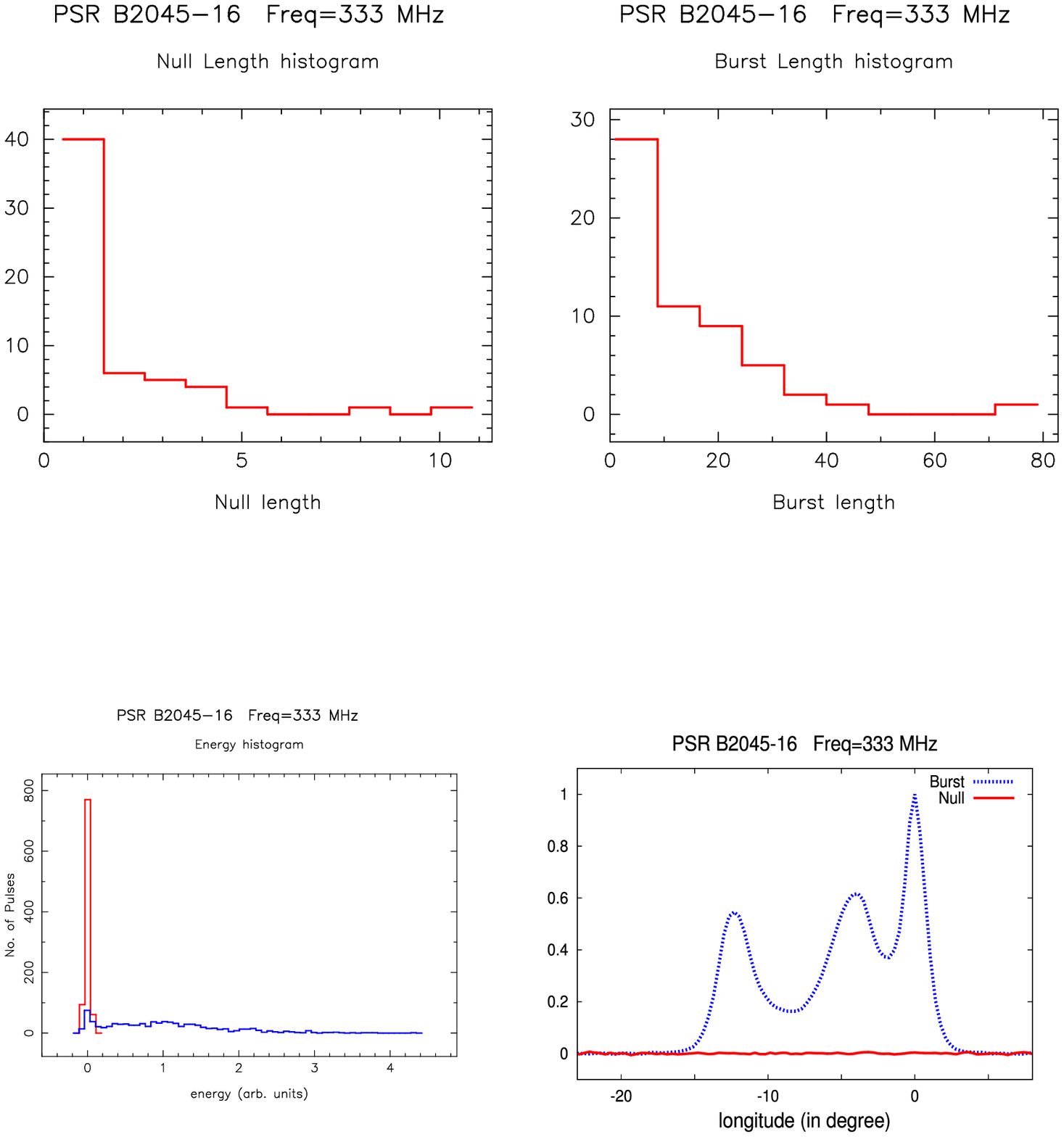}}
\end{center}
\caption{Null length histogram (top left); Burst length histogram (top right); the average energy distribution (bottom left) for on-pulse window (blue hisogram) and off-pulse window (red hisogram); the folded profile (bottom right) for the null pulses (red line, noise like characteristics) and burst pulses (blue line).}
\end{figure*}

\clearpage

\begin{figure*}
\begin{center}
\mbox{\includegraphics[angle=0,scale=0.9]{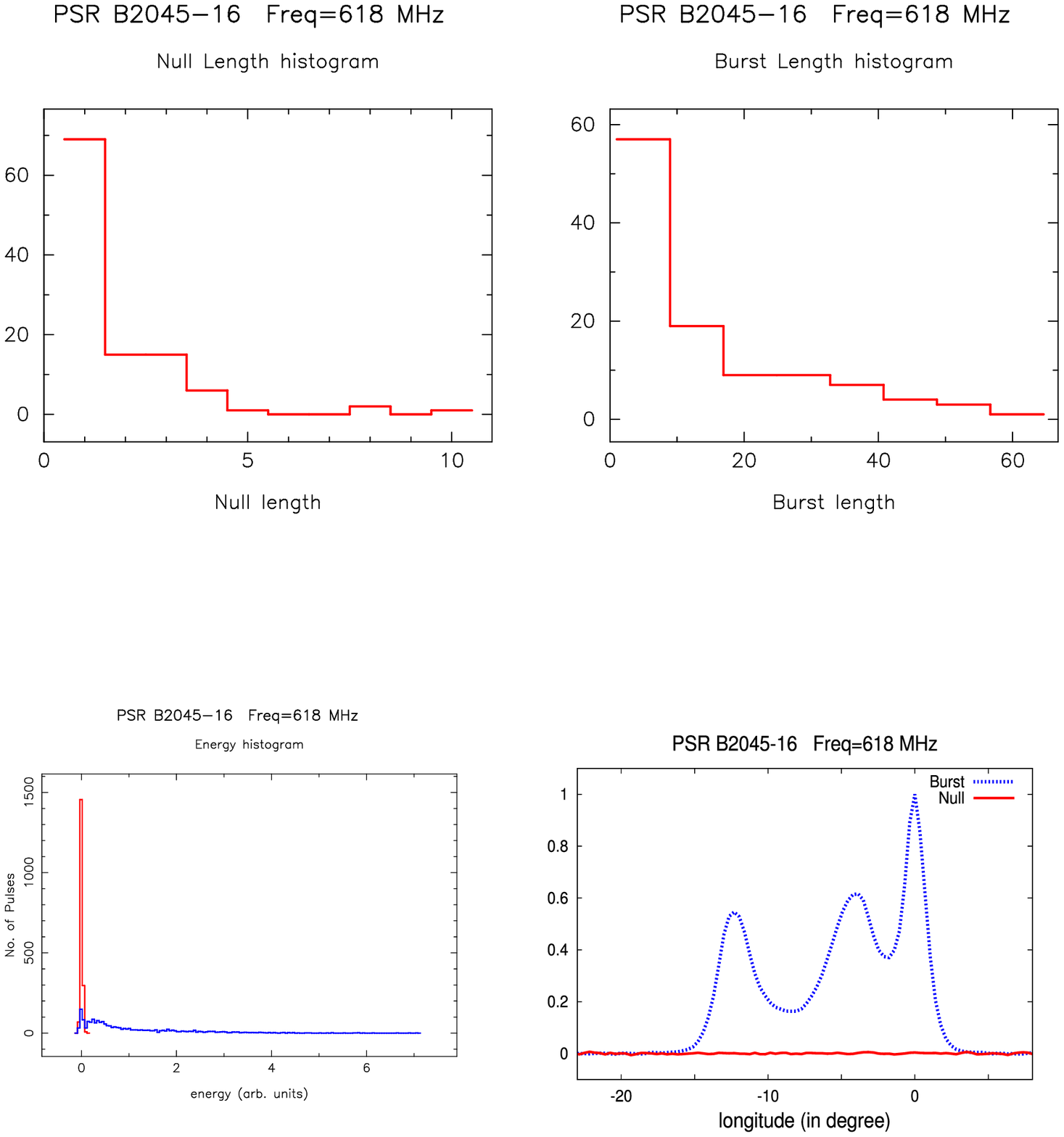}}
\end{center}
\caption{Null length histogram (top left); Burst length histogram (top right); the average energy distribution (bottom left) for on-pulse window (blue hisogram) and off-pulse window (red hisogram); the folded profile (bottom right) for the null pulses (red line, noise like characteristics) and burst pulses (blue line).}
\end{figure*}

\clearpage

\begin{figure*}
\begin{center}
\mbox{\includegraphics[angle=0,scale=0.9]{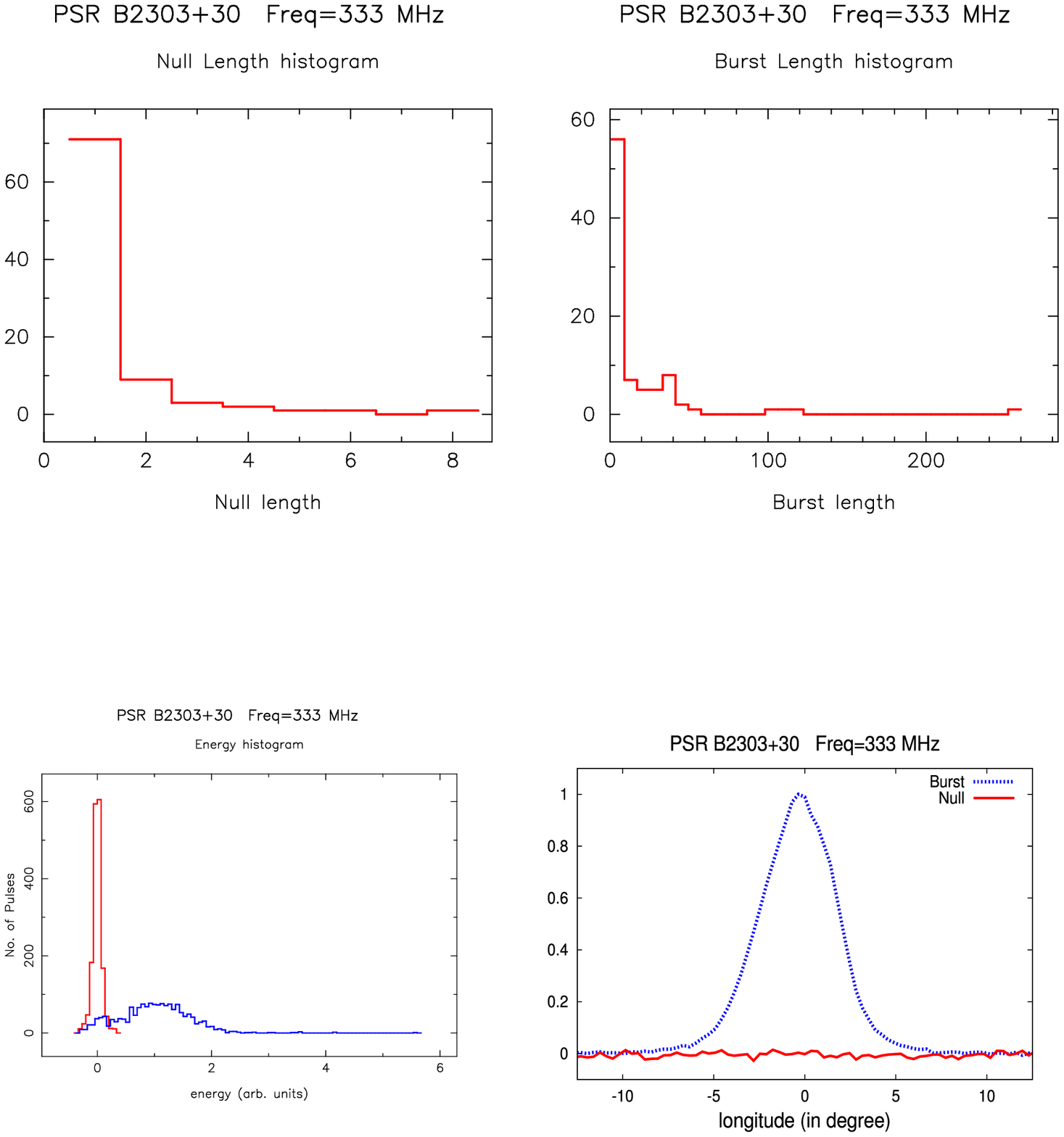}}
\end{center}
\caption{Null length histogram (top left); Burst length histogram (top right); the average energy distribution (bottom left) for on-pulse window (blue hisogram) and off-pulse window (red hisogram); the folded profile (bottom right) for the null pulses (red line, noise like characteristics) and burst pulses (blue line).}
\end{figure*}

\clearpage

\begin{figure*}
\begin{center}
\mbox{\includegraphics[angle=0,scale=0.9]{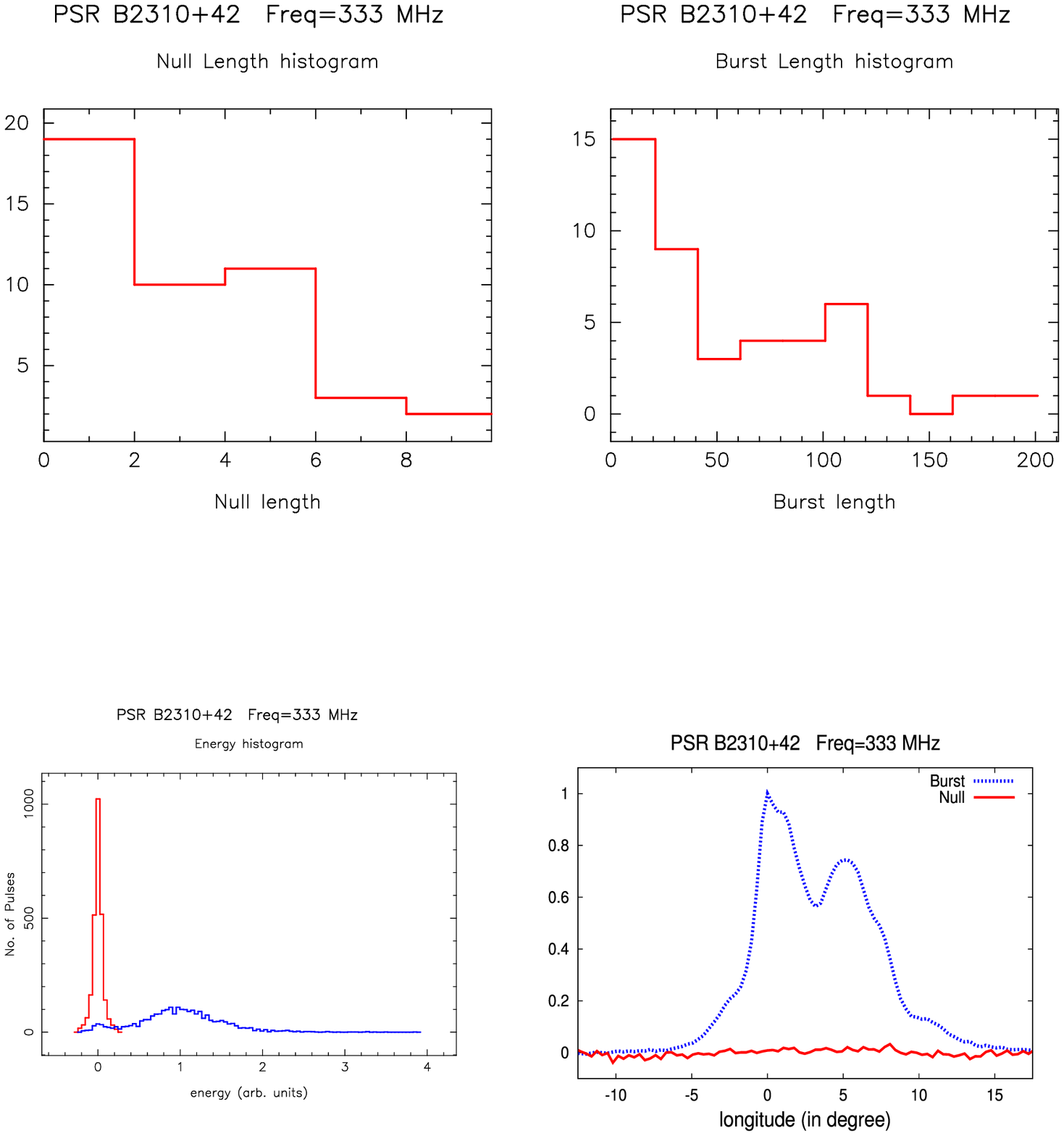}}
\end{center}
\caption{Null length histogram (top left); Burst length histogram (top right); the average energy distribution (bottom left) for on-pulse window (blue hisogram) and off-pulse window (red hisogram); the folded profile (bottom right) for the null pulses (red line, noise like characteristics) and burst pulses (blue line).}
\end{figure*}

\clearpage

\begin{figure*}
\begin{center}
\mbox{\includegraphics[angle=0,scale=0.9]{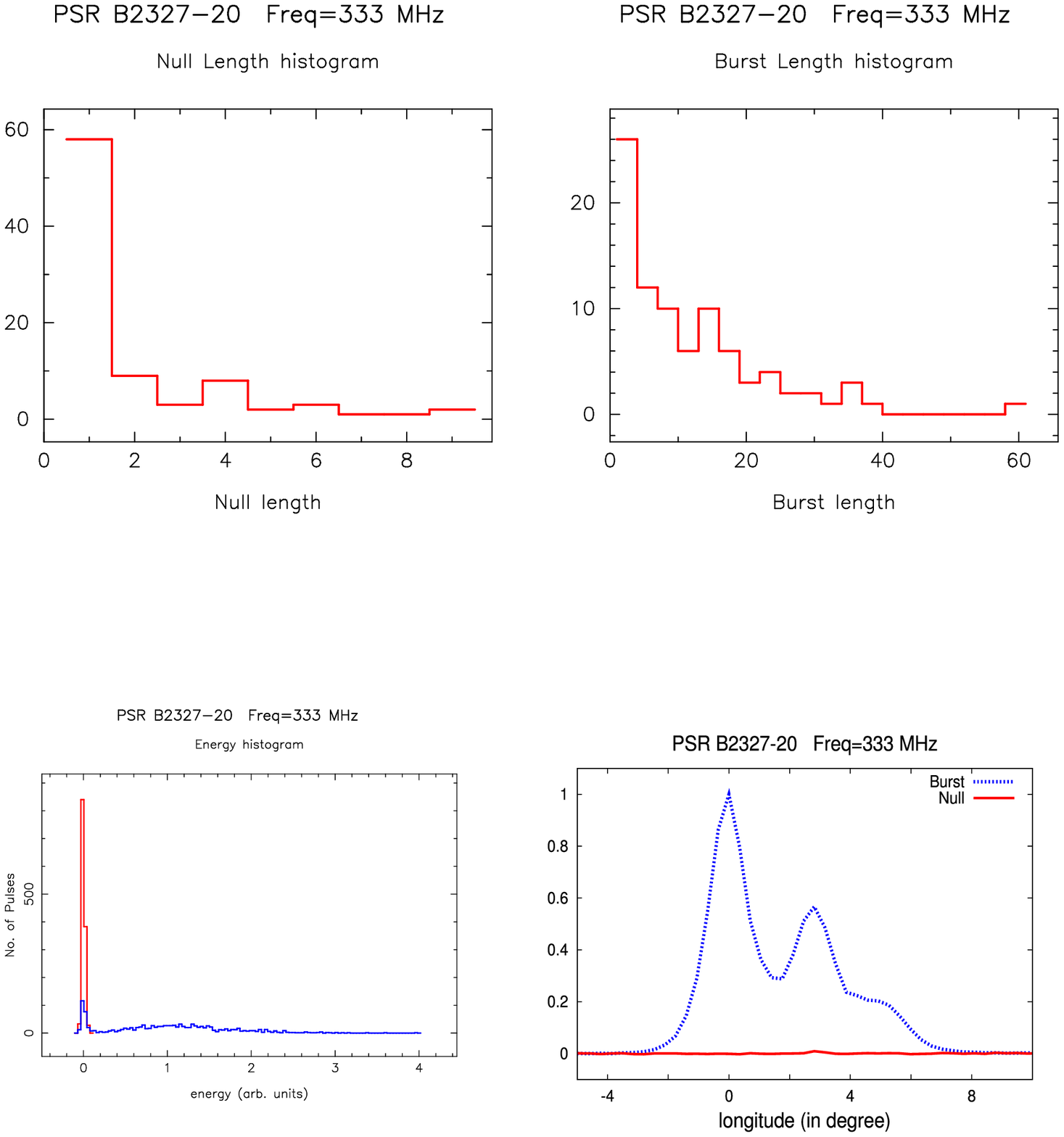}}
\end{center}
\caption{Null length histogram (top left); Burst length histogram (top right); the average energy distribution (bottom left) for on-pulse window (blue hisogram) and off-pulse window (red hisogram); the folded profile (bottom right) for the null pulses (red line, noise like characteristics) and burst pulses (blue line).}
\end{figure*}

\clearpage

\begin{figure*}
\begin{center}
\mbox{\includegraphics[angle=0,scale=0.9]{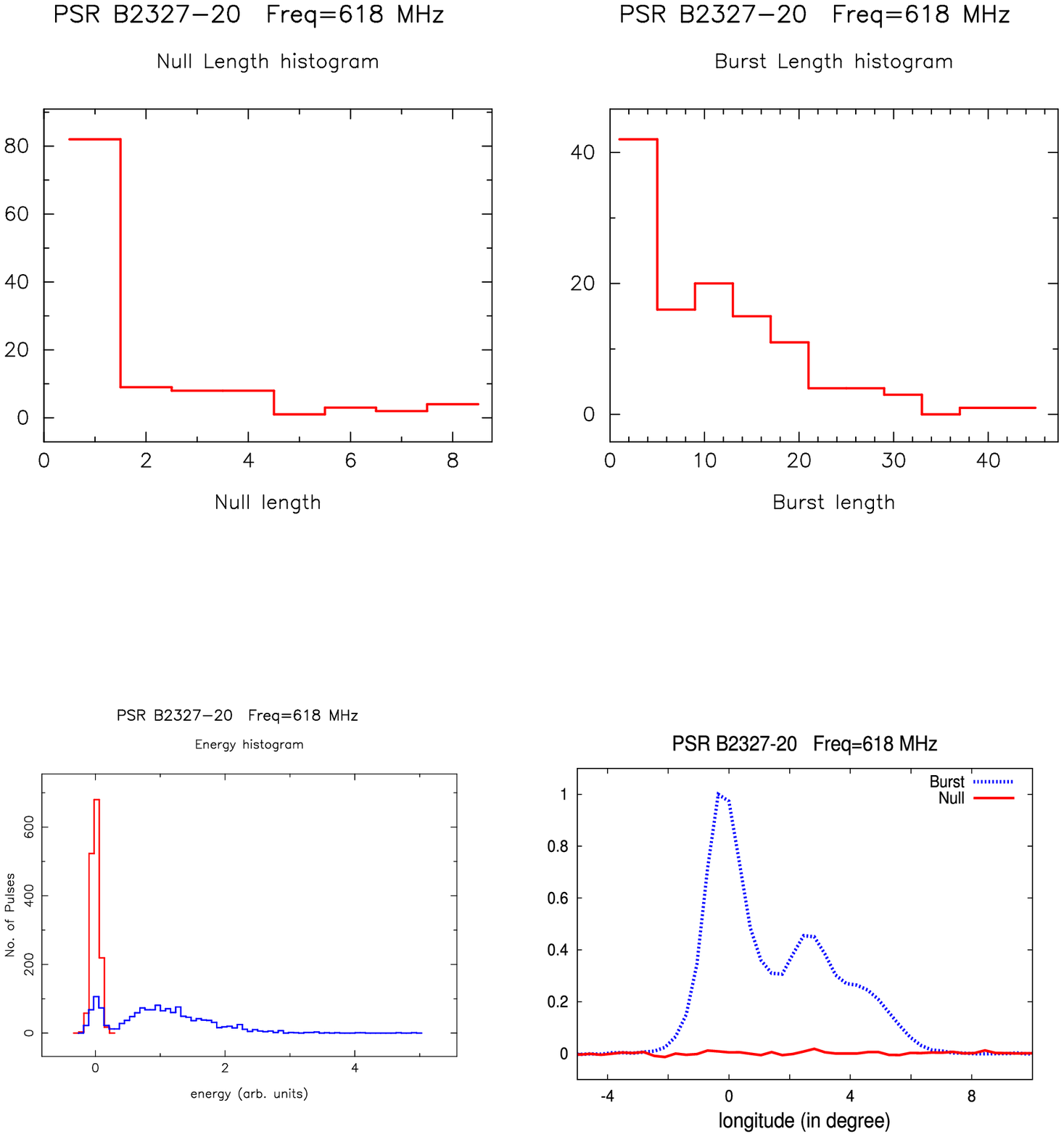}}
\end{center}
\caption{Null length histogram (top left); Burst length histogram (top right); the average energy distribution (bottom left) for on-pulse window (blue hisogram) and off-pulse window (red hisogram); the folded profile (bottom right) for the null pulses (red line, noise like characteristics) and burst pulses (blue line).}
\end{figure*}

%% file: appendix2.tex
\clearpage

\begin{figure*}
\begin{center}
\begin{tabular}{@{}lr@{}}
{\mbox{\includegraphics[angle=0,scale=0.4]{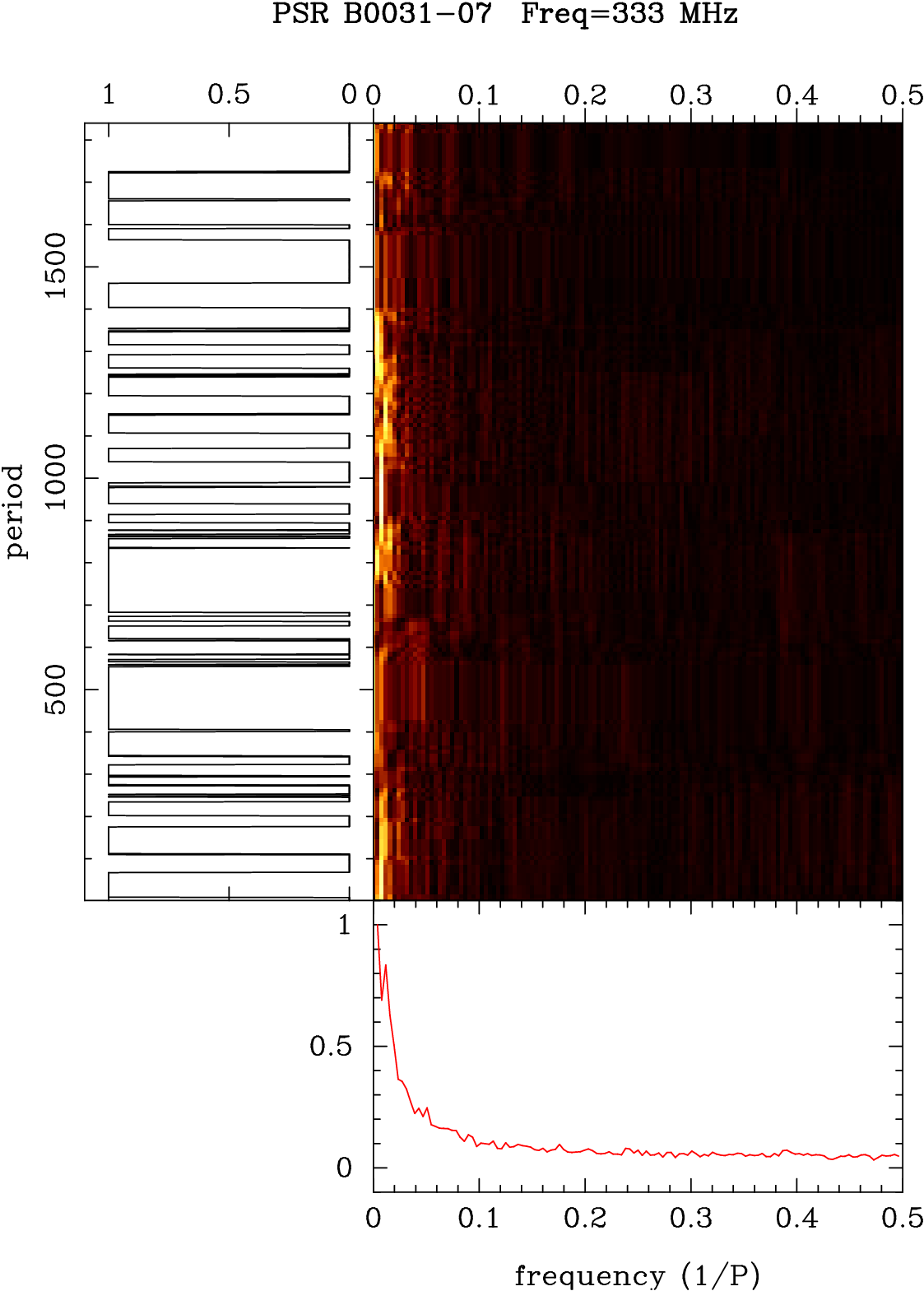}}}&
{\mbox{\includegraphics[angle=0,scale=0.4]{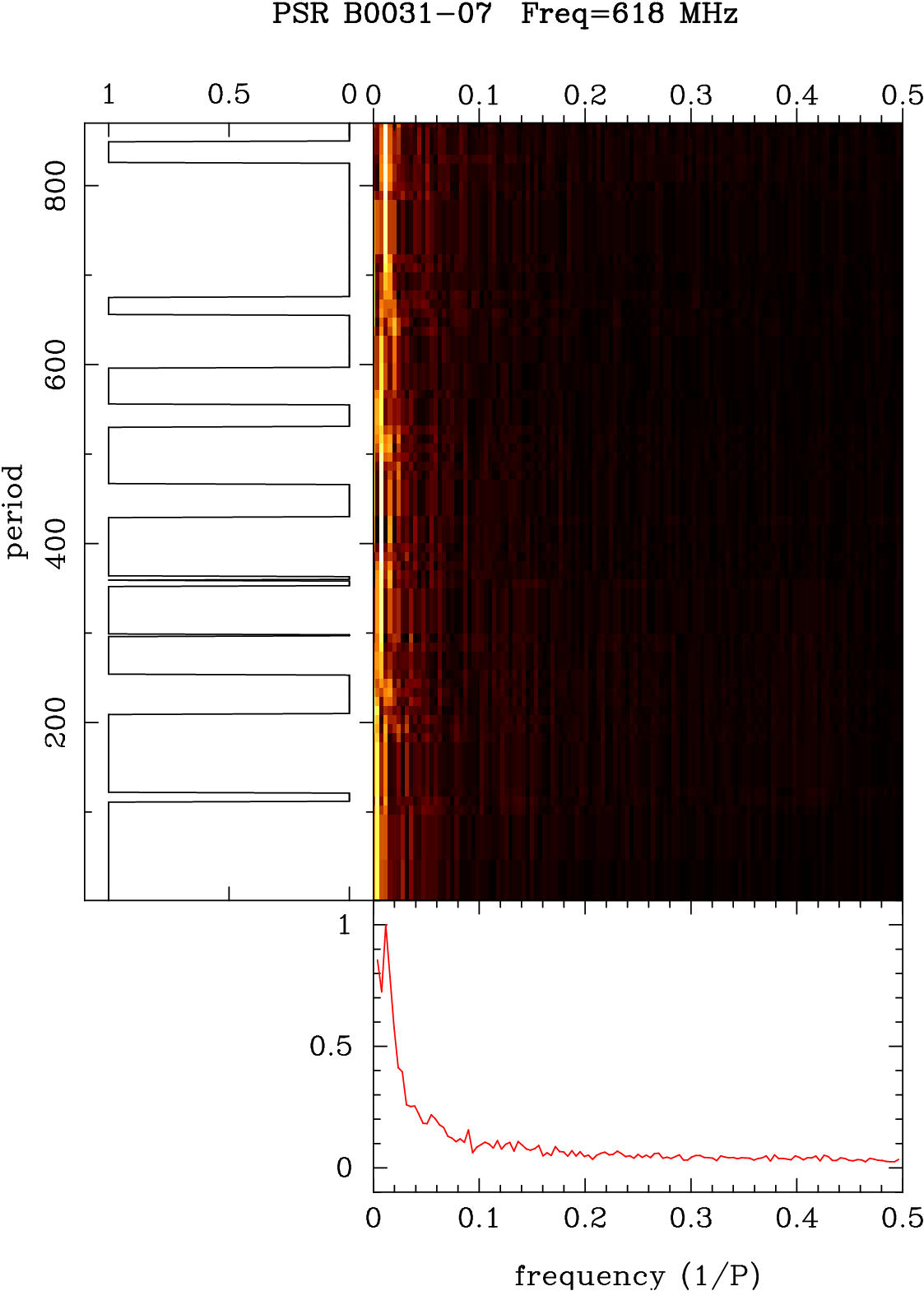}}}\\
\end{tabular}
\end{center}
\caption{The figure shows the time varying Fourier transform of the null/burst time series data at 333 MHz (left panel) and 618 MHz (right panel).}
\end{figure*}

\clearpage

\begin{figure*}
\begin{center}
\begin{tabular}{@{}lr@{}}
{\mbox{\includegraphics[angle=0,scale=0.4]{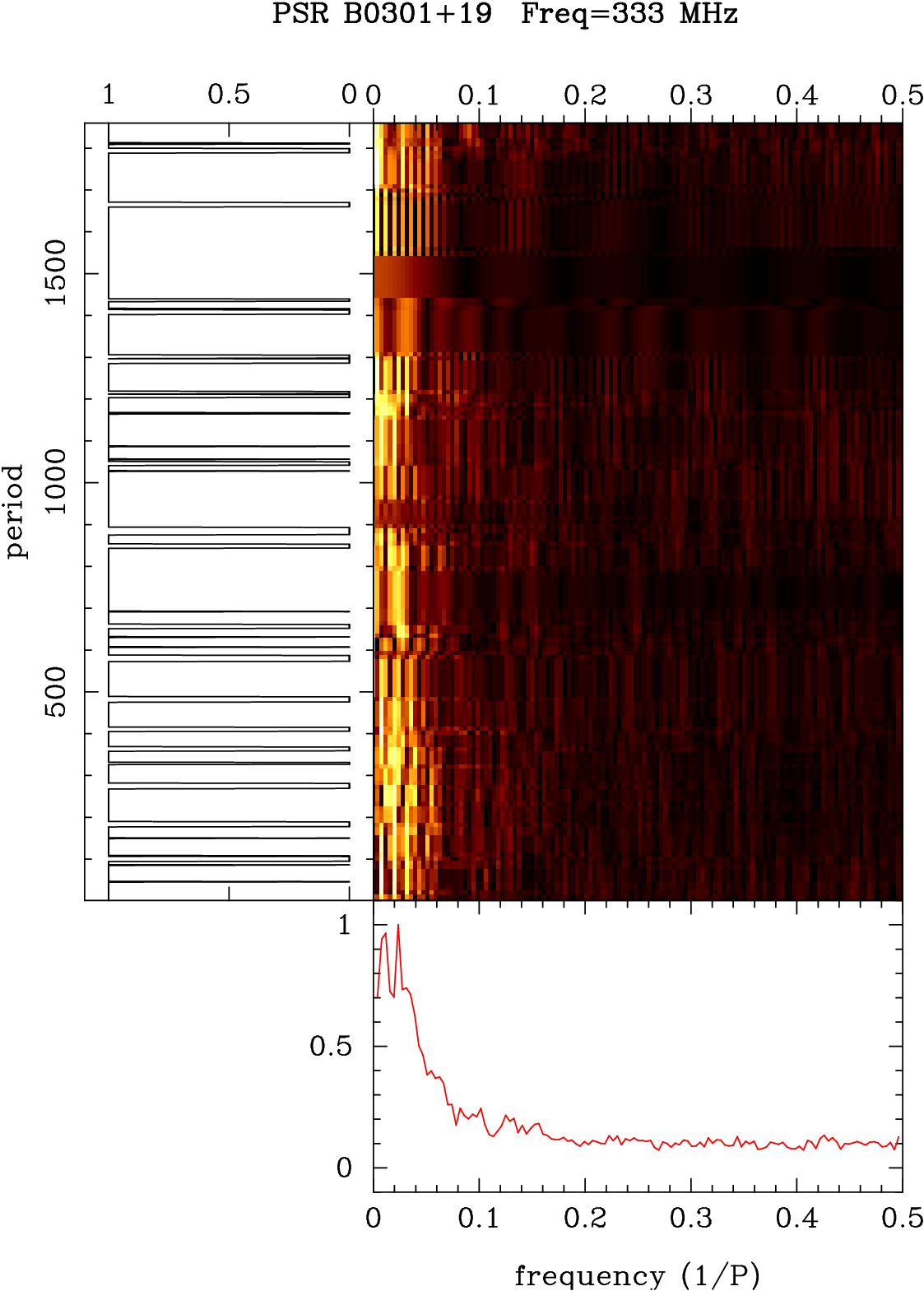}}}&
{\mbox{\includegraphics[angle=0,scale=0.4]{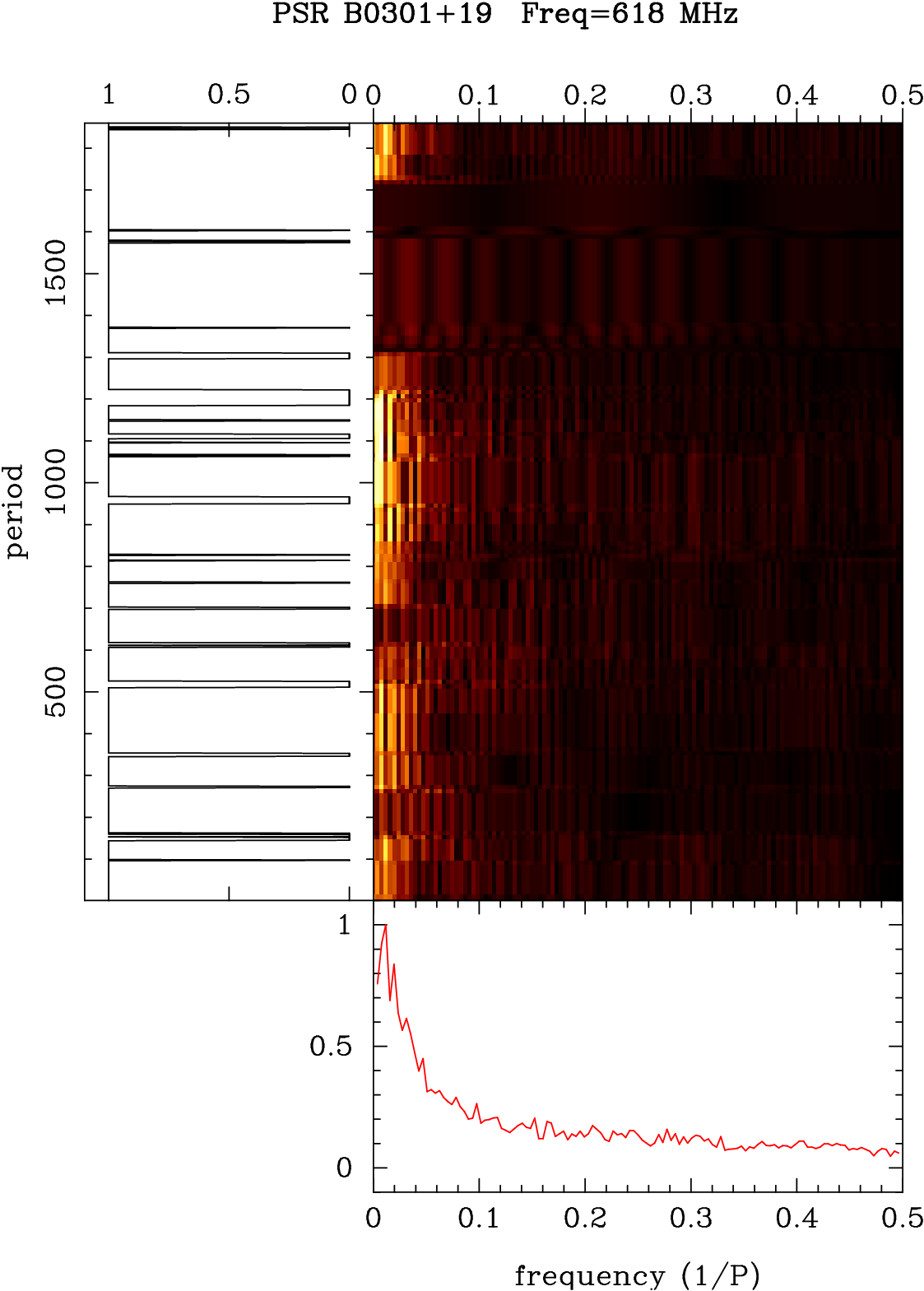}}}\\
\end{tabular}
\end{center}
\caption{The figure shows the time varying Fourier transform of the null/burst time series data at 333 MHz (left panel) and 618 MHz (right panel).}
\end{figure*}

\clearpage

\begin{figure*}
\begin{center}
\begin{tabular}{@{}lr@{}}
{\mbox{\includegraphics[angle=0,scale=0.4]{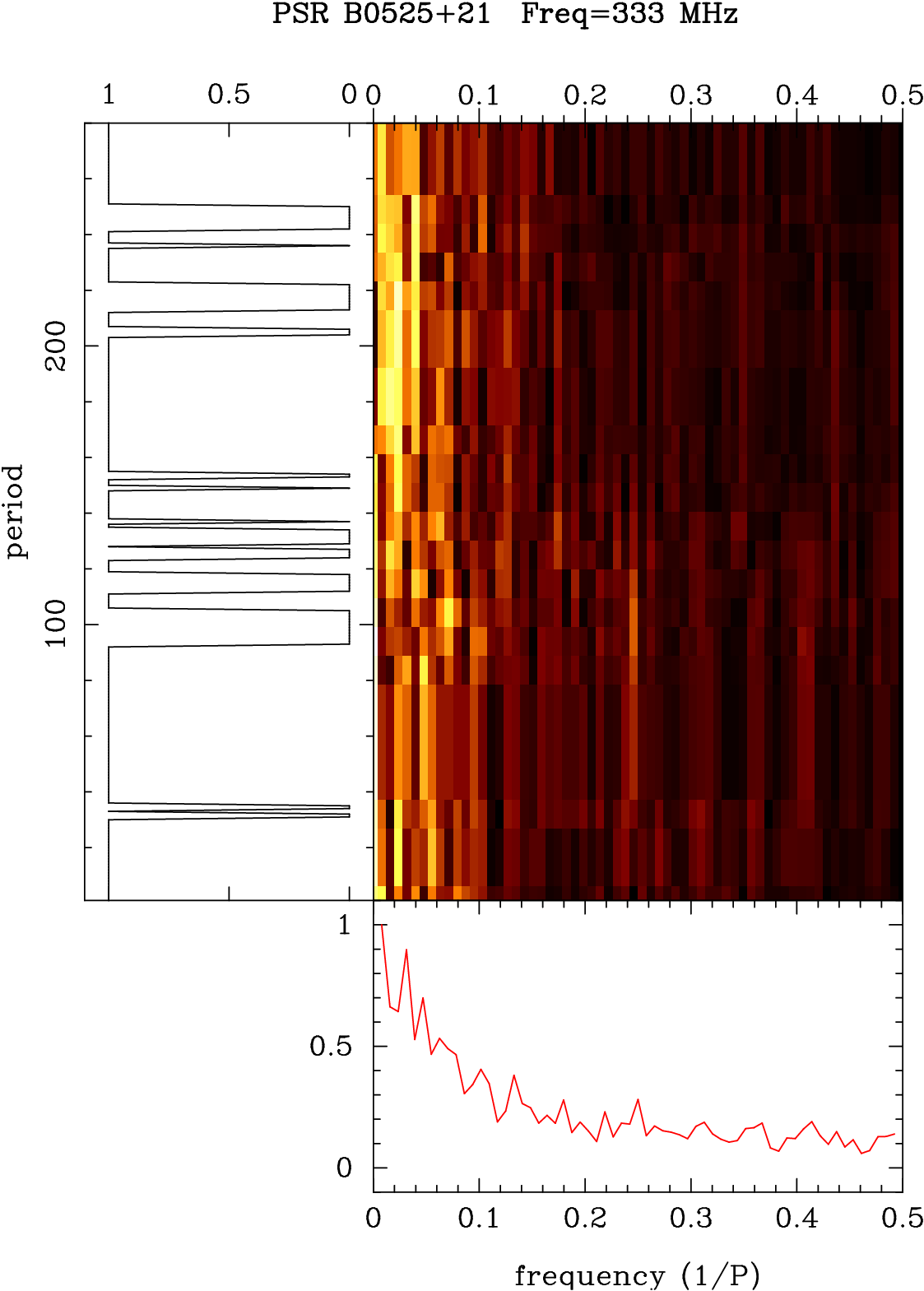}}}&
 \\
\end{tabular}
\end{center}
\caption{The figure shows the time varying Fourier transform of the null/burst time series data at 333 MHz.}
\end{figure*}

\clearpage

\begin{figure*}
\begin{center}
\begin{tabular}{@{}lr@{}}
{\mbox{\includegraphics[angle=0,scale=0.4]{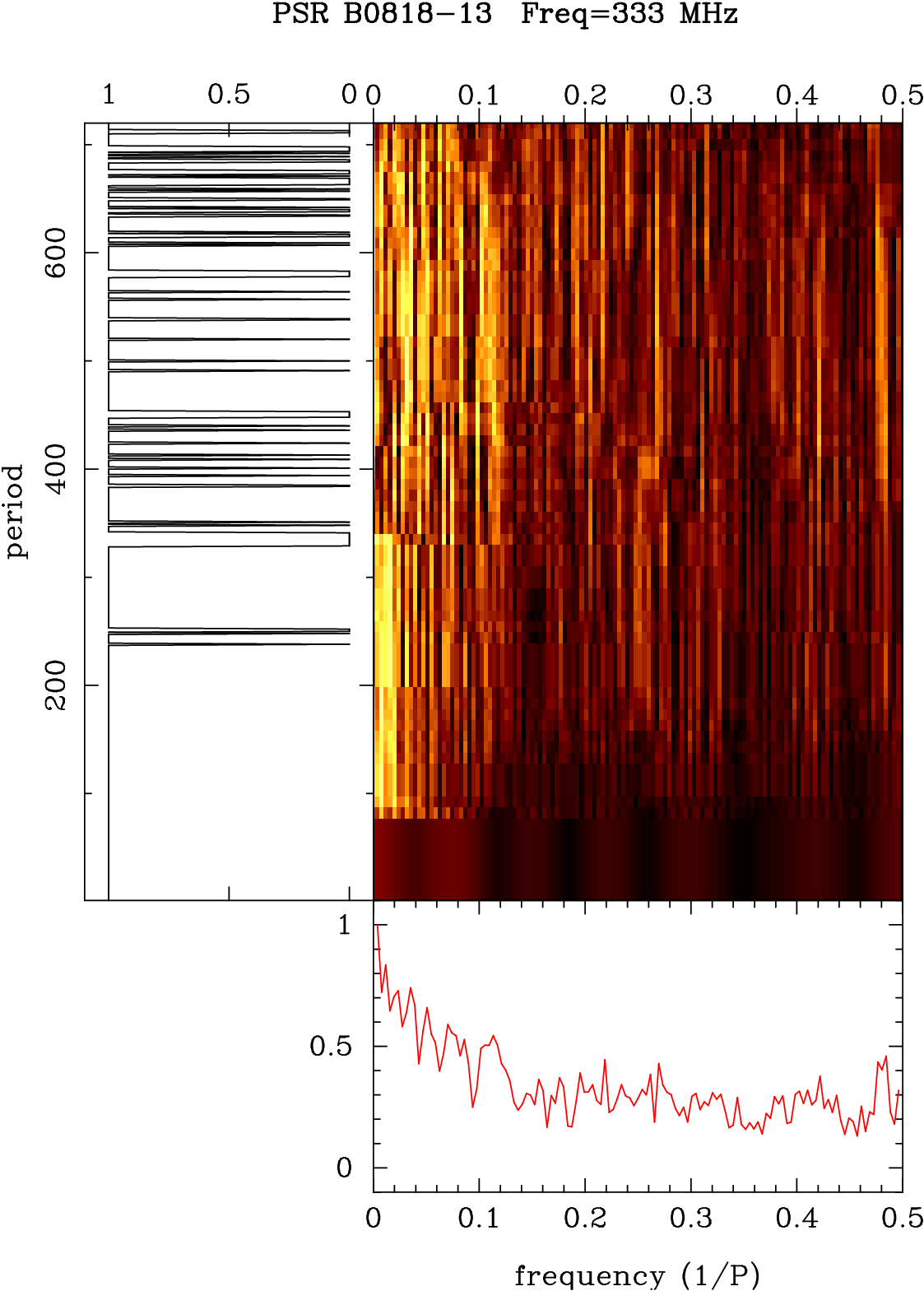}}}&
{}\\
\end{tabular}
\end{center}
\caption{The figure shows the time varying Fourier transform of the null/burst time series data at 333 MHz.}
\end{figure*}

\clearpage

\begin{figure*}
\begin{center}
\begin{tabular}{@{}lr@{}}
{\mbox{\includegraphics[angle=0,scale=0.4]{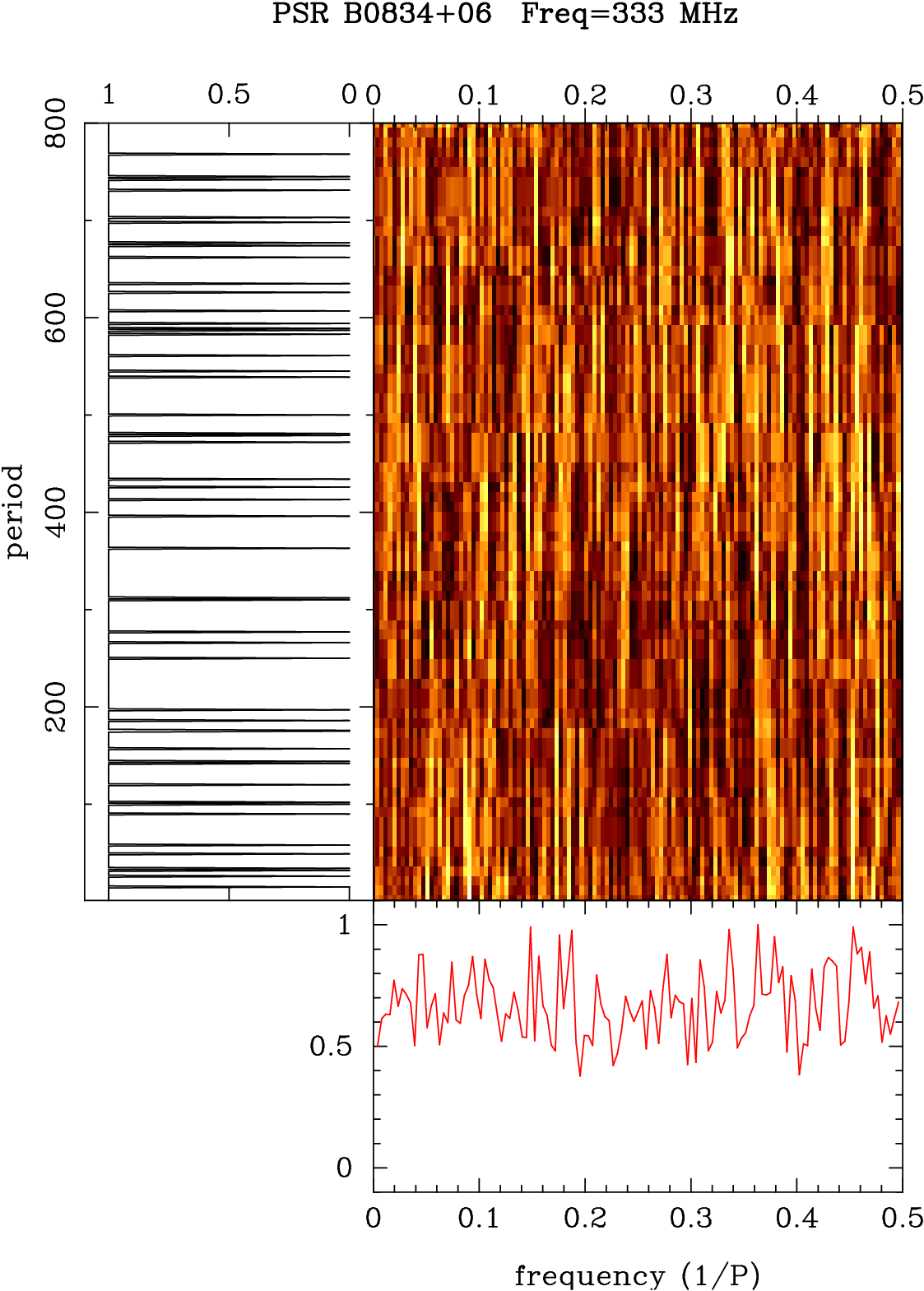}}}&
{\mbox{\includegraphics[angle=0,scale=0.4]{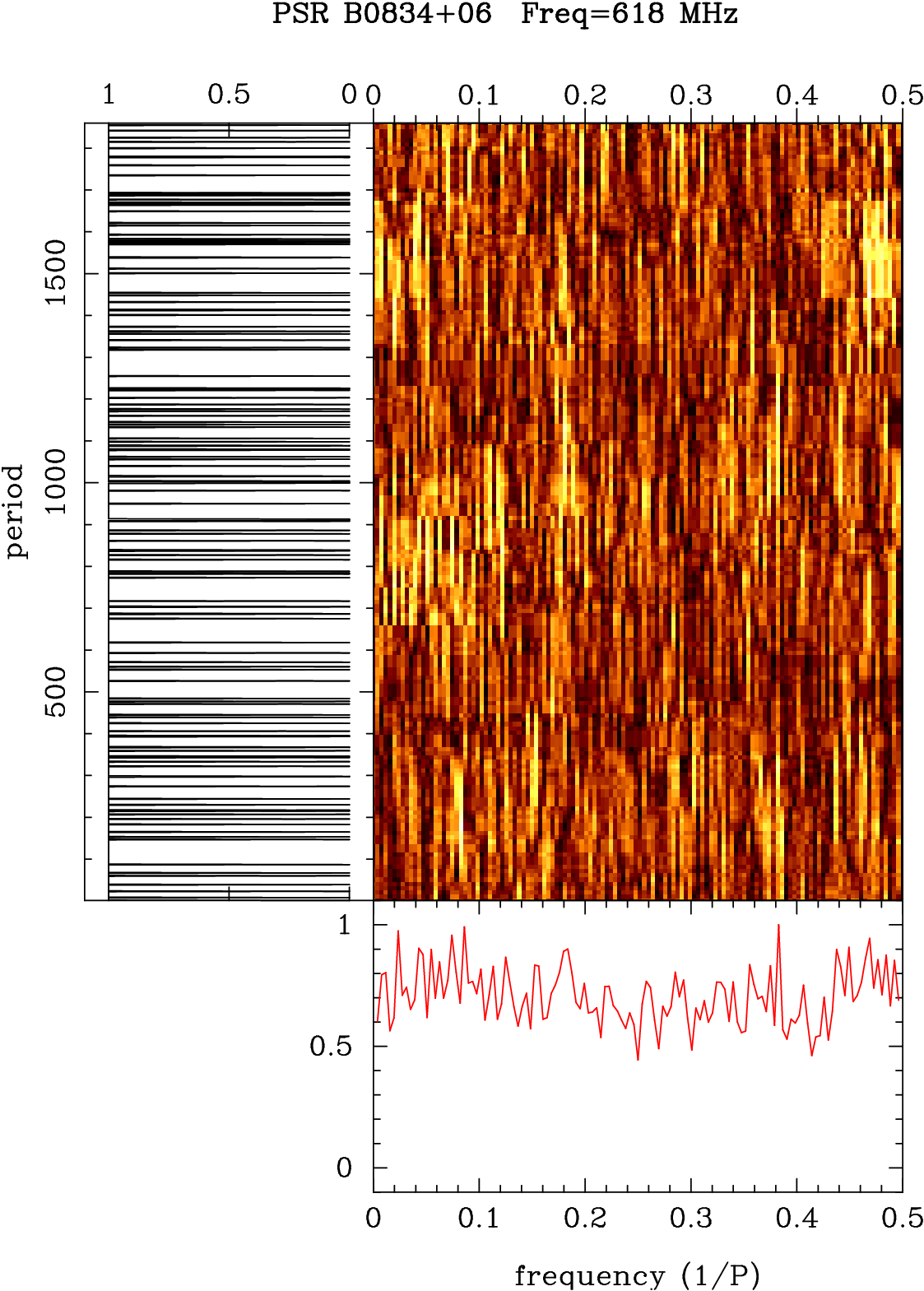}}}\\
\end{tabular}
\end{center}
\caption{The figure shows the time varying Fourier transform of the null/burst time series data at 333 MHz (left panel) and 618 MHz (right panel).}
\end{figure*}

\clearpage

\begin{figure*}
\begin{center}
\begin{tabular}{@{}lr@{}}
 &
{\mbox{\includegraphics[angle=0,scale=0.4]{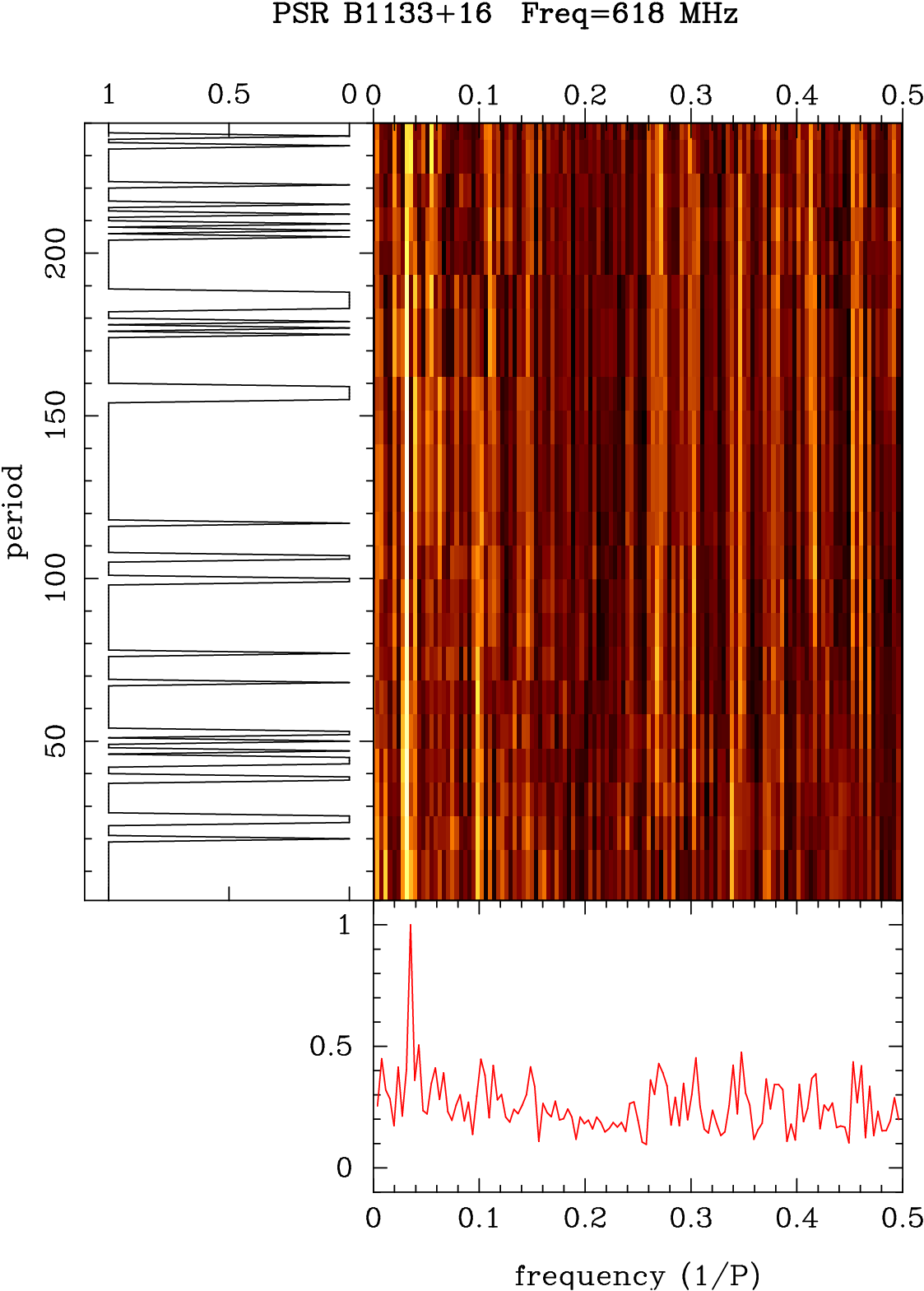}}}\\
\end{tabular}
\end{center}
\caption{The figure shows the time varying Fourier transform of the null/burst time series data at 618 MHz.}
\end{figure*}

\clearpage

\begin{figure*}
\begin{center}
\begin{tabular}{@{}lr@{}}
{\mbox{\includegraphics[angle=0,scale=0.4]{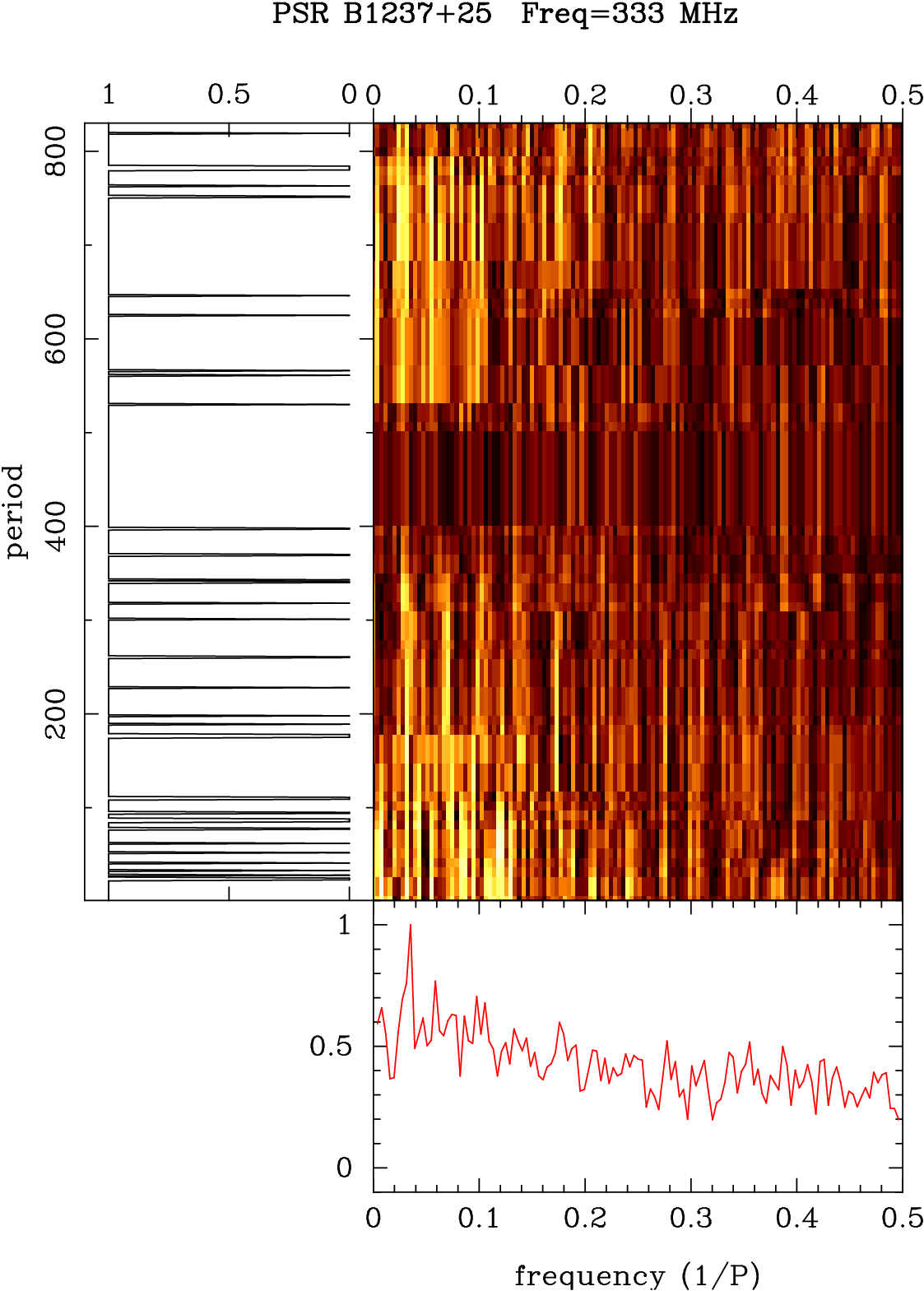}}}&
{\mbox{\includegraphics[angle=0,scale=0.4]{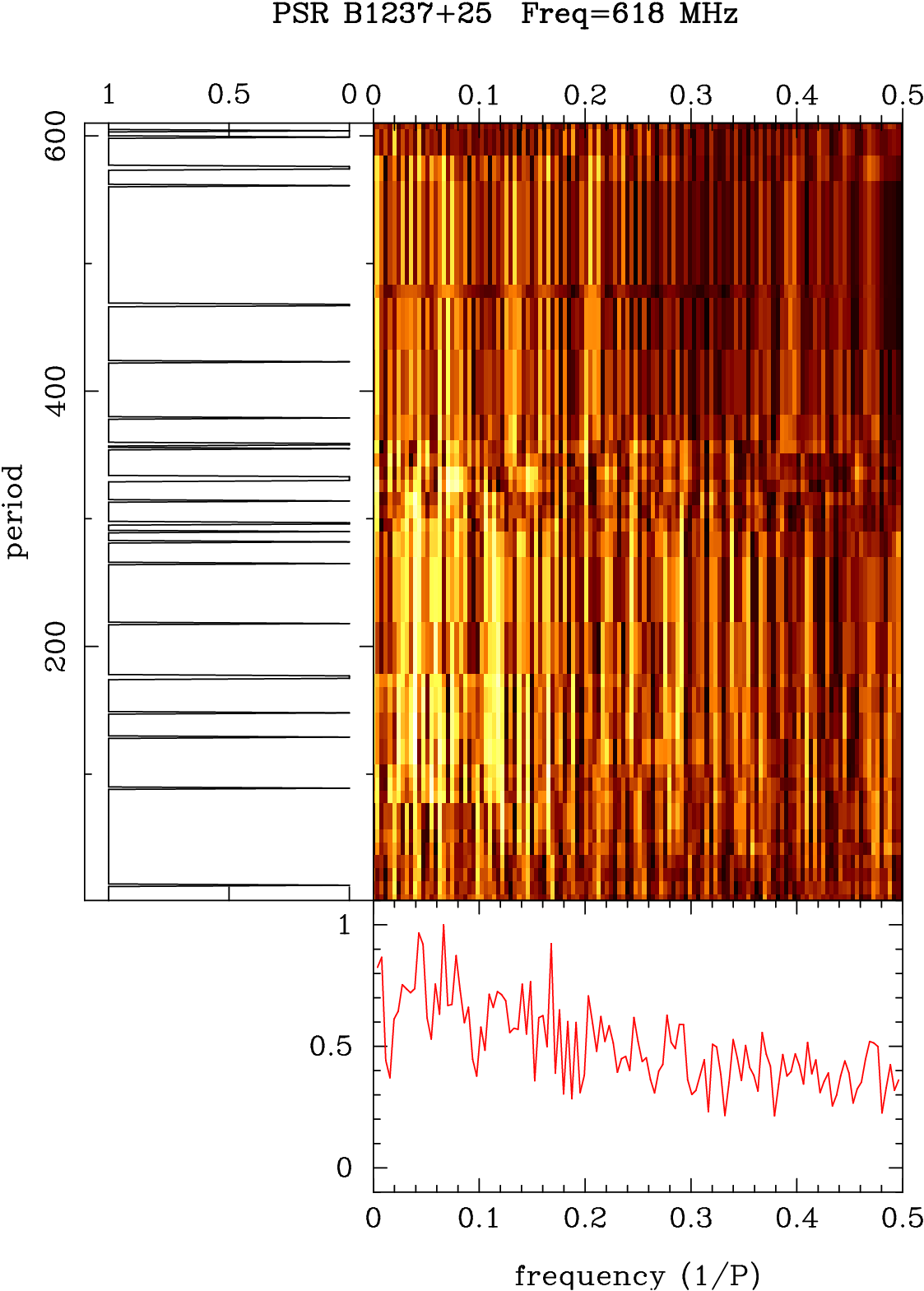}}}\\
\end{tabular}
\end{center}
\caption{The figure shows the time varying Fourier transform of the null/burst time series data at 333 MHz (left panel) and 618 MHz (right panel).}
\end{figure*}

\clearpage

\begin{figure*}
\begin{center}
\begin{tabular}{@{}lr@{}}
{\mbox{\includegraphics[angle=0,scale=0.4]{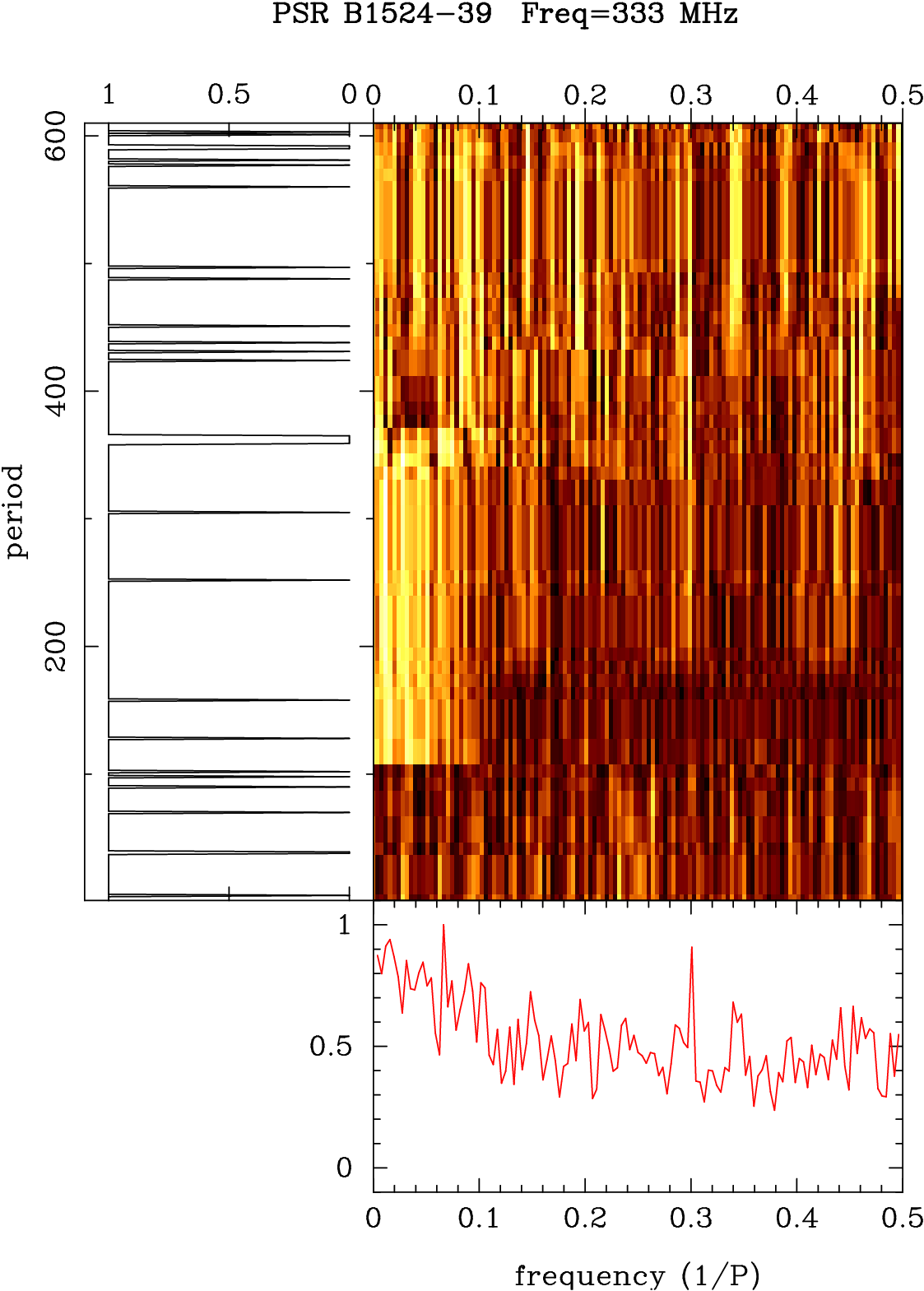}}}&
 \\
\end{tabular}
\end{center}
\caption{The figure shows the time varying Fourier transform of the null/burst time series data at 333 MHz.}
\end{figure*}

\clearpage

\begin{figure*}
\begin{center}
\begin{tabular}{@{}lr@{}}
{\mbox{\includegraphics[angle=0,scale=0.4]{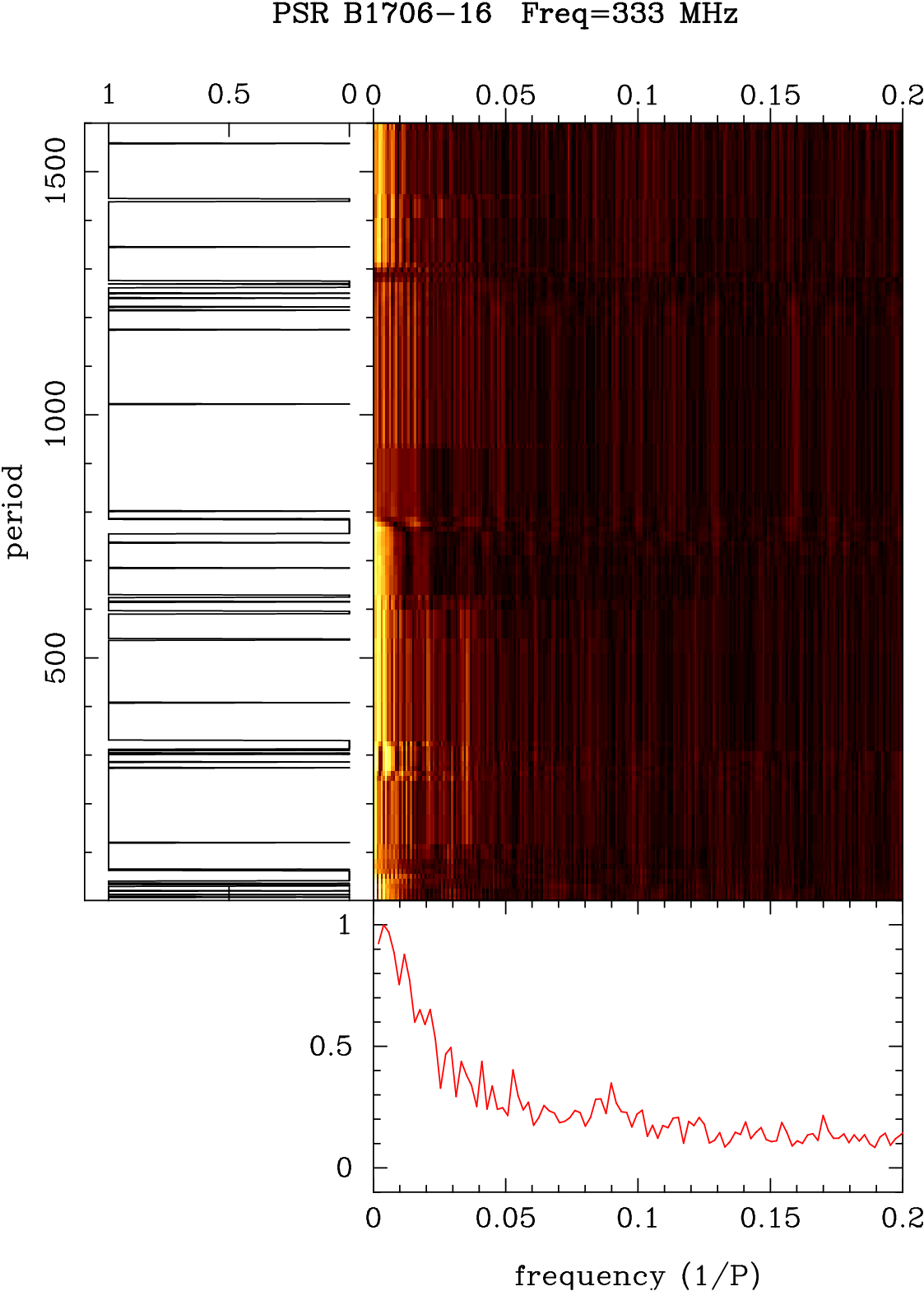}}}&
{\mbox{\includegraphics[angle=0,scale=0.4]{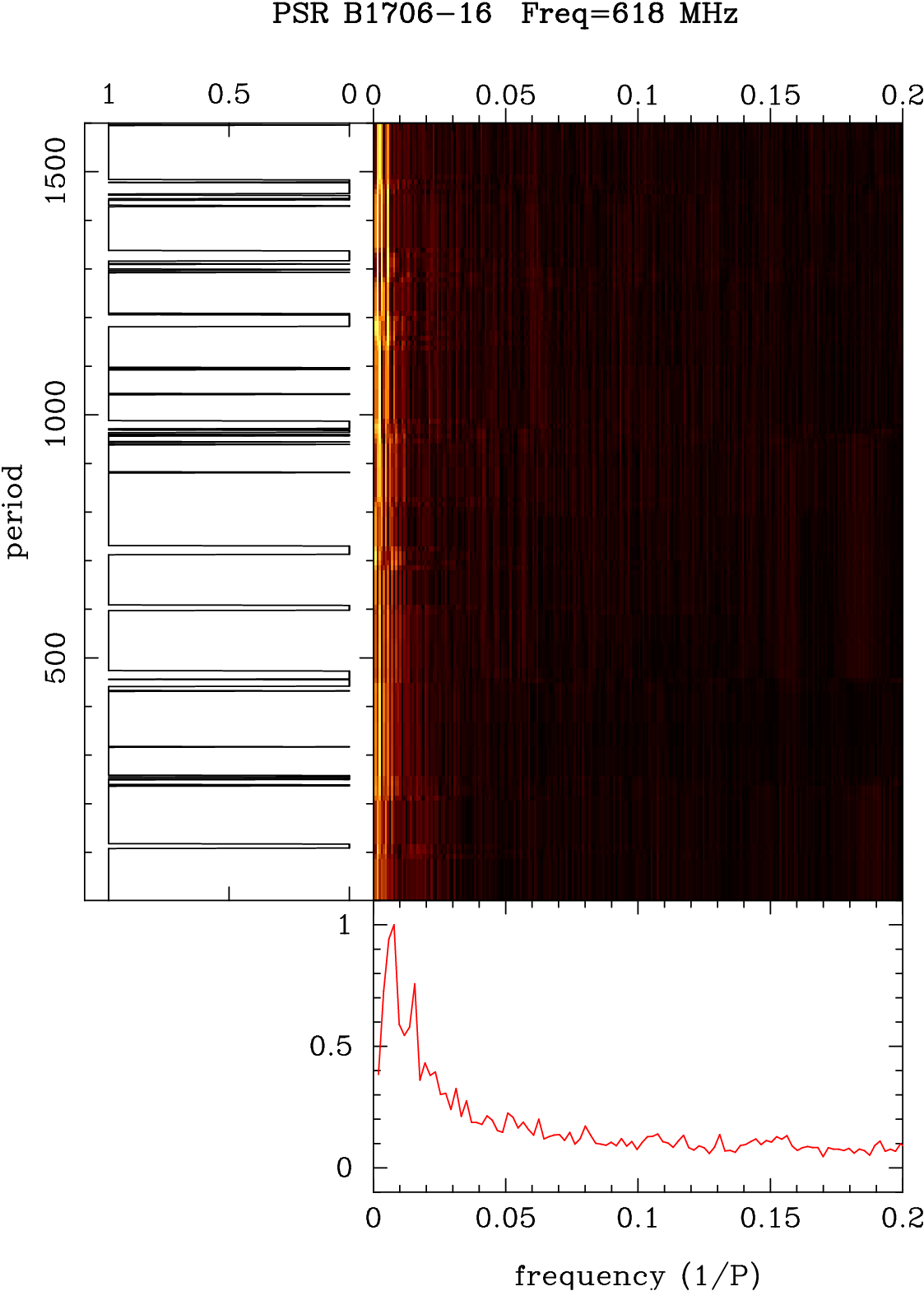}}}\\
\end{tabular}
\end{center}
\caption{The figure shows the time varying Fourier transform of the null/burst time series data at 333 MHz (left panel) and 618 MHz (right panel).}
\end{figure*}

\clearpage

\begin{figure*}
\begin{center}
\begin{tabular}{@{}lr@{}}
 &
{\mbox{\includegraphics[angle=0,scale=0.4]{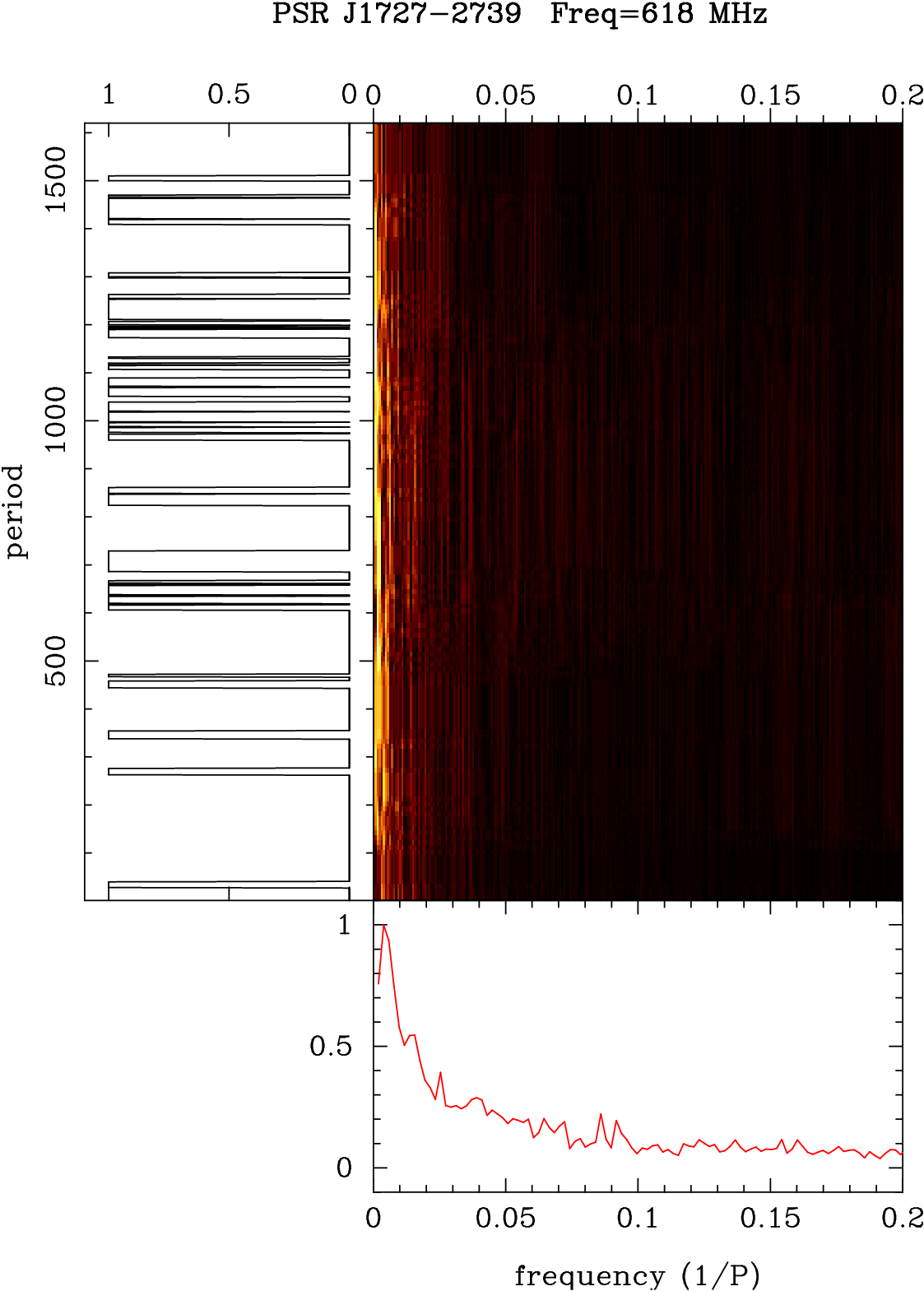}}}\\
\end{tabular}
\end{center}
\caption{The figure shows the time varying Fourier transform of the null/burst time series data at 618 MHz.}
\end{figure*}

\clearpage

\begin{figure*}
\begin{center}
\begin{tabular}{@{}lr@{}}
{\mbox{\includegraphics[angle=0,scale=0.4]{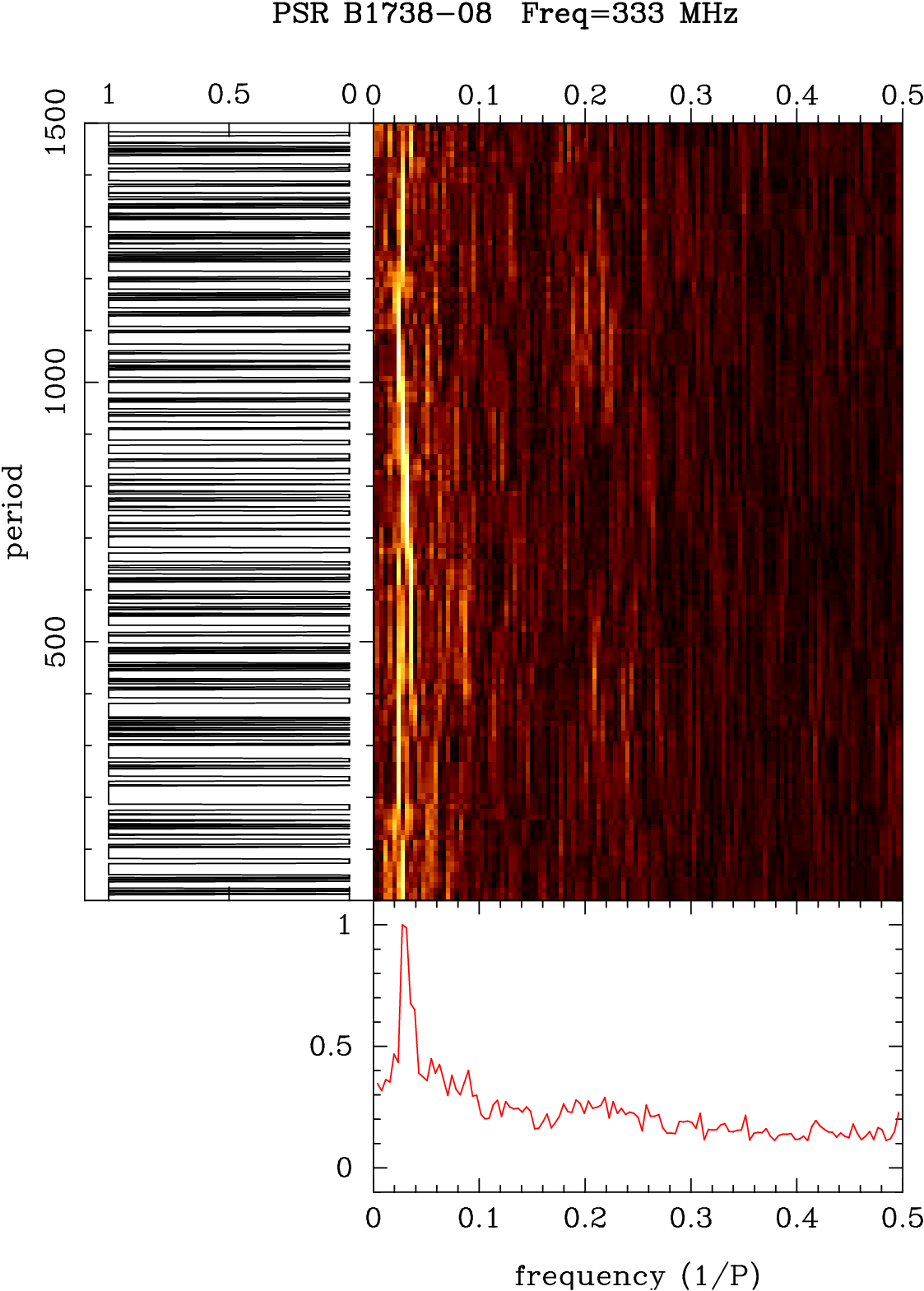}}}&
{\mbox{\includegraphics[angle=0,scale=0.4]{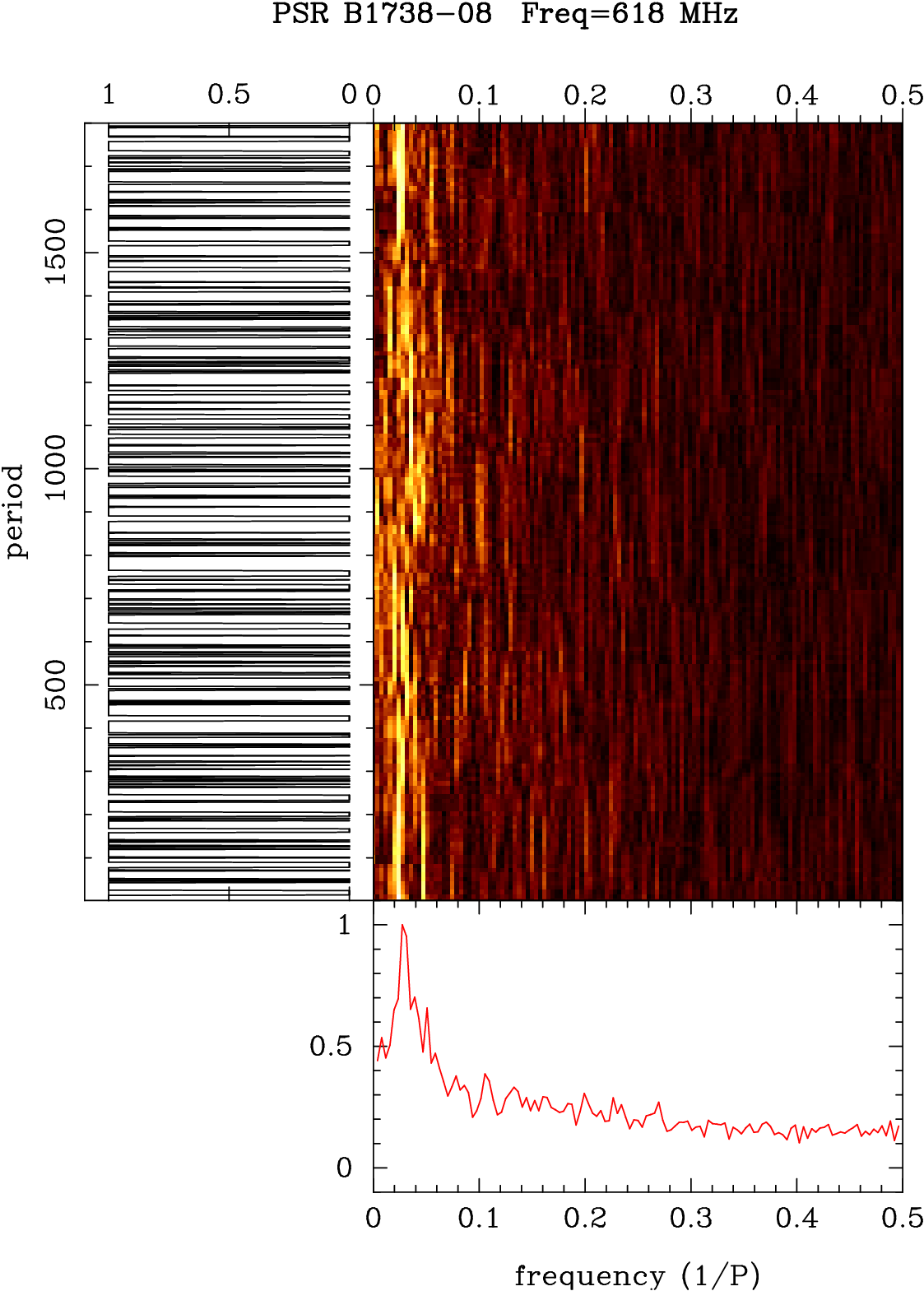}}}\\
\end{tabular}
\end{center}
\caption{The figure shows the time varying Fourier transform of the null/burst time series data at 333 MHz (left panel) and 618 MHz (right panel).}
\end{figure*}

\clearpage

\begin{figure*}
\begin{center}
\begin{tabular}{@{}lr@{}}
{\mbox{\includegraphics[angle=0,scale=0.4]{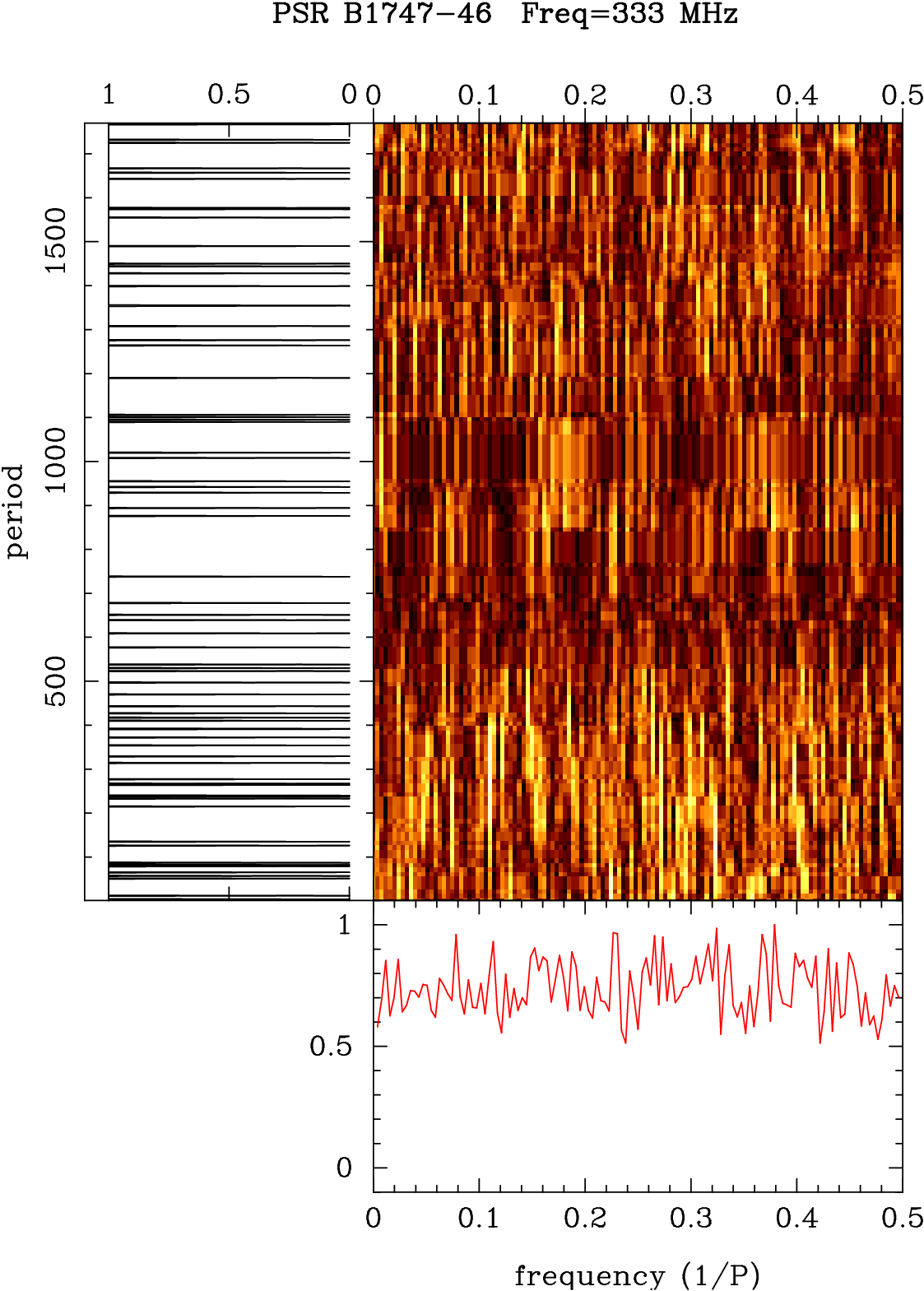}}}&
{\mbox{\includegraphics[angle=0,scale=0.4]{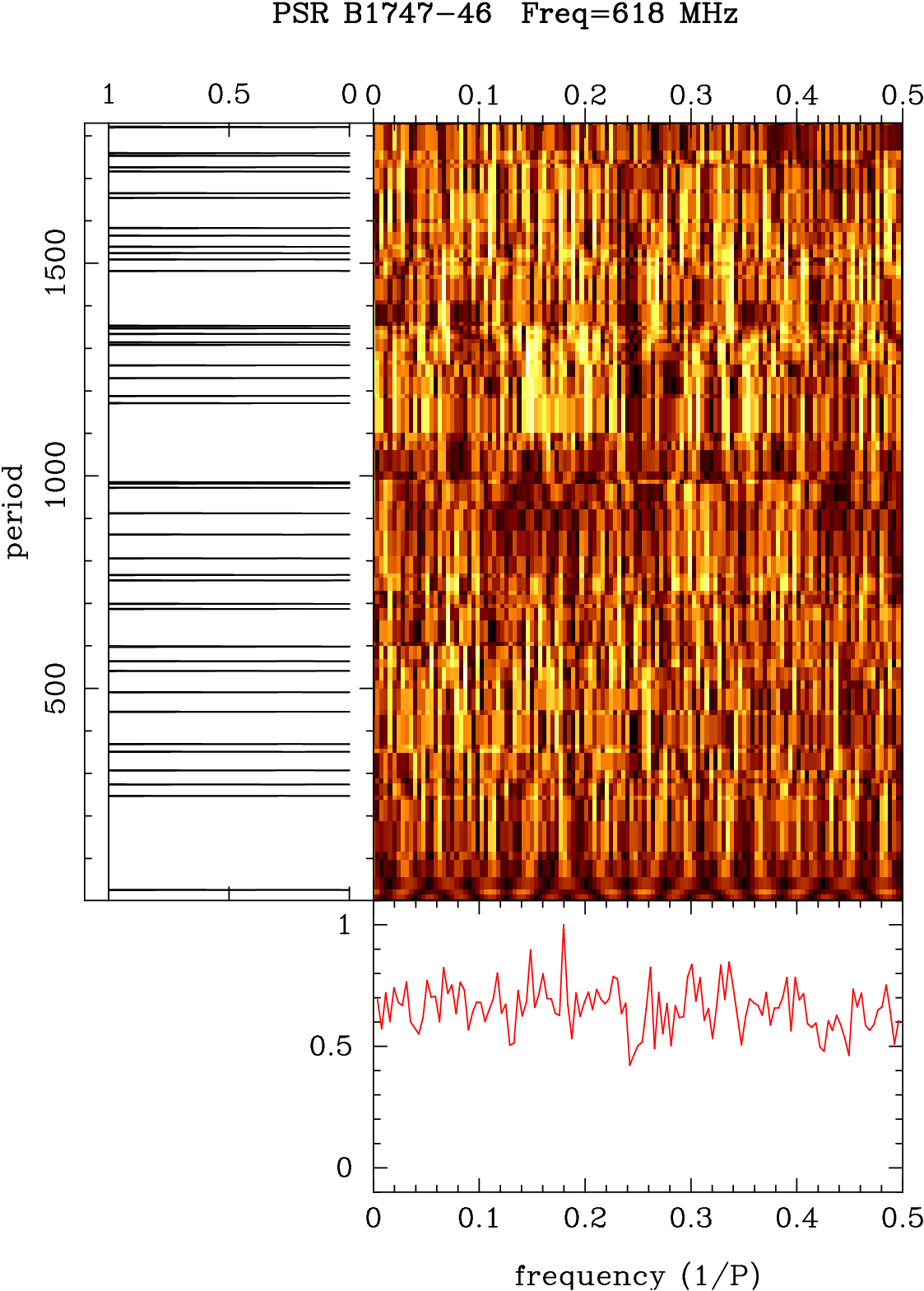}}}\\
\end{tabular}
\end{center}
\caption{The figure shows the time varying Fourier transform of the null/burst time series data at 333 MHz (left panel) and 618 MHz (right panel).}
\end{figure*}

\clearpage

\begin{figure*}
\begin{center}
\begin{tabular}{@{}lr@{}}
{\mbox{\includegraphics[angle=0,scale=0.4]{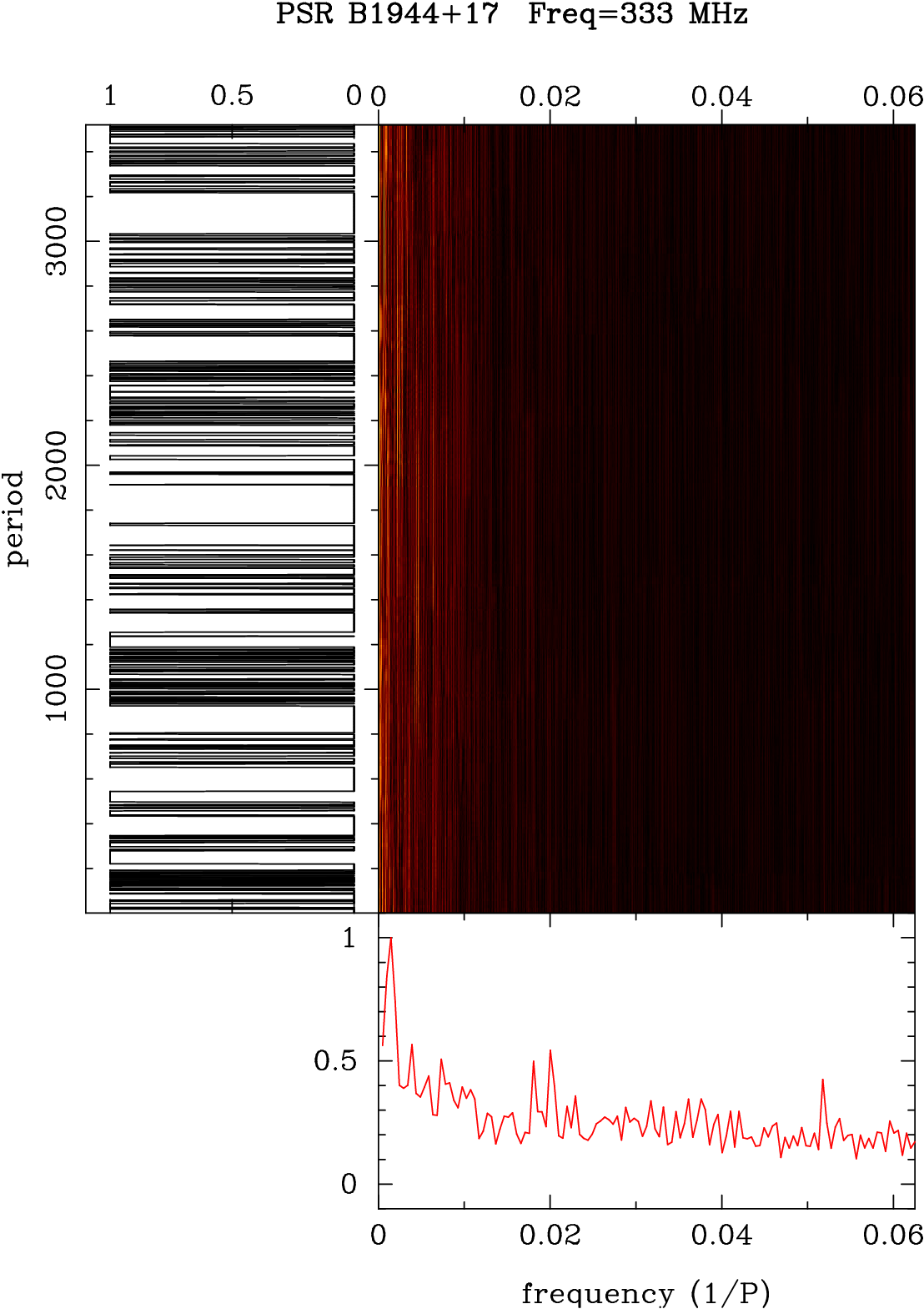}}}&
{\mbox{\includegraphics[angle=0,scale=0.4]{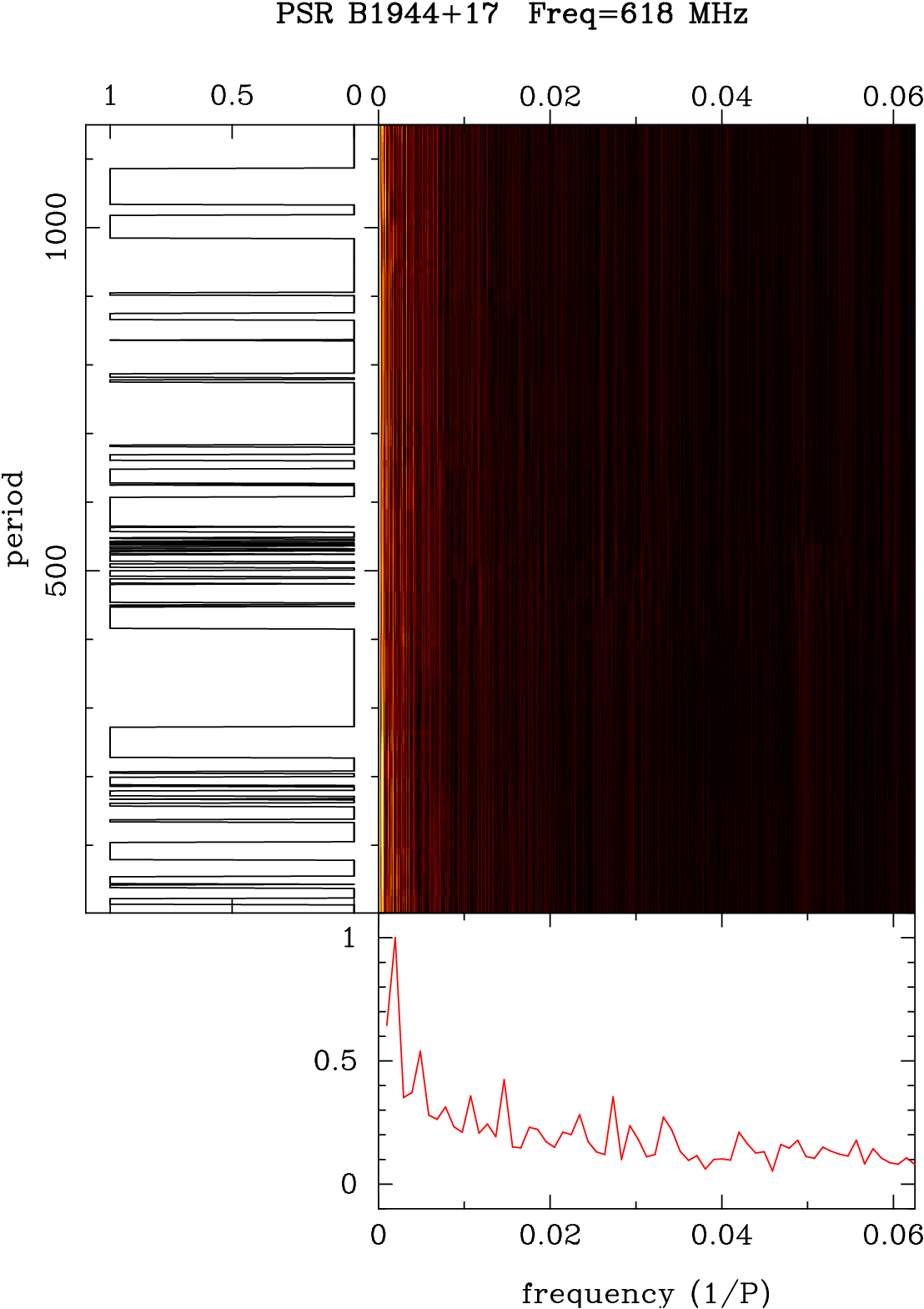}}}\\
\end{tabular}
\end{center}
\caption{The figure shows the time varying Fourier transform of the null/burst time series data at 333 MHz (left panel) and 618 MHz (right panel).}
\end{figure*}

\clearpage

\begin{figure*}
\begin{center}
\begin{tabular}{@{}lr@{}}
{\mbox{\includegraphics[angle=0,scale=0.4]{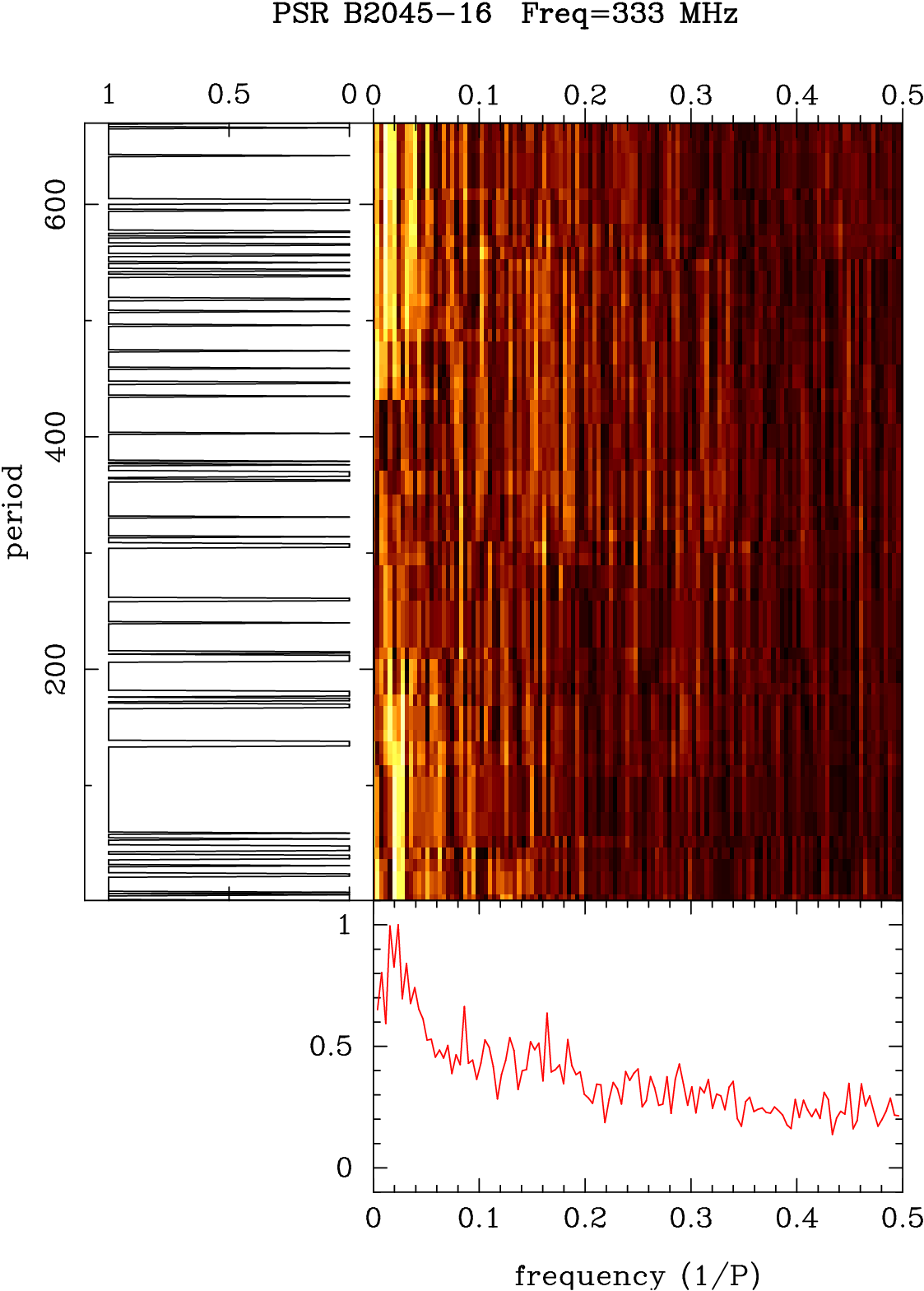}}}&
{\mbox{\includegraphics[angle=0,scale=0.4]{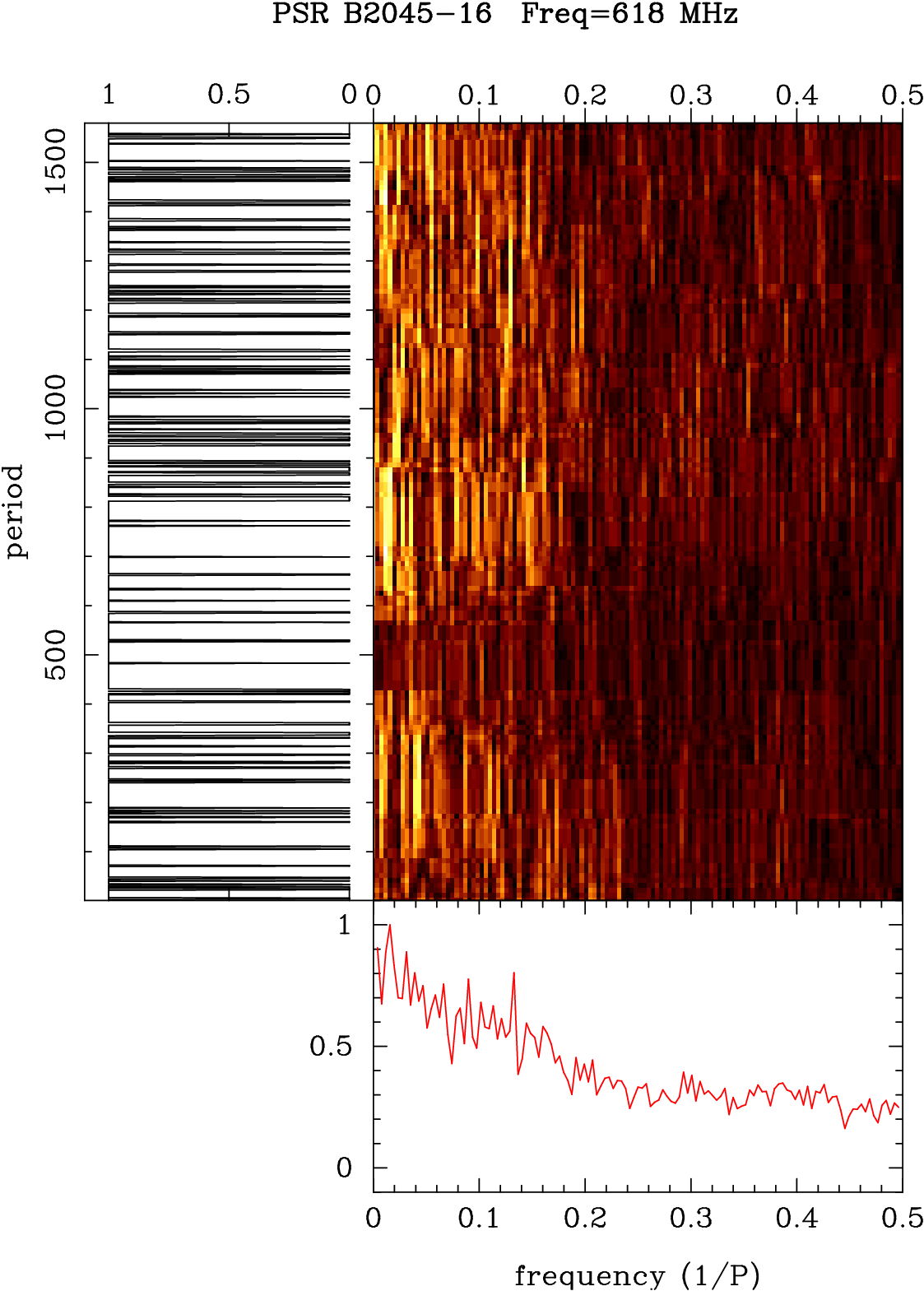}}}\\
\end{tabular}
\end{center}
\caption{The figure shows the time varying Fourier transform of the null/burst time series data at 333 MHz (left panel) and 618 MHz (right panel).}
\end{figure*}

\clearpage

\begin{figure*}
\begin{center}
\begin{tabular}{@{}lr@{}}
{\mbox{\includegraphics[angle=0,scale=0.4]{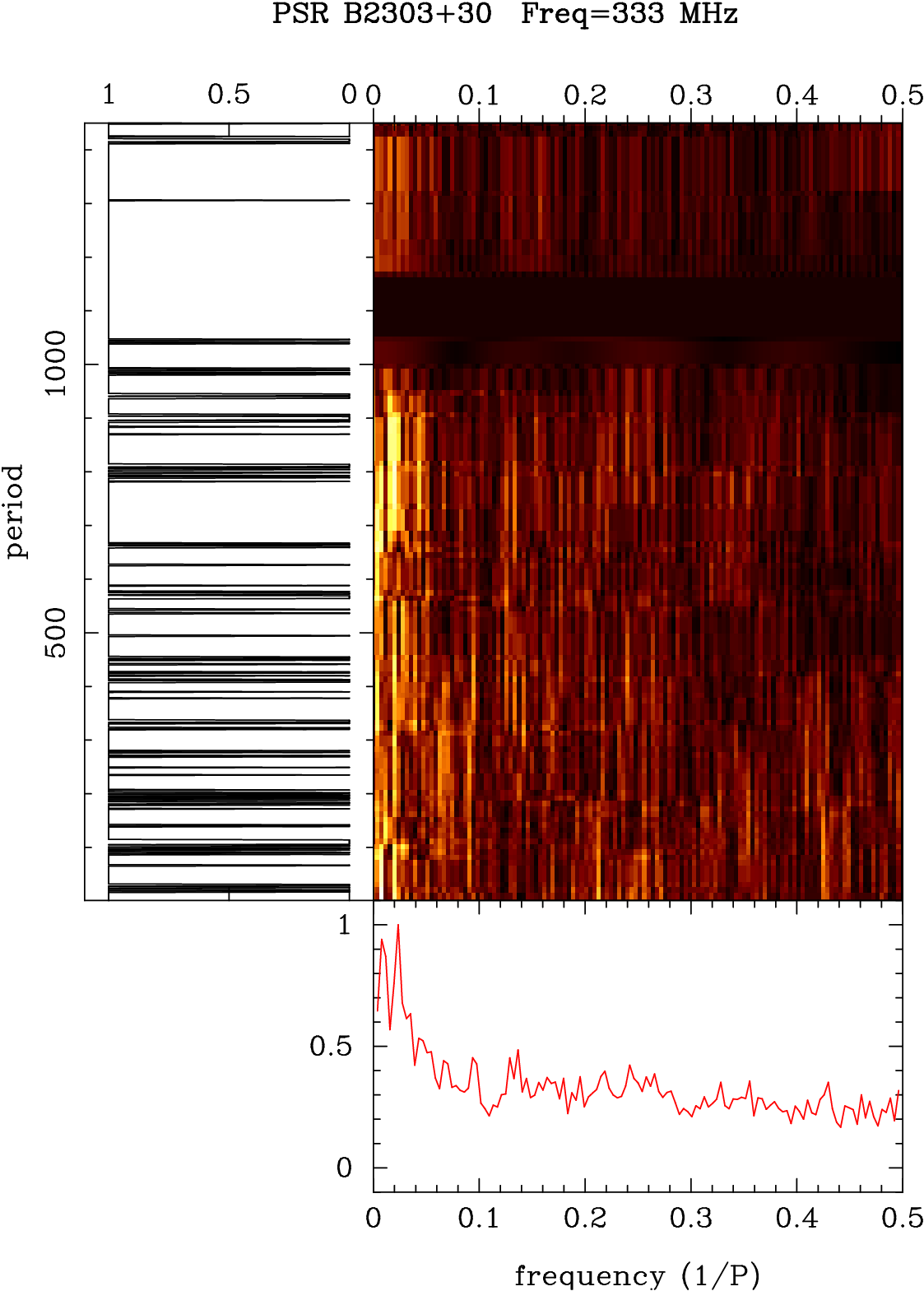}}}&
 \\
\end{tabular}
\end{center}
\caption{The figure shows the time varying Fourier transform of the null/burst time series data at 333 MHz.}
\end{figure*}

\clearpage

\begin{figure*}
\begin{center}
\begin{tabular}{@{}lr@{}}
{\mbox{\includegraphics[angle=0,scale=0.4]{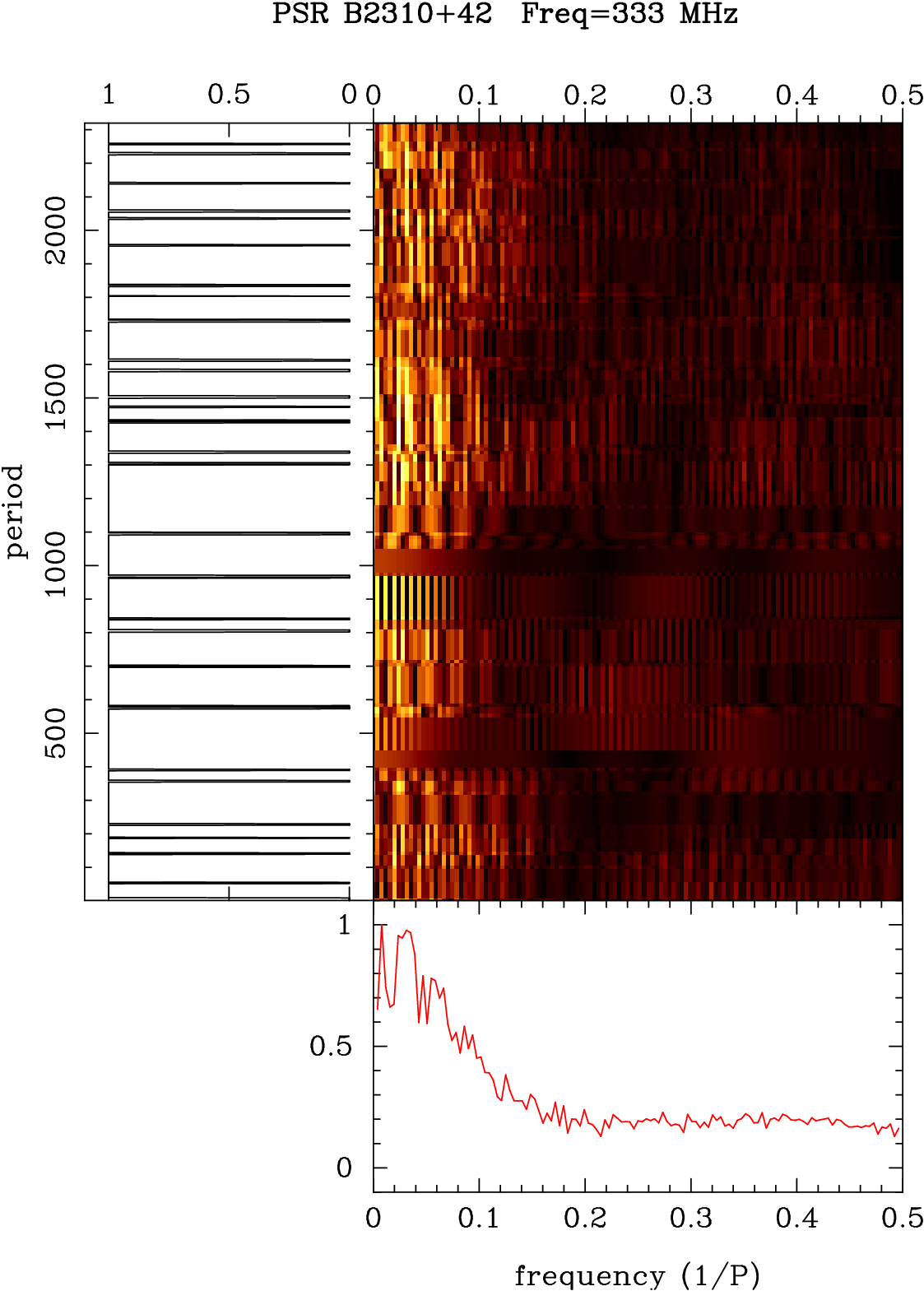}}}&
 \\
\end{tabular}
\end{center}
\caption{The figure shows the time varying Fourier transform of the null/burst time series data at 333 MHz.}
\end{figure*}

\clearpage

\begin{figure*}
\begin{center}
\begin{tabular}{@{}lr@{}}
{\mbox{\includegraphics[angle=0,scale=0.4]{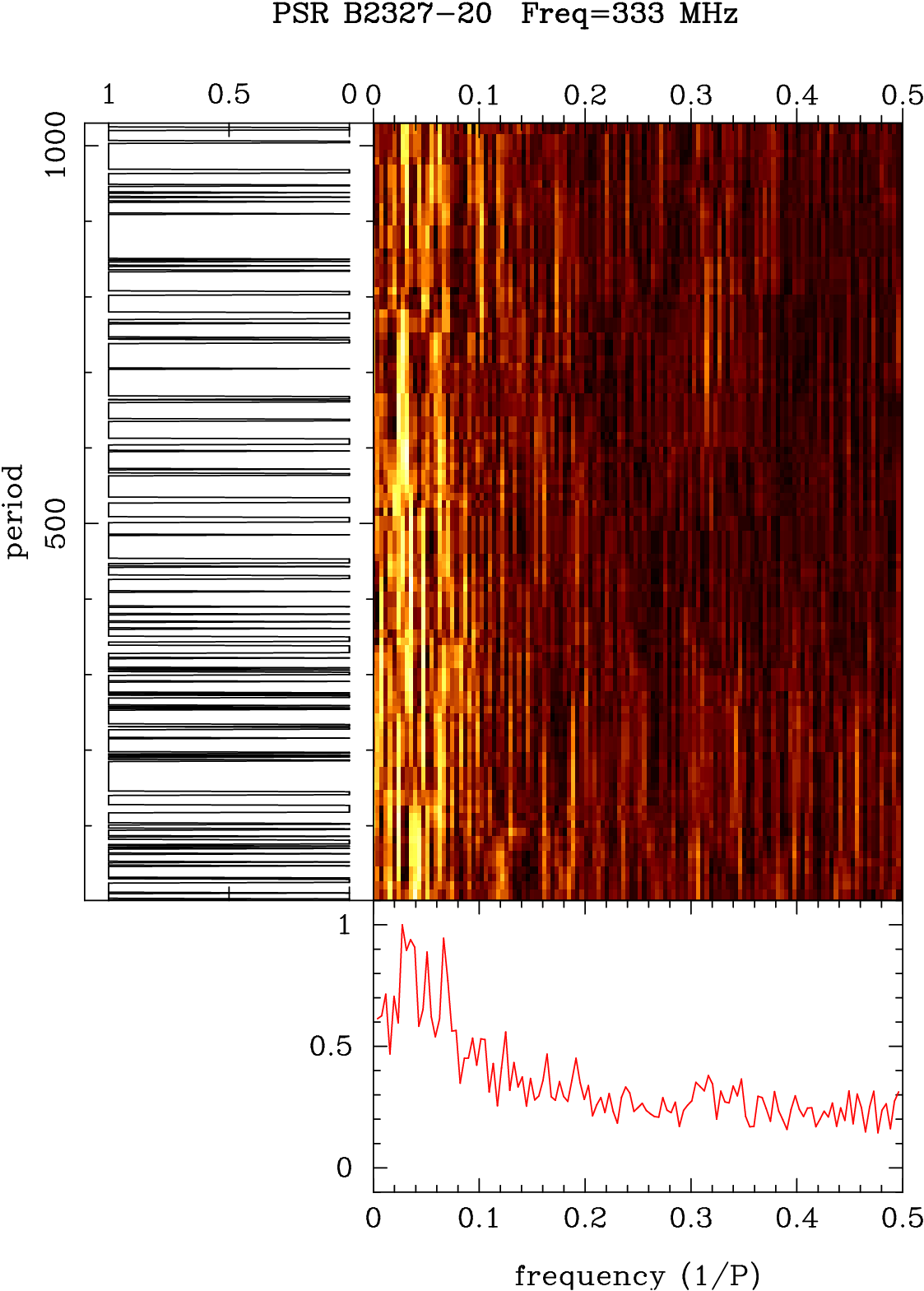}}}&
{\mbox{\includegraphics[angle=0,scale=0.4]{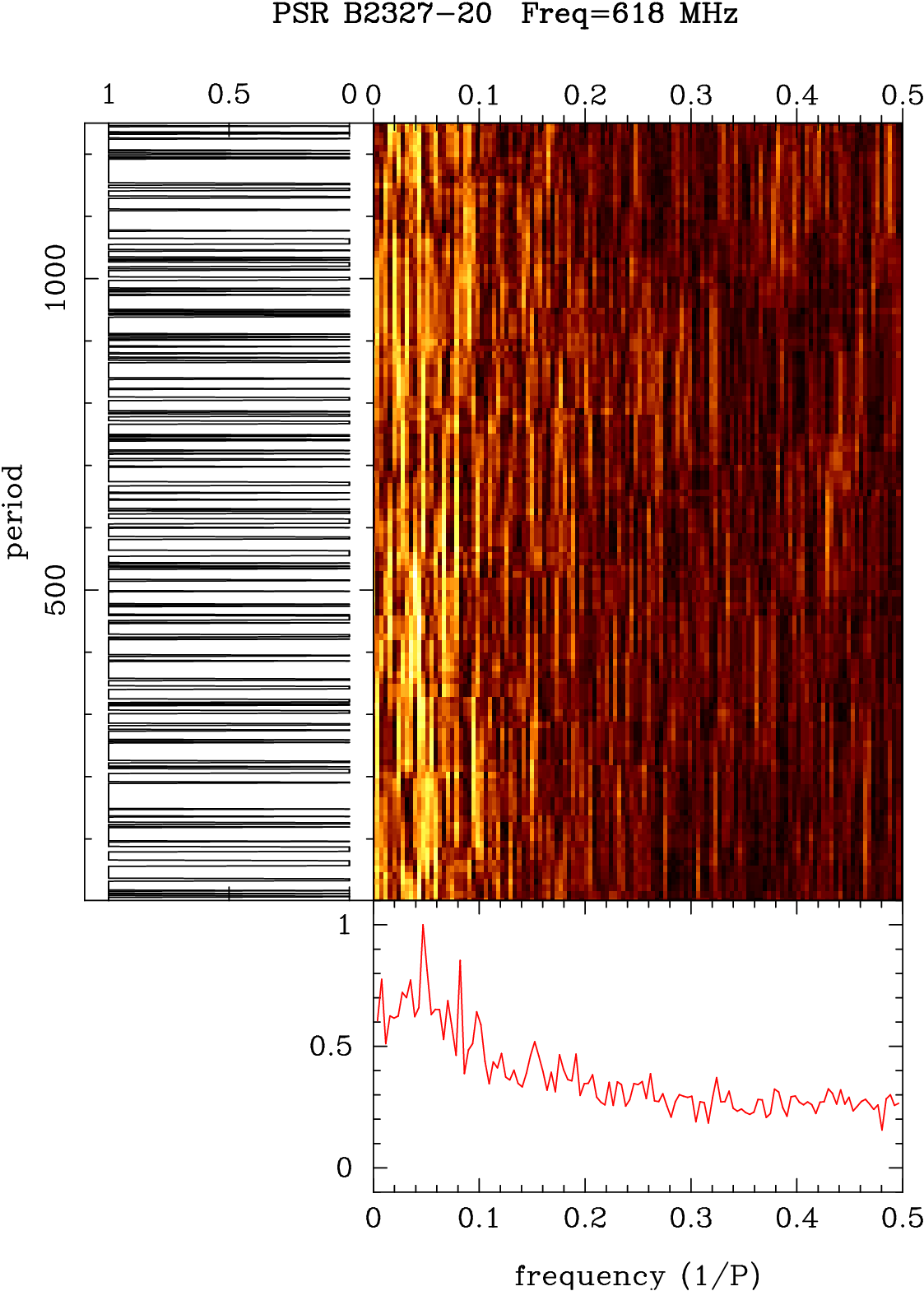}}}\\
\end{tabular}
\end{center}
\caption{The figure shows the time varying Fourier transform of the null/burst time series data at 333 MHz (left panel) and 618 MHz (right panel).}
\end{figure*}